\documentclass{LMCS}

\def\dOi{11(2:14)2015}
\lmcsheading%
{\dOi}
{1--44}
{}
{}
{Oct.~14, 2013}
{Jun.~25, 2015}
{}

\ACMCCS{[{\bf Theory of computation}]: Logic\,/\,Models of computation}
\subjclass{F.1.1, F.4.1}

\usepackage[all]{xy}
\usepackage{hyperref}
\usepackage{amssymb, latexsym, amscd, amsmath,stmaryrd,wrapfig}
\usepackage{wasysym}
\newcommand{\rightdcirc}{\makebox[1.1\width][l]{\ensuremath{%
\longrightarrow%
\makebox{$\mkern-24mu\color{white}{\bullet}\mkern+12mu$}%
\makebox{$\mkern-21mu\circ\mkern+10mu$}%
\ignorespacesafterend}}}%
\numberwithin{equation}{section}

\begin{document}
\title[Weak bisimulation for coalgebras over order enriched monads]{Weak bisimulation for coalgebras over order enriched monads}
\author[T. Brengos]{Tomasz Brengos}
\email{t.brengos@mini.pw.edu.pl}
\keywords{coalgebra, bisimulation, weak bisimulation, delay bisimulation, logic}
\thanks{This work has been supported by the grant of Warsaw University of Technology no.  504P/1120/0136/000}

\address{Faculty of Mathematics and Information Sciences\\
         Warsaw University of Technology\\ 
         Koszykowa~75 \\       
         00-662 Warszawa, Poland}

\keywords{coalgebra, bisimulation, weak bisimulation, saturation, monad, ordered saturation monad, logic, topology, internal transition, Kleisli algebra, silent transition}

\begin{abstract}
The paper introduces the notion of  a weak bisimulation for coalgebras whose type is a monad satisfying some extra properties. In the first part of the paper we argue that systems with silent moves should be modelled coalgebraically as coalgebras whose type is a monad. We show that the visible and invisible part of the functor can be handled internally inside a monadic structure. In the second part we introduce the notion of an ordered saturation monad, study its properties, and show that it allows us to present two approaches towards defining weak bisimulation for coalgebras and compare them.  We support the framework presented in this paper by two main examples of models: labelled transition systems and simple Segala systems. 
\end{abstract}

%%%%%%%%%%TITLE%%%%%%%%%%%%%%%%%%%%

\maketitle
\tableofcontents
%%%%%%%%%%%%%%%%%%%%%%%%%%%%%%%%%%%
\section{Introduction}\label{section:introduction}
%
%The notion of strong bisimulation for different transition systems plays an important role in theoretical computer science. However, sometimes it is useful to consider a part of computation branch of a process that is invisible and is allowed to take several steps. In this case the aforementioned relation proves to be too strong an equivalence. is allowed to take several steps and in some sense remain neutral to its structure. For instance, 

The notion of strong bisimilarity for different transition systems plays an important role in theoretical computer science. However, in some cases this relation proves to be too strong an equivalence. For instance in  Milner's Calculus of Communicating Systems \cite{Milner3,Sangiorgi11}, which is among the most widely studied process calculi, one considers a special computation branch that is silent. This special branch, sometimes also called {invisible} or {internal}, is allowed to take several steps and in some sense remain neutral to the structure of a process. Strong bisimulation treats all parts of computation equally and does not distinguish between visible and invisible steps.  The desirable behavioural equivalence should take the nature of internal activities into account.  There are several non-equivalent approaches towards defining weaker versions of bisimulation for transition systems with silent steps \cite{Sangiorgi11}. In this paper we focus on weak bisimulation proposed by Milner \cite{Milner,Milner3, Sangiorgi11} and its generalization.  Analogues of Milner's weak bisimulation are established for different deterministic and probabilistic transition systems (e.g. \cite{BaierHermanns,Segala}). From now on the term weak bisimulation refers to Milner's relation and its analogues. 

%For instance, in Milner's Calculus of Communicating Systems \cite{Milner} unobservable transitions are used in the description of semantics of parallel composition between processes. 

We have witnessed a rapid development of the theory of coalgebras as a unifying theory of state-based systems \cite{GummEl,HasJacSok,Rutt2000}. Coalgebras to some extent are one-step entities in their nature. They can be thought of, and understood, as a representation of a single step of computation of a given process.  Strong bisimulation, unlike weak bisimulation, has been well captured coalgebraically (see e.g. \cite{GummEl,Rutt2000,Staton}).  Different approaches to defining weak bisimulations for coalgebras have been presented in the literature. The earliest paper is \cite{Rutten}, where Rutten studies weak bisimulations for while programs. In \cite{Rothe} Rothe introduces a definition of weak bisimulation for coalgebras by translating a coalgebraic structure into a labelled transition system. This construction works for coalgebras of different types but does not cover the distribution functor, hence it is not applicable to different types of probabilistic systems. In \cite{RotheMas} weak bisimulations are introduced via weak homomorphisms. As noted by Sokolova \emph{et al.} in \cite{SokViWor} the construction from \cite{RotheMas} does not lead to intuitive results for probabilistic systems. In \cite{SokViWor} Sokolova \emph{et al.} present a definition of weak bisimulation for classes of coalgebras of type functors obtained from bifunctors.  In that paper, weak bisimulation of a system is defined as a strong bisimulation of a transformed system. However, a drawback of the proposed approach is that the transformation of a system does not follow from any general categorical construction.  Finally, in \cite{Brengos12} we present a new approach to defining weak bisimulation in two different ways, proposed in the setting of coalgebras whose type is an ordered functor.  The key ingredient of the definitions is the notion of a saturator. As noted by us in \cite{Brengos12} the saturator is sometimes too general to model only weak bisimulation and may be used to define other known equivalences, e.g. delay bisimulation \cite{Sangiorgi11}. Moreover, the saturators from \cite{Brengos12} do not arise in any natural way.

The aim of this paper is to present a general coalgebraic setting in which it is possible to introduce weak bisimulation via saturation for coalgebras whose type is a monad satisfying additional properties. Here, the saturation of a coalgebra $\alpha:~X\to~TX$ is reduced to taking its abstract reflexive and transitive closure $\alpha^{*}$ in the so-called Kleisli category for the given monad. Unlike in our previous work \cite{Brengos12}, where the saturator was an arbitrary closure operator, the saturation in this paper is uniquely determined by  the monadic structure of the type of coalgebras taken into consideration.  The framework for defining weak bisimulation presented here is supported by the following two examples of models:
\begin{itemize}
\item labelled transition systems,
\item simple Segala systems.
\end{itemize}
Unfortunately, some examples of transition systems with a well established coalgebraic treatment and a notion of weak bisimulation fail to fit the proposed framework directly. Fully probabilistic systems \cite{BaierHermanns} (see also \cite{Sokolova} for a coalgebraic perspective on these systems) are among such examples. We discuss these limitations briefly in the last section of the paper.

The paper is organized as follows. In Section \ref{section:basics} we present basic notions and facts in the theory of algebra, coalgebra, and category theory. In Section \ref{section:LTS_coalgebraically} we recall different but equivalent ways to define weak bisimulation for labelled transition systems. Moreover, we show how to view labelled transition systems from the coalgebraic perspective and how to deal with silent and visible transition labels by introducing a monadic structure on the labelled transition systems functor or embedding it into a monad. In Section ~\ref{section:hiding_invisible_transitions} we generalize the idea of handling the invisible part of computation to any functor of the form $T(F+\mathcal{I}d)$, where $T$ is a monad and $F$ an endofunctor satisfying some extra properties. Section \ref{section:saturation} develops the theory of ordered saturation monads. These monads prove to be useful in Section~\ref{section:weak_bisimulation}, where we introduce weak bisimulation for coalgebras whose type is an ordered saturation monad on the category of sets and mappings. In Section \ref{section:weak_bisimulation_final_semantics} we present a general approach towards defining weak bisimulation semantics via final semantics for coalgebras of a monadic type whose Kleisli category is order enriched. Section~\ref{section:segala} is devoted to (simple) Segala systems and their weak bisimulation. Finally, Section \ref{section:summary} is a summary.

%%%%%%%%%%%%%%%%%%SECTION%%%%%%%%%%%%%%%%%%%%%%%%%%%%%%%%%%%
\section{Basic notions and properties} 
\label{section:basics}

We assume the reader is familar with the following basic category theory notions and their properties: a category, a functor, a (co)limit and an adjunction (see e.g.  \cite{MacLane} for an introduction to category theory).
 
\subsection{Algebras and coalgebras} 
Let $\mathsf{C}$ be a category and let $F \colon \mathsf{C}\rightarrow \mathsf{C}$ be a functor. An \emph{$F$-algebra} is a morphism $a:FA\to A$ in $\mathsf{C}$. For two $F$-algebras $a:FA\to A$ and $b:FB\to B$ a morphism $f:A\to B$ in $\mathsf{C}$ is called \emph{homomorphism} provided that $f \circ a = b \circ Ff$. The category of all $F$-algebras and homomorphisms between them is denoted by $\mathsf{C}^F$.
Dually, an \emph{$F$-coalgebra}, is a morphism $\alpha:X\to FX$ in $\mathsf{C}$.  The domain $X$ of $\alpha$ is
called \emph{carrier} and the morphism  $\alpha$ is sometimes also called  \emph{structure}. A \emph{homomorphism} from an $F$-coalgebra $\alpha:X\to FX$ to an $F$-coalgebra $\beta:Y\to FY$  is a morphism $f \colon X\rightarrow Y$ in $\mathsf{C}$ such that $ F(f)\circ \alpha =  \beta \circ f$. The category of all $F$-coalgebras and homomorphisms between them is denoted by $\mathsf{C}_F$. Many important transition systems can be captured using coalgebras.  
Here are some examples. In this subsection $\Sigma$ is a fixed set.  

\subsubsection*{Kripke frames} Kripke frames \cite{Kripke}, one of the most widely studied semantics of modal logic (see e.g. \cite{Blackburn}), are modelled as coalgebras of the type $\mathcal{P}$, where  $\mathcal{P}:\mathsf{Set}\to\mathsf{Set}$ denotes the powerset functor. 
\subsubsection*{Labelled transition systems} Labelled transition systems (see e.g. \cite{Sangiorgi11}) can be viewed as coalgebras of the type $\mathcal{P}(\Sigma\times \mathcal{I}d):\mathsf{Set}\to \mathsf{Set}$ \cite{Rutt2000}.  See also Section~\ref{section:LTS_coalgebraically} for a more detailed description of the coalgebraic perspective on these systems.

\subsubsection*{Non-deterministic automata} Non-deterministic automata (e.g. \cite{HopUll}) are modelled coalgebraically as systems of the type  $\mathcal{P}(\Sigma \times \mathcal{I}d+1)$, where $1= \{\checked\}$ (e.g. \cite{HasJacSok}). Given a coalgebra $\alpha:X\to \mathcal{P}(\Sigma \times X +1)$ a state $x\in X$ is \emph{final} whenever $\checked \in \alpha(x)$. We will elaborate more on these systems in Section \ref{section:weak_bisimulation_final_semantics} (see also Section \ref{section:hiding_invisible_transitions} and \ref{section:saturation}) where we present an interesting example of final weak bisimulation semantics for coalgebras whose base category is different from $\mathsf{Set}$.  

\subsubsection*{Fully probabilistic systems} Fully probabilistic systems \cite{BaierHermanns}  are modelled as $\mathcal{D}(\Sigma~\times~\mathcal{I}d)$-coalgebras \cite{Sokolova}.  Here,  $\mathcal{D}$ denotes the distribution functor which assigns to any set $X$ the set $$\mathcal{D}X :=\{ \phi:X\to [0,1]\mid \sum_{x\in X} \phi(x) = 1\}$$ of discrete measures and to any mapping $f:X\to Y$ the mapping $\mathcal{D}f:~\mathcal{D}X\to \mathcal{D}Y$ defined for $\phi\in \mathcal{D}X$ by $$\mathcal{D}f(\phi)(y) =  \sum_{f(x) = y}\phi(x) \text{ for any }y\in Y.$$
In this paper we also work with the subdistribution functor $\mathcal{D}_{\leqslant 1}$ which extends the functor $\mathcal{D}$. This functor is defined in the same way as the distribution functor $\mathcal{D}$, but the equality $\sum_{x\in X} \phi(x) = 1$ is replaced in its definition by the inequality $\sum_{x\in X} \phi(x)~\leqslant~1$.

\subsubsection*{ Segala and simple Segala systems}  (Simple) probabilistic systems \cite{Segala,SegalaThesis}, known in the coalgebraic literature under the name of (simple) Segala systems, are modelled as coalgebras of the type $\mathcal{P}(\Sigma \times \mathcal{D})$ and $\mathcal{P}\mathcal{D}(\Sigma\times\mathcal{I}d)$ respectively \cite{Sokolova}. In this paper we will deviate from this approach and consider Segala systems as coalgebras of the type $\mathcal{CM}(\Sigma\times \mathcal{I}d)$. This treatment is highly inspired by \cite{Jacobs08}. For a detailed definition of $\mathcal{CM}$ and a thorough discussion about consequences of this treatment the reader is referred to Section \ref{section:segala}.

For an endofunctor $F:\mathsf{C}\to\mathsf{C}$ and an $F$-coalgebra $\alpha:X\to FX$ we define a relation $X\stackrel{\pi_1}{\leftarrow} R\stackrel{\pi_2}{\to} X$ (i.e. a jointly monic span in $\mathsf{C}$) to be  \emph{Aczel-Mendler bisimulation} or simply \emph{bisimulation} \cite{AczMen} provided that there is a structure $\gamma:R\to FR$ making $\pi_1:R\to X$ and $\pi_2:R\to X$ homomorphisms between $\gamma$ and $\alpha$. In other words, the following diagram commutes:
$$
	\xymatrix{\ar @{} [dr] |{  }
		X\ar[d]_{\alpha} & R\ar @{} [dr] |{  } \ar[d]_\gamma \ar[l]_{\pi_1} \ar[r]^{\pi_2} & X \ar[d]^{\alpha} \\
		FX & FR\ar[l]^{F\pi_1} \ar[r]_{F\pi_2} & FX
	}
$$
It is worth noting that there are other approaches to capture the notion of bisimulation coalgebraically which are summarized in \cite{Staton}. We choose  to define strong bisimulation in Aczel-Mendler style because our notion of weak bisimulation introduced in this paper can be easily related to this definition. See Section \ref{section:weak_bisimulation} for details. 

\subsection{Monads} \label{subsection:monads} A \emph{monad} on $\mathsf{C}$ is a triple $(T,\mu,\eta)$, where $T:\mathsf{C}\to \mathsf{C}$ is an endofunctor and $\mu:T^2\implies T$, $\eta:\mathcal{I}d\implies T$ are two natural transformations for which the following diagrams commute:
$$
\xymatrix{
T^3\ar[d]_{T\mu} \ar[r]^{\mu_T} & T^2\ar[d]^{\mu} & & T\ar[d]_{\eta_T}\ar[r]^{T\eta}\ar@{=}[dr] & T^2\ar[d]^{\mu} \\
T^2 \ar[r]_{\mu} & T & &T^2 \ar[r]_\mu & T
}
$$
The transformation $\mu$ is called  \emph{multiplication} and $\eta$ \emph{unit}.
Any monad gives rise to the Kleisli category for $T$. To be more precise, if $(T,\mu,\eta)$ is a monad on a category $\mathsf{C}$ then the \emph{Klesli category} $\mathcal{K}l(T)$ for $T$ has the class of objects equal to the class of objects of $\mathsf{C}$ and for two objects $X,Y$ in $\mathcal{K}l(T)$ we have $Hom_{\mathcal{K}l(T)}(X,Y) = Hom_{\mathsf{C}}(X,TY)$
with the composition $\cdot$ in $\mathcal{K}l(T)$ defined between two morphisms $f:X\to TY$ and $g:Y\to TZ$ by
$g\cdot f := \mu_Z \circ T(g) \circ f$.

\begin{exa} \label{example:kleisli_powerset}
The powerset endofunctor $\mathcal{P}:\mathsf{Set}\to \mathsf{Set}$ is a monad whose multiplication $\mu:\mathcal{P}^2\implies\mathcal{P}$ and unit $\eta:\mathcal{I}d\implies \mathcal{P}$ are given on their $X$-components by: 
\begin{align*}
& \mu_X:\mathcal{P}\mathcal{P}X\to \mathcal{P}X; S \mapsto \bigcup S \text{ and }\eta_X:X\to \mathcal{P}X; x\mapsto \{x\}.
\end{align*}
The category $\mathcal{K}l(\mathcal{P})$ consists of sets as objects and maps of the form $X\to \mathcal{P}Y$ as morphisms. For $f:X\to \mathcal{P}Y$ and $g:Y\to \mathcal{P}Z$ the composition $g\cdot f:X\to \mathcal{P}Z$ is as follows:
\begin{align*}
& g\cdot f(x) = \{z\in Z \mid z\in \bigcup g(f(x))\} = \{z \mid z\in g(y) \text{ and }y\in f(x) \text{ for some }y\in Y\}. 
\end{align*} 
For any two sets $X,Y$ there is a bijective correspondence between maps $X\to \mathcal{P}Y$ and binary relations between elements of $X$ and $Y$. 
Indeed, for $f:X\to \mathcal{P}Y$ we put $R_f\subseteq X\times Y$, $(x,y)\in R_f \iff y\in f(x)$ and for $R\subseteq X\times Y$ we define $f_R:X\to \mathcal{P}Y;x\mapsto \{y\mid x R y\}$. It is now easy to see that the category $\mathcal{K}l(\mathcal{P})$ is isomorphic to the category $\mathsf{Rel}$ of sets as objects, binary relations as morphisms and relation composition as morphism composition.  
\end{exa}

\begin{exa}
The distribution functor $\mathcal{D}$ carries a monadic structure, with $\mu$ and $\eta$ given as follows. For any set $X$ define the $X$-component of $\eta$ by 
$\eta_X:X\to \mathcal{D}X; x\mapsto \delta_x$, where $\delta_x$ is the Dirac distribution for $x$ i.e. $\delta_x (x) = 1$ and $\delta_x(y)=0$ if $y\neq x$. The $X$-component of $\mu$ is given by:
$$
\mu_X: \mathcal{D}^2 X\to \mathcal{D}X; \mu_X(\phi)(x)= \sum_{\psi\in \mathcal{D}X} \phi(\psi)\cdot \psi(x).
$$
The subdistribution functor $\mathcal{D}_{\leqslant 1}$ is also a monad with $\mu$ and $\eta$ defined in the same manner.
\end{exa}

\begin{exa}
A \emph{filter} on a set $X$ is a collection $\mathcal{G}$ of subsets of $X$ such that 
\begin{itemize}
\item $X\in \mathcal{G}$,
\item if $U_1,U_2\in \mathcal{G}$ then $U_1\cap U_2\in \mathcal{G}$,
\item if $U\in \mathcal{G}$ and $U\subseteq V\subseteq X$ then $V\in \mathcal{G}$.
\end{itemize}
We define the filter functor $\mathcal{F}$ which assigns to any set $X$ the set $\mathcal{F}X$ of all filters on $X$ and to any map $f:X\to Y$ the map $\mathcal{F}f:\mathcal{F}X\to \mathcal{F}Y$ which assigns to any filter $\mathcal{G}$ the smallest filter containing the family $\{f(G)\mid G\in \mathcal{G}\}$. The functor $\mathcal{F}$ carries a monadic structure $(\mathcal{F},\mu,\eta)$ given as follows (see e.g. \cite{Gahler}):
\begin{align*}
&\mu_X:\mathcal{F}\mathcal{F}X\to \mathcal{F}X; \mathcal{G}\mapsto \mu_X(\mathcal{G}),\\
&\eta_X:X\to \mathcal{F}X; x\mapsto \{U\subseteq X\mid x\in U\},
\end{align*}
where $\mu_X(\mathcal{G}) = \{A\subseteq X\mid A^\mathcal{F} \in \mathcal{G}\}$ with $A^\mathcal{F} = \{\mathcal{H}\in\mathcal{F}X\mid A\in \mathcal{H} \}$ defined as the set of all filters on $X$ containing $A$.
\end{exa}

Since most of the time we work with two categories at once: $\mathsf{C}$ and $\mathcal{K}l(T)$, morphisms in $\mathsf{C}$ will be denoted using standard arrows $\to$, whereas for morphisms in $\mathcal{K}l(T)$ we will use the symbol $\rightdcirc$.
For any object $X$ in $\mathsf{C}$ (or equivalently in $\mathcal{K}l(T)$) the identity map from $X$ to itself in $\mathsf{C}$  will be denoted by $id_X$ and in $\mathcal{K}l(T)$ by $1_X$ or simply $1$ if the domain can be deduced from the context. 
The category $\mathsf{C}$ is a subcategory of $\mathcal{K}l(T)$ where the inclusion functor $^{\sharp}$ sends each object $X\in \mathsf{C}$ to itself and each morphism $f:X\to Y$ in $\mathsf{C}$ to the morphism $f^{\sharp}:X\rightdcirc Y$ given by
$
f^{\sharp}:X\to TY; f^{\sharp} = \eta_Y\circ f.
$
Every monad $(T,\mu,\eta)$ on a category $\mathsf{C}$ arises from the composition of a left and a right adjoint given by:
$$
\xymatrix{
\mathsf{C}\ar@<1.5ex>[r]^-{^{\sharp}}\ar@{}[r]|-\perp & \mathcal{K}l(T)\ar@<1.5ex>[l]^-{U_T},
}
$$
where $U_T:\mathcal{K}l(T)\to \mathsf{C}$ is a functor defined as follows. For any object $X\in \mathcal{K}l(T)$ (i.e. $X\in \mathsf{C}$) the object $U_T X$ is given by $U_T X := TX$ and for any morphism $f:X\rightdcirc Y$ in $\mathcal{K}l(T)$ (i.e. $f:X\to TY$ in $\mathsf{C}$) the morphism $U_T f:TX\to TY$ is given by $U_T f = \mu_Y\circ Tf$.

We say that a functor $F:\mathsf{C}\to\mathsf{C}$ \emph{lifts to} an endofunctor $\overline{F}:\mathcal{K}l(T)\to\mathcal{K}l(T)$ provided that the following diagram commutes:
$$
\xymatrix{
\mathcal{K}l(T) \ar@{-->}[r]^{\overline{F}} & \mathcal{K}l(T)  \\
\mathsf{C}\ar[u]^{\sharp} \ar[r]_F & \mathsf{C}\ar[u]_{\sharp}
}
$$
Given a functor $F:\mathsf{C}\to\mathsf{C}$ there is a one-to-one correspondence between its liftings $\overline{F}$ and \emph{distributive laws}  $\lambda: FT\implies TF$ between the functor $F$ and the monad $T$ (see e.g. \cite{JacSilSok,Mulry} for a detailed definition and properties). Given a distributive law $\lambda:FT\implies TF$ we define $\overline{F}:\mathcal{K}l(T)\to\mathcal{K}l(T)$ by:
\begin{align*}
& \overline{F}X := FX \text{ for any object } X\in \mathcal{K}l(T),\\
& \overline{F}f:FX\to TFY; \overline{F}f = \lambda_Y \circ Ff \text{ for any morphism } f:X\to TY.
\end{align*}
Conversely, a lifting $\overline{F}:\mathcal{K}l(T)\to \mathcal{K}l(T)$ gives rise to $\lambda:FT\implies TF$ defined on its $X$-component by $\lambda_X:FTX\to TFX; \lambda_X = \overline{F}(id_{TX})$.

A monad $(T,\mu,\eta)$ on a cartesian closed category $\mathsf{C}$ is called \emph{strong} if there is a natural transformation $t_{X,Y}:X\times TY\to T(X\times Y)$ called \emph{tensorial strength} satisfying the strength laws listed in e.g. \cite{Kock}. Existence of strength guarantees that for any object $\Sigma$ the functor $\Sigma\times \mathcal{I}d:\mathsf{C}\to \mathsf{C}$ admits a lifting  $\overline{\Sigma}:\mathcal{K}l(T)\to\mathcal{K}l(T)$ defined as follows. For any object $X\in \mathcal{K}l(T)$ we put
$
\overline{\Sigma}X := \Sigma\times X,
$
and for any morphism $f:X\rightdcirc Y$ (i.e. $f:X\to TY$ in $\mathsf{C}$) we define
$$
\overline{\Sigma}f:\Sigma\times X\to T(\Sigma\times Y);\overline{\Sigma}f:= t_{\Sigma,Y}\circ ( id_{\Sigma}\times f).
$$
Existence of the transformation ${t}$ is not a strong requirement. For instance all monads on $\mathsf{Set}$ are strong. 

\begin{exa}\label{example:lifting_of_powerset}
Take $\mathsf{C}=\mathsf{Set}$ and $T = \mathcal{P}$. The strength $t_{X,Y}:X\times \mathcal{P}Y \to \mathcal{P}(X\times Y)$ is given by the following formula:
$$
t_{X,Y}(x,Y') = \{(x,y')\mid y'\in Y'\} \text{ for }(x,Y')\in X\times \mathcal{P}Y.
$$
Hence, the functor $\Sigma \times \mathcal{I}d:\mathsf{Set}\to \mathsf{Set}$ lifts to $\overline{\Sigma}:\mathcal{K}l(\mathcal{P})\to \mathcal{K}l(\mathcal{P})$, where for any object $X\in \mathcal{K}l(\mathcal{P})$  and any morphism $f:X\rightdcirc Y$ in $\mathcal{K}l(\mathcal{P})$ we have:
\begin{align*}
&\overline{\Sigma} X := \Sigma\times X \text{ and } \overline{\Sigma}f:\overline{\Sigma} X\rightdcirc \overline{\Sigma} Y; (\sigma,x) \mapsto \{(\sigma,y)\mid y\in f(x)\}.
\end{align*}
\end{exa}

\subsection{Monads on Kleisli categories} \label{subsection:monads_on_kleisli}
In this paper we will often work with monads on Kleisli categories. Here, we list basic properties of such monads. Everything presented below follows easily by elementary properties of adjunctions and liftings (see e.g. \cite{MacLane,ManMul,Mulry} and the previous subsection). For a monad $(T,\mu,\eta)$ on $\mathsf{C}$ assume $S:\mathsf{C}\to\mathsf{C}$ is a functor that lifts to $\overline{S}:\mathcal{K}l(T)\to \mathcal{K}l(T)$ by the corresponding distributive law $\lambda:ST\implies TS$.
Moreover, let $(\overline{S},m,e)$ be a monad on $\mathcal{K}l(T)$ (note that we do not assume $S:\mathsf{C}\to \mathsf{C}$ to carry a monadic structure but only require its lifting $\overline{S}$ to be a monad).  We have the following two adjunctions:
$$
\xymatrix{
\mathsf{C}\ar@<1.5ex>[r]^-{^{\sharp}}\ar@{}[r]|-\perp & \mathcal{K}l(T)\ar@<1.5ex>[r]^-{^{\sharp}}\ar@<1.5ex>[l]^{U_T}\ar@{}[r]|\perp & \mathcal{K}l(\overline{S})\ar@<1.5ex>[l]^{U_{\overline{S}}}.
}
$$
Because the composition of two adjunctions is an adjunction, the functor $^\sharp\circ ^\sharp$ is a left adjoint to $U_T\circ U_{\overline{S}}$. Since $U_T \circ U_{\overline{S}}\circ ^\sharp\circ ^\sharp = TS$, this yields a monadic structure on the functor $TS:\mathsf{C}\to \mathsf{C}$. 
%The $X$-components of the multiplication $\mathfrak{m}$ and the unit $\mathfrak{e}$ of the monad $TS$ are given by:
%$$
%\mathfrak{m}_X = \mu_{SX}\circ T\mu_{SX} \circ TT m_X \circ T\lambda_{SX} \quad \text{and}\quad \mathfrak{e}_X = e_X.
%$$
The composition $\cdot$ in $\mathcal{K}l(TS) = \mathcal{K}l(\overline{S})$ is given in terms of the composition in $\mathsf{C}$ as follows. For $f:X\to TSY$ and $g:Y\to TSZ$ we have:
$$
\xymatrix{
X\ar@{-->}[d]_{g\cdot f} \ar[r]^f & TSY \ar[r]^-{TSg} &
TSTSZ \ar[r]^{T\lambda_{SZ}} & T^2S^2Z \ar[d]^{T^2(m_Z)} \\
 TSZ & &T^2SZ\ar[ll]^{\mu_{SZ}} & T^3SZ\ar[l]^{T\mu_{SZ}} 
}
$$

\subsection{Order enriched categories}
A category $\mathsf{C}$ is \emph{order enriched} if each hom-set is a poset with the order preserved by the composition. An endofunctor on an order enriched category is called \emph{locally monotonic} if it preserves the order.

We say that  \emph{arbitrary cotupling in $\mathsf{C}$ is monotonic}  if for any family of objects $\{X_i\}_{i\in I}$ and two families $\{f_i:X_i\to Y\}_{i\in I} \text{ and }\{g_i:X_i\to Y\}_{i\in I}$ of morphisms if the coproduct $\coprod_i X_i$ of $\{X_i\}_{i\in I}$ exists in $\mathsf{C}$  and $f_i\leqslant g_i$ for any $i\in I$ then the cotuples $[\{f_i\}]:\coprod_i X_i \to Y$ and $[\{ g_i\}]:\coprod_i X_i \to Y$ satisfy $[\{ f_i\} ]\leqslant [\{ g_i\} ]$.

\begin{exa}\label{example:order_enriched}
Kleisli category for any monad $T\in\{\mathcal{D}_{\leqslant 1},\mathcal{F}, \mathcal{P}\}$ is order enriched and arbitrary cotupling in $\mathcal{K}l(T)$ is monotonic.
The order on hom-sets in $\mathcal{K}l(T)$ is imposed by the natural pointwise order summarized in the table below. For $T\in \{\mathcal{D}_{\leqslant 1},\mathcal{F}, \mathcal{P}\}$ and $f,g:X\to TY$ we have:
\begin{table}[h]
\begin{tabular}{|p{1.2cm}|p{8cm}|}
\hline 
Monad & $f\leqslant g$ if and only if  \\\hline \hline
$\mathcal{D}_{\leqslant 1}$ & $f(x)(y) \leqslant g(x)(y)$ for any $x\in X$, $y\in Y$  \\ \hline
$\mathcal{F}$ & $f(x)\supseteq g(x)$ for any $x\in X$ \\\hline
$\mathcal{P}$ & $f(x)\subseteq g(x)$ for any $x\in X$ \\ \hline 
\end{tabular}
\end{table}
\end{exa}

\section{Labelled transition systems coalgebraically}  
\label{section:LTS_coalgebraically} 

  Labelled transition systems have been defined and thoroughly studied in the computer science literature (see e.g. \cite{Milner,Milner3, Sangiorgi11}). We will now briefly recall some classical definitions and properties from the theory of these systems.  Let $\Sigma$ be a fixed set called \emph{set of alphabet letters} or simply \emph{alphabet}.  \emph{Labelled transition system} over the alphabet $\Sigma$ (or \emph{LTS} in short) is a triple $\left<X,\Sigma,\to\right>$, where $X$ is a set, called  \emph{set of states}, and $\to\subseteq X\times \Sigma\times X$ is the transition relation. For an LTS $\left<X,\Sigma,\to\right>$ instead of writing $(x,\sigma,x')\in \to$ we write $x\stackrel{\sigma}{\to}x'$. 

\begin{defi}
\label{definition:LTS_bisimulation}
 A symmetric relation $R\subseteq X\times X$ is called a \emph{bisimulation} on $\left<X,\Sigma,\to\right>$ if the following condition holds:
$$
(x,y)\in R \text{ and } x\stackrel{\sigma}{\to}x' \text{ implies } y\stackrel{\sigma}{\to}y' \text{ for some }y'\in X \text{ s.t. }(x',y')\in R.
$$ 
\end{defi}

From now on we assume that the alphabet we consider contains a special label, called \emph{internal} or \emph{invisible} label, which is usually denoted by the letter $\tau$. To be more precise, we put $\Sigma_\tau:=\Sigma+\{\tau\}$ and consider a labelled transition system $\left<X,\Sigma_\tau,\to\right>$ over the alphabet $\Sigma_\tau$. Letters in $\Sigma$ are called \emph{visible}.

\begin{defi}\cite{Milner, Milner3, Sangiorgi11}
\label{definition:LTS_weak_bisimulation}
A symmetric relation $R\subseteq X\times X$ is called  \emph{a weak bisimulation} on $\left<X,\Sigma_\tau,\to\right>$ if the following condition holds:
$$
(x,y)\in R \text{ and } x\stackrel{\sigma}{\to}x' \text{ implies } y\stackrel{\sigma}{\implies }y' \text{ for some }y'\in X \text{ s.t. }(x',y')\in R,
$$
where $\stackrel{\sigma}{\implies}\subseteq X\times X$ is defined by
$$
\stackrel{\sigma}{\implies} = \left \{\begin{array}{cc} (\stackrel{\tau}{\to})^{*}& \text{ if }\sigma = \tau \\
(\stackrel{\tau}{\to})^{*}\circ \stackrel{\sigma}{\to}\circ (\stackrel{\tau}{\to})^{*} & \text{ otherwise.}\end{array}\right.
$$
Here, $(-)^{*}$ denotes the reflexive and transitive closure of a relation. 
\end{defi}
%We can extend the definition of transitions $\stackrel{\sigma}{\to}$ and $\stackrel{\sigma}{\implies}$ as follows. For a word $s\in \Sigma^{*}$ over the set of visible labels  put
%$$
%\stackrel{s}{\to}  =\left \{\begin{array}{cc} \stackrel{\tau}{\to}& \text{ if }s \text{ is the empty word,} \\
%\stackrel{\sigma_1}{\to}\circ \ldots \circ \stackrel{\sigma_n}{\to} & \text{for }s=\sigma_1\ldots \sigma_n,
%\end{array}\right.
%$$ 
%and

We can extend the definition of the transition $\stackrel{\sigma}{\implies}$ as follows. For a word $s\in \Sigma^{*}$ over the set of visible labels  put
$$
\stackrel{s}{\implies}  =\left \{\begin{array}{cc} (\stackrel{\tau}{\to})^{*}& \text{ if }s \text{ is the empty word} \\
(\stackrel{\tau}{\to})^{*}\circ \stackrel{\sigma_1}{\to}\circ (\stackrel{\tau}{\to})^{*}\circ \ldots \circ (\stackrel{\tau}{\to})^{*}\circ \stackrel{\sigma_n}{\to}\circ (\stackrel{\tau}{\to})^{*}  & \text{for }s=\sigma_1\ldots \sigma_n.\end{array}\right.
$$ 

It is an easy exercise to prove the following (see e.g. \cite{Milner3, Sangiorgi11} for details).

\begin{fact}
\label{fact:lts_weak_bisimulation_equiv}
A symmetric relation $R\subseteq X\times X$ is a weak bisimulation on $\left<X,\Sigma_\tau,\to\right>$ if and only if the following condition holds:
$$
(x,y)\in R \text{ and } x\stackrel{\sigma}{\implies }x' \text{ implies } y\stackrel{\sigma}{\implies }y' \text{ for some }y'\in A \text{ s.t. }(x',y')\in R.
$$
If we replace all occurrences of $\stackrel{\sigma}{\implies }$ by $\stackrel{s}{\implies}$ for any $s\in \Sigma^{*}$ we also get a true statement. \qed
\end{fact}
The fact above suggests that weak bisimulation on $\left<X,\Sigma_\tau,\to\right>$ can be defined as a strong bisimulation on a \emph{saturated model} $\left<X,\Sigma_\tau,\implies\right>$. It is worth noting that from the point of view of computation and automated reasoning the former approach to defining weak bisimulation is better since, unlike the condition in Fact~\ref{fact:lts_weak_bisimulation_equiv}, it does not require the knowledge of the full saturated transition. Indeed, in order to show that two states $x,y\in X$ of a labelled transition system are weakly bisimilar in the sense of the equivalent condition from Fact \ref{fact:lts_weak_bisimulation_equiv} one needs to consider all states $x'\in X$ reachable from $x$ via the saturated transitions  and compare them with similar states reachable from $y$. Whereas, to prove that two states $x,y\in X$ are weakly bisimilar in the sense of Definition \ref{definition:LTS_weak_bisimulation} one needs to consider all states  reachable from $x$ via single step transitions $\stackrel{\sigma}{\to}$ and compare them with \emph{some} states reachable from $y$ via the saturated transitions.
We see that the key ingredient in defining weak bisimulation for LTS is the saturation. In order to describe it categorically we have to understand the nature of invisible transitions first.

Before we discuss the nature of internal steps, we recall how labelled transition systems are modelled coalgebraically. Any LTS over the alphabet $\Sigma_\tau$ can be viewed as a coalgebra of the type $\mathcal{P}(\Sigma_\tau\times \mathcal{I}d)$. Given an LTS $\left<X,\Sigma_\tau,\to\right>$ we turn it into a $\mathcal{P}(\Sigma_\tau\times \mathcal{I}d)$-coalgebra $\alpha:X\to \mathcal{P}(\Sigma_\tau\times X)$ as follows:
$$
\alpha(x) =\{(\sigma,x')\mid x\stackrel{\sigma}{\to}x'\}.
$$
In this case, any coalgebraic bisimulation which is a symmetric relation satisfies the conditions from  Definition \ref{definition:LTS_bisimulation}. Conversely, any relation which is a bisimulation in the sense of Definition  \ref{definition:LTS_bisimulation} is a symmetric coalgebraic bisimulation  (see e.g. \cite{Rutten} for details). It should be noted here that the assumption about symmetry in Definition~\ref{definition:LTS_bisimulation} serves only one purpose: it makes the definition more succinct. 

\subsection{Monadic structure on
  \texorpdfstring{$\mathcal{P}(\Sigma_\tau\times
    \mathcal{I}d)$}{P(Sigma-tau x Id)}}
\label{subsection:LTS_monad}

Our aim will be now to introduce a monadic structure on the LTS functor $\mathcal{P}(\Sigma_\tau \times \mathcal{I}d)$ which leads to a definition of saturation of an LTS and which internally handles visible and invisible labels. This observation together with a similar one in Subsection \ref{subsection:LTS_extended_monad} and their generalizations in Section \ref{section:hiding_invisible_transitions} lead us to a conclusion that weak bisimulation for coalgebras should not focus on specifying and handling visible and invisible parts of the functor explicitly. Instead, it should assume the type of coalgebras taken into consideration is a monad with internal transitions being a part of its unit.

Consider the functor $\Sigma_\tau\times \mathcal{I}d:\mathsf{Set}\to \mathsf{Set}$ and its lifting $\overline{\Sigma_\tau}:\mathcal{K}l(\mathcal{P})\to \mathcal{K}l(\mathcal{P})$ as in Example \ref{example:lifting_of_powerset}. Define two natural transformations: 
$$m: \overline{\Sigma_\tau}\overline{\Sigma_\tau}\implies \overline{\Sigma_\tau}\text{ and }e:\mathcal{I}d \implies \overline{\Sigma_\tau}$$ in $\mathcal{K}l(\mathcal{P})$ as follows. For any object $X\in \mathcal{K}l(\mathcal{P})$ define the $X$-components 
\begin{align*}
&m_X:\overline{\Sigma_\tau}\overline{\Sigma_\tau} X\rightdcirc \overline{\Sigma_\tau} X \text{ (i.e. } m_X:\Sigma_\tau\times \Sigma_\tau \times X\to \mathcal{P}(\Sigma_\tau\times X)) \text{ and }\\
&e_X:X\rightdcirc \overline{\Sigma_\tau}X \text{ (i.e. }e_X:X\to \mathcal{P}(\Sigma_\tau \times X)) \text{ to be:}
\end{align*}
$$m_X(\sigma_1,\sigma_2,x)  = \left\{ \begin{array}{cc} \{(\sigma_1,x)\} & \text{if } \sigma_2 = \tau, \\
\{(\sigma_2,x)\} & \text{if } \sigma_1 = \tau, \\
\varnothing & \text{otherwise}\end{array}\right.
\qquad 
e_X(x) = \{(\tau,x)\}.
$$
\begin{fact}\label{fact:LTS_label_monad}
The triple $(\overline{\Sigma_\tau},m,e)$ is a monad on $\mathcal{K}l(\mathcal{P})$. \qed
\end{fact}
\noindent The above fact is a consequence of a more general statement, namely Theorem \ref{theorem:main_monad_theorem} below, and hence is left without a proof. We have the following two adjunctions:
$$
\xymatrix{
\mathsf{Set}\ar@<1.5ex>[r]^-{^{\sharp}}\ar@{}[r]|-\perp & \mathcal{K}l(\mathcal{P})\ar@<1.5ex>[r]^-{^{\sharp}}\ar@<1.5ex>[l]^-{U_\mathcal{P}}\ar@{}[r]|-\perp & \mathcal{K}l(\overline{\Sigma_\tau})\ar@<1.5ex>[l]^-{U_{\overline{\Sigma_\tau}}}.
}
$$
Following the guidelines of Subsection \ref{subsection:monads_on_kleisli} we obtain a monadic structure on the functor $\mathcal{P}(\Sigma_\tau\times \mathcal{I}d):\mathsf{Set}\to\mathsf{Set}$. The composition $\cdot$ in  $\mathcal{K}l(\mathcal{P}(\Sigma_\tau \times \mathcal{I}d))$ is given by the following formula. For $f:X\to \mathcal{P}(\Sigma_\tau \times Y)$ and $g:Y\to~\mathcal{P}(\Sigma_\tau \times Z)$ we have 
$$
\xymatrix{
X\ar@{-->}[d]_{g\cdot f} \ar[r]^f & \mathcal{P}(\Sigma_\tau \times Y) \ar[r]^-{\mathcal{P}(\Sigma_\tau \times g)} &
\mathcal{P}(\Sigma_\tau \times \mathcal{P}(\Sigma_\tau \times Z)) \ar[r]^{\mathcal{P}t} & \mathcal{P}^2(\Sigma_\tau \times \Sigma_\tau \times Z) \ar[d]^{\mathcal{P}^2(m_Z)} \\
 \mathcal{P}(\Sigma_\tau \times Z) & &\mathcal{P}^2(\Sigma_\tau \times Z)\ar[ll]^{\bigcup} & \mathcal{P}^3(\Sigma_\tau \times Z)\ar[l]^{\mathcal{P}(\bigcup)} 
}
$$
It is easy to verify that the formula for the composition is explicitly given by:
$$
g\cdot f (x) = \{(\sigma,z)\mid x\stackrel{\sigma}{\to}_f y \stackrel{\tau}{\to}_g z \text{ or }x\stackrel{\tau}{\to}_f y \stackrel{\sigma}{\to}_g z \text{ for some }y\in Y\}.
$$

The construction above which imposes a monadic structure on the LTS functor allows us to handle $\tau$-steps internally by the monadic multiplication. In the subsection below we present a second approach of handling the $\tau$-steps inside a monad whose functor extends the LTS functor.  

\subsection{Monadic structure on
  \texorpdfstring{$\mathcal{P}(\Sigma^{*}\times
    \mathcal{I}d)$}{P(Sigma* x Id)}}
\label{subsection:LTS_extended_monad}
There is a second approach to handle silent steps internally in a monadic structure of the functor $\mathcal{P}(\Sigma^{*}\times \mathcal{I}d)$ which extends the LTS functor $\mathcal{P}(\Sigma_\tau \times \mathcal{I}d)$. It is easy to see that there is a natural transformation $\nu$ from the functor $\Sigma_\tau \times \mathcal{I}d$ to $\Sigma^{*}\times \mathcal{I}d$ whose components are injective maps. The $X$-component $\nu_X$ is given by:
 $$\nu_X:\Sigma_\tau \times X\to \Sigma^{*}\times X; (\sigma,x) \mapsto \left \{ \begin{array}{cc} (\epsilon,x) & \text{ if } \sigma=\tau, \\ (\sigma,x) & \text{otherwise.} \end{array} \right.$$
Hence, the family of maps $\{\mathcal{P}(\nu_X):\mathcal{P}(\Sigma_\tau\times X)\to \mathcal{P}(\Sigma^{*}\times X)\}_{X\in \mathsf{Set}}$ is an injective natural transformation from the LTS functor $\mathcal{P}(\Sigma_\tau \times \mathcal{I}d)$ to $\mathcal{P}(\Sigma^{*}\times \mathcal{I}d)$. Any $\mathcal{P}(\Sigma_\tau \times \mathcal{I}d)$-coalgebra can be turned into a $\mathcal{P}(\Sigma^{*}\times \mathcal{I}d)$-coalgebra by post-composing it with a suitable component of $\mathcal{P}(\nu)$. Moreover, the functor $\mathcal{P}(\Sigma^{*}\times \mathcal{I}d)$ comes equipped with a monadic structure which is a consequence of existence of a monadic structure on $\mathcal{P}$ and $\Sigma^{*}\times \mathcal{I}d$. We leave all the details for Section \ref{section:hiding_invisible_transitions}.  Here, we only present the explicit formula for the composition $\cdot$ in  $\mathcal{K}l(\mathcal{P}(\Sigma^{*}\times \mathcal{I}d))$. For $f:X\to \mathcal{P}(\Sigma^{*}\times Y)$ and $g:Y\to \mathcal{P}(\Sigma^{*}\times Z)$ we have $g\cdot f:X\to \mathcal{P}(\Sigma^{*}\times Z)$:
$$
g\cdot f (x) = \{(s_1s_2,z)\mid x\stackrel{s_1}{\to}_f y \stackrel{s_2}{\to}_g z \text{ for some }y\in Y\}.
$$

\subsection{LTS saturation coalgebraically}\label{subsection:saturation_LTS}
In the previous two subsections we showed two ways of dealing with an invisible label by encoding it as a part of the unit of a monad. 
Let us for now assume that $\cdot$ is the composition in $\mathcal{K}l(\mathcal{P}(\Sigma_\tau \times \mathcal{I}d))$ as in Subsection \ref{subsection:LTS_monad}. Given an LTS coalgebra $\alpha:X\to \mathcal{P}(\Sigma_\tau \times X)$ the saturated LTS $\alpha^{*}:X\to \mathcal{P}(\Sigma_\tau \times X)$ is obtained as follows:
$$
\alpha^{\ast} = 1_X \vee \alpha \vee \alpha^2 \vee \ldots = \bigvee_{n=0,1,2\ldots} \alpha^n,
$$
where $\bigvee$ denotes the supremum in the complete lattice $(\mathcal{P}(\Sigma_\tau \times X)^X,\leqslant)$ with the relation $\leqslant$  given by:
$\alpha\leqslant\beta \iff \alpha(x)\subseteq \beta(x) \text{ for any }x\in X$.
We see that for $(\sigma,y)\in \Sigma_\tau \times X$:
$$
(\sigma,y)\in \alpha^{*}(x) \text{ if and only if }x\stackrel{\sigma}{\implies}_\alpha y.
$$
Now consider $\cdot$ to be the composition in $\mathcal{K}l(\mathcal{P}(\Sigma^{*}\times \mathcal{I}d))$ as in Subsection~\ref{subsection:LTS_extended_monad}. For an LTS coalgebra $\alpha:X\to \mathcal{P}(\Sigma_\tau\times X)$ define 
$$
\underline{\alpha} = \mathcal{P}(\nu_X)\circ \alpha:X\to \mathcal{P}(\Sigma^{*}\times X),
$$
where $\nu$ is given in the previous subsection. Put $\underline{\alpha}^{\ast}:X\to \mathcal{P}(\Sigma^{*}\times X)$ to be
$$
\underline{\alpha}^{\ast} = 1_X \vee \underline{\alpha} \vee \underline{\alpha}^2 \vee \ldots = \bigvee_{n=0,1,2\ldots} \underline{\alpha}^n,
$$
where $\bigvee$ denotes the supremum in $(\mathcal{P}(\Sigma^{*}\times X)^X,\leqslant)$. Then for $(s,y)\in \Sigma^{*}\times X$ we have:
$$
(s,y)\in \underline{\alpha}^{\ast}(x) \text{ if and only if } x\stackrel{s}{\implies}_\alpha y.
$$

Weak bisimulation on $\alpha:X\to \mathcal{P}(\Sigma_\tau \times X)$ is then any symmetric strong bisimulation on $\alpha^{*}$ or on $\underline{\alpha}^{\ast}$ (by Fact \ref{fact:lts_weak_bisimulation_equiv}). As we will see in Section~\ref{section:weak_bisimulation} it will also be possible to give a coalgebraic definition of weak bisimulation which is a generalization of Definition \ref{definition:LTS_weak_bisimulation} and compare the two approaches. Note here that both maps $\alpha^{*}$ and $\underline{\alpha}^{\ast}$ are defined so that they abstractly represent reflexive and transitive closure of $\alpha$ w.r.t. the suitable Kleisli compositions and order. Both $\alpha^{\ast}$ and $\underline{\alpha}^\ast$ are the least fixed points of the assignments $x\mapsto 1\vee x\cdot \alpha$ and $x\mapsto 1\vee x\cdot \underline{\alpha}$ respectively. This observation forms foundations for the theory developed in Section \ref{section:saturation}, \ref{section:weak_bisimulation} and \ref{section:weak_bisimulation_final_semantics}. 

%\begin{rem}
%Note that a similar approach cannot be applied to model e.g. delay bisimulation for LTS's. Let $\alpha:X\to \mathcal{P}(\Sigma\times X)$ be an LTS. To be more precise, a symmetric relation $R\subseteq X\times X$  is a \emph{delay bisimulation} on $\alpha$ \cite{Milner,Sangiorgi11} provided that the following condition is satisfied. The fact $(x,y)\in R$ implies:
%\begin{align*}
%&x\stackrel{\sigma}{\to}x' \implies y(\stackrel{\tau}{\to})^{*}\circ \stackrel{\sigma}{\to} y' \text{ and }(x',y')\in R \text{ for }\sigma\neq \tau,\\
%&x\stackrel{\tau}{\to}x' \implies y(\stackrel{\tau}{\to})^{*} y' \text{ and }(x',y')\in R.
%\end{align*}
%We see that in the process of "delay-saturating" a labelled transition system we only allow invisible transitions to take place \emph{before} the visible label. If now adopt the approach from Subsection \ref{subsection:LTS_monad} and try to encode "delay-composition" into a structure on $\overline{\Sigma}:\mathcal{K}l(\mathcal{P})\to \mathcal{K}l(\mathcal{P})$ we end up with two natural transformations $\mu'$ and $\eta'$ given by:
% $$
%\mu_X'(\sigma_1,\sigma_2,x)  = \left\{ \begin{array}{cc}
%\{(\sigma_2,x)\} & \text{if } \sigma_1 = \tau, \\
%\varnothing & \text{otherwise}\end{array}\right. \text{ and } \eta_X'(x) = \{(\tau,x)\}.
%$$
%This triple $(\overline{\Sigma},\mu',\eta')$ is \emph{not} a monad as the first diagram from definition of a monad does not commute for $\mu'$ and $\eta'$. 
%\end{rem}

%%%%%%%%%%%%%%%%%%%%%%%%%%%%%%%%%%%%%%%
%
% SECTION: Order enriched monads
%
%%%%%%%%%%%%%%%%%%%%%%%%%%%%%%%%%%%%%%%
\section{Hiding invisible transitions inside a monadic structure}
\label{section:hiding_invisible_transitions}
Throughout this paper we denote the coproduct operator by $+$ and the coprojection into the first and the second component of a coproduct by $\mathsf{inl}$ and $\mathsf{inr}$ respectively.

In this section we assume that $(T,\mu,\eta)$ is a monad on a category $\mathsf{C}$ with binary coproducts and $F:\mathsf{C}\to \mathsf{C}$ is a functor.  Since in this paper we are interested in coalgebras with internal moves we adopt the approach from \cite{HasJacSok_jap,SilWester} and for now consider the type of coalgebras to be 
$TF_\tau :\mathsf{C}\to \mathsf{C}$ with $F_\tau := F+\mathcal{I}d$. 
The LTS functor $\mathcal{P}(\Sigma_\tau \times \mathcal{I}d)$ studied in the previous section is of this form since:
$$
\mathcal{P}(\Sigma_\tau \times \mathcal{I}d) = \mathcal{P}((\Sigma+\{\tau\})\times \mathcal{I}d) \cong \mathcal{P}(\Sigma\times \mathcal{I}d +\{\tau\}\times \mathcal{I}d)\cong \mathcal{P}(\Sigma\times \mathcal{I}d+\mathcal{I}d).
$$
Coalgebras of the type $TF_\tau$ were studied in \cite{HasJacSok_jap,SilWester} from the perspective of trace semantics for systems with internal moves. The motivation for considering the type $TF_\tau$ to model systems with silent transitions is the following: the monad $T$ represents the branching type, the functor $F$ represents the visible part of the transition (just like $\Sigma\times \mathcal{I}d$ for LTS), and $\mathcal{I}d$ represents the invisible label transition \cite{HasJacSok_jap,SilWester}. 

The aim of this section is to generalize both strategies demonstrated in Section~\ref{section:LTS_coalgebraically} for the LTS functor for handling invisible transitions internally in a monadic structure of the type we consider. To be more precise, given some mild assumptions on $T$ and $F$ we will turn $TF_\tau $ into a monad or embed it into one, where the monadic multiplication deals with visible and silent part of the functor internally.

In the rest of this section we assume the following.  
\begin{itemize}
\item The functor $F:\mathsf{C}\to \mathsf{C}$ lifts to $\overline{F}:\mathcal{K}l(T)\to\mathcal{K}l(T)$ by the corresponding distributive law $\lambda:FT\implies TF$. 
\end{itemize}
As a direct consequence of this assumption the functor $F_\tau = F +  \mathcal{I}d$ lifts to the functor $\overline{F_\tau} = \overline{F+\mathcal{I}d}=\overline{F} +  \mathcal{I}d$ on $\mathcal{K}l(T)$. This follows by the fact that the coproducts in $\mathcal{K}l(T)$ come from coproducts in the base category. The $X$-component of the distributive law $\lambda^\tau$ associated with the lifting $\overline{F_\tau}$ of $F_\tau$ is given by the composition of the following morphisms in $\mathsf{C}$:
$$
\lambda^\tau_X: F_\tau TX = FTX + TX \stackrel{\lambda_X+id_{TX}}{\to} TFX+TX \stackrel{[T\mathsf{inl},T\mathsf{inr}]}{\to}  T(FX+X) = TF_\tau X.
$$
See also e.g. \cite{HasJacSok} for a discussion on liftings of coproducts of functors.

\subsection{Monadic structure on \texorpdfstring{$TF_\tau$}{TF-tau}}  
\label{subsec:monad_on_TF}
We say that $\mathsf{K}$ is \emph{a category with zero morphisms} provided that for any $X,Y\in \mathsf{K}$ there is a morphism $0_{X,Y}:X\to Y$ such that
$$
f\circ 0_{X,Y} = 0_{Y,Z}\circ g = 0_{X,Z} \text{ for any } f:Y\to Z \text{ and } g:X\to Y.
$$
The proof of the lemma below follows directly by the definition of zero morphisms.

\begin{lem}
For any endofunctor  $G:\mathsf{K}\to\mathsf{K}$ the family $0=\{0_{GX,X}\}_{X\in \mathsf{K}}$ is a natural transformation from $G$ to $\mathcal{I}d$. \qed
\end{lem}

\begin{lem}\cite{Milius}\label{lemma:milius_monad}
Let $(S,\mu,\eta)$ be a monad on a category $\mathsf{K}$ with binary coproducts and let $s:H\to S$ be a natural transformation from an endofunctor $H$ on $\mathsf{K}$. Define $(\tilde{S},m,e)$ as follows:
\begin{itemize}
\item $\tilde{S} = HS +\mathcal{I}d$,
\item $e = \mathsf{inr}:\mathcal{I}d \implies HS +\mathcal{I}d$,
\item $m:\tilde{S}^2 \implies \tilde{S}$,
\begin{align*}
&m:\tilde{S}^2 = HS(HS+\mathcal{I}d)+HS+\mathcal{I}d\stackrel{HS(s S+id)+id}{\implies}HS(S^2+\mathcal{I}d)+HS+\mathcal{I}d\\
&\stackrel{HS[\mu,\eta] +id}{\implies} HS^2+HS+\mathcal{I}d\stackrel{[H\mu,\mathsf{inl}]+id}{\implies}HS+\mathcal{I}d = \tilde{S}.
\end{align*}

\end{itemize} 
Then the triple $(\tilde{S},m,e)$ is a monad. \qed

\end{lem}

We are now ready to formulate the main result of this subsection.
\begin{thm}\label{theorem:main_monad_theorem}
Assume that $\mathcal{K}l(T)$ is a category with zero morphisms. 
Define a triple $(\overline{F_\tau},m,e)$, where 
$$
e:\mathcal{I}d\implies \overline{F_\tau} = \overline{F}+\mathcal{I}d; e = \mathsf{inr}
$$
and 
\begin{align*}
&m:\overline{F}(\overline{F}+\mathcal{I}d)+(\overline{F}+\mathcal{I}d)\stackrel{\overline{F}([0,id])+id}{\implies} \overline{F} + (\overline{F}+\mathcal{I}d) \stackrel{[\mathsf{inl},id]}{\implies} \overline{F}+\mathcal{I}d.
\end{align*}
Then the triple $(\overline{F_\tau},m,e)$ is a monad on $\mathcal{K}l(T)$.
\end{thm}
\proof
This theorem follows directly by Lemma \ref{lemma:milius_monad}. Indeed, if we put $S$ to be the identity monad on $\mathcal{K}l(T)$, $H=\overline{F}$ and $s=0:\overline{F}\implies \mathcal{I}d$ then we get $\tilde{S} = \overline{F}+\mathcal{I}d$,
$$e=\mathsf{inr}:\mathcal{I}d \implies \overline{F}+\mathcal{I}d$$ and
$$
\xymatrix{
\overline{F}(\overline{F}+\mathcal{I}d)+(\overline{F}+\mathcal{I}d) \ar@{==>}[d]_{m} \ar@{=>}[drr]|{\overline{F}[0,id] + id }\ar@{=>}[rr]^{\overline{F}(0+id)+id} && \overline{F}(\mathcal{I}d+\mathcal{I}d)+(\overline{F} + \mathcal{I}d)\ar@{=>}[d]^{\overline{F}[id,id] + id} \\
\overline{F}+\mathcal{I}d & & \overline{F}+(\overline{F} + \mathcal{I}d) \ar@{=>}[ll]^{[\mathsf{inl},id]}
}
$$
\qed
Hence, if $\mathcal{K}l(T)$ admits zero morphisms then the functor $TF_\tau:\mathsf{C}\to\mathsf{C}$ carries a monadic structure which is obtained by composing two adjunctions: 
$$
\xymatrix{
\mathsf{C}\ar@<1.5ex>[r]^-{^{\sharp}}\ar@{}[r]|-\perp & \mathcal{K}l(T)\ar@<1.5ex>[r]^-{^{\sharp}}\ar@<1.5ex>[l]^-{U_\mathcal{T}}\ar@{}[r]|-\perp & \mathcal{K}l(\overline{F_\tau})\ar@<1.5ex>[l]^-{U_{\overline{F_\tau}}}.
}
$$
Following the guidelines of Subsection \ref{subsection:monads_on_kleisli} we derive the formula for the composition in $\mathcal{K}l(TF_\tau)=\mathcal{K}l(\overline{F_\tau})$. For $f:X\to TF_\tau Y$, $g:Y\to TF_\tau Z$ we have the following:
$$
\xymatrix@-0.5pc{
TF_\tau Y \ar[r]^-{TF_\tau g} & TF_\tau TF_\tau Z \ar@{=}[r] & T(FTF_\tau Z + TF_\tau Z)\ar[d]_{T(\lambda_{F_\tau Z}+id_{TF_\tau Z})}  
\ar@/^5pc/[dd]^{T\lambda^\tau_{F_\tau Z}}&  \\
X\ar@{-->}[dd]_{g\cdot f} \ar[u]^f  &  &
 T(TF F_\tau Z + T F_\tau Z) \ar[d]_{T[T\mathsf{inl},T\mathsf{inr}]}  \\
 && TT(FF_\tau Z+F_\tau Z)=T^2 F_\tau ^2 Z \ar[d]^{T^2(m_Z)}\\
 TF_\tau Z & T^2F_\tau Z\ar[l]^{\mu_{F_\tau Z}} & T^3F_\tau Z\ar[l]^{T\mu_{F_\tau Z}}  & 
}
$$
\begin{rem}
Before we list some examples of application of this theorem we want to make a remark concerning the assumption about $\mathcal{K}l(T)$ admitting zero morphisms. It is clear that not all monads satisfy this property.  A simple example is the non-empty powerset monad $\mathcal{P}_{\neq \varnothing}$ on $\mathsf{Set}$ whose Kleisli category fails to admit zero morphisms. In order to deal with this problem in the next subsection we present a second strategy towards handling internal moves inside a monadic structure which does not require the Kleisli category to satisfy any extra requirements.
\end{rem}

An example of an application of Theorem \ref{theorem:main_monad_theorem} has already been witnessed in the previous section, namely Fact \ref{fact:LTS_label_monad}. There $T=\mathcal{P}$, $F=\Sigma\times \mathcal{I}d$ and the zero morphisms in $\mathcal{K}l(\mathcal{P})$ are given by 
$$0_{X,Y}:X\to \mathcal{P}Y; x\mapsto \varnothing.$$ To see a second example of an application of this theorem the reader is referred to Section \ref{section:segala}. There,  $T$ is taken to be the convex distribution monad $\mathcal{CM}$ and $F=\Sigma\times \mathcal{I}d$. Since $\mathcal{K}l(\mathcal{CM})$ admits zero morphisms the construction from Theorem~\ref{theorem:main_monad_theorem} will yield a monadic structure on a lifting $\overline{\Sigma_\tau}:\mathcal{K}l(\mathcal{CM})\to\mathcal{K}l(\mathcal{CM})$ of the functor $\Sigma_\tau\times \mathcal{I}d:\mathsf{Set}\to \mathsf{Set}$.  Below, we present a third example which will be used in Section~\ref{section:weak_bisimulation_final_semantics} in order to show  an interesting case of final weak bisimulation semantics for the base category different from $\mathsf{Set}$. 

\begin{exa}\label{example:NA_monad}
Let $T=\mathcal{P}$ and $F=\Sigma\times \mathcal{I}d+1$.  For the sake of simplicity and clarity of notation we identify $F_\tau = F+\mathcal{I}d = \Sigma\times \mathcal{I}d+1+\mathcal{I}d$ with the functor $\Sigma_\tau \times \mathcal{I}d+1:\mathsf{Set}\to \mathsf{Set}$ which is naturally isomorphic to it. The functor $F_\tau$ lifts 
to $\overline{F_\tau}:\mathcal{K}l(\mathcal{P})\to \mathcal{K}l(\mathcal{P})$ given by  \cite{HasJacSok}: 
\begin{align*}
&\overline{F_\tau} X:= F_\tau X = \Sigma_\tau \times X+1 \text{ for any object }X\in \mathcal{K}l(\mathcal{P}),\\
&\overline{F_\tau}f:F_\tau X \rightdcirc F_\tau Y;
\left \{ \begin{array}{ccc} (\sigma,x) & \mapsto & \{(\sigma,y)\mid y\in f(x)\} \\
\checked & \mapsto & \{\checked\} \end{array} \right.\\ 
& \text{ for any morphism }f:X\rightdcirc Y \text{ in } \mathcal{K}l(\mathcal{P}) \text{ (i.e. } f:X\to \mathcal{P}Y).
\end{align*}
The distributive law $\lambda:\Sigma_\tau \times \mathcal{P} + 1 \implies \mathcal{P}(\Sigma_\tau\times \mathcal{I}d +1)$ associated with this lifting is given on the $X$-component by:
$$
\lambda_X:\Sigma_\tau \times \mathcal{P}X +1 \to \mathcal{P}(\Sigma_\tau \times X+1); \left \{ \begin{array}{ccc} (\sigma,X') & \mapsto & \{(\sigma,x)\mid x\in X'\}, \\
\checked & \mapsto & \{\checked\}. \end{array} \right.
$$
In this case, it is not difficult to verify that the monadic structure from Theorem~\ref{theorem:main_monad_theorem} is the following:
\begin{align*}
& e_X:X\rightdcirc F_\tau X; x\mapsto \{(\tau,x)\},  \\
& m_X:F_\tau F_\tau X\rightdcirc F_\tau X; \left \{ \begin{array}{cccc} (\sigma,\tau,x) & \mapsto & \{(\sigma,x)\} &   \\
(\tau,\sigma,x) & \mapsto & \{(\sigma,x)\} &\\
(\sigma_1,\sigma_2,x) & \mapsto & \varnothing & \text{ for }\sigma_1 \neq \tau, \sigma_2\neq \tau,\\
(\tau,\checked) & \mapsto & \{\checked\} & \\
(\sigma,\checked) & \mapsto & \varnothing & \text{ for }\sigma \neq \tau, \\
\checked & \mapsto & \{\checked\}.   \end{array} \right.
\end{align*}
By Subsection \ref{subsection:monads_on_kleisli} we get a monadic structure on $\mathcal{P}F_\tau = \mathcal{P}(\Sigma_\tau \times \mathcal{I}d+1)$. Given two morphisms $f:X\to \mathcal{P}(\Sigma_\tau \times Y+1)$ and $g:Y\to \mathcal{P}(\Sigma_\tau \times Z+1)$ their composition $g\cdot f$ in $\mathcal{K}l(\mathcal{P}(\Sigma_\tau \times \mathcal{I}d+1))$ is:
\begin{align*}
g\cdot f(x) =& \{(\sigma,z)\mid x\stackrel{\sigma}{\to}_f y \stackrel{\tau}{\to}_g z \text{ or }x\stackrel{\tau}{\to}_f y \stackrel{\sigma}{\to}_g z \text{ for some }y\in Y\} \cup \\
 &\{\checked \mid \checked\in f(x) \text{ or } x\stackrel{\tau}{\to}_f y \text{ and }  \checked\in g(y) \text{ for some }y\in Y\}.
\end{align*}
\end{exa}

\begin{rem}
The intuition behind introducing a monadic structure on $\overline{F_\tau}$ as above is the following. If we carefully study the definition of the multiplication $m$ in Theorem \ref{theorem:main_monad_theorem} we see that it kills all visible-visible transitions using the zero morphisms and leaves the rest somewhat intact. By ``visible-visible transitions" we mean the part of the  composition $\overline{F_\tau} \overline{F_\tau}$ which is underlined in the following formula:
$$
\overline{F_\tau}\overline{F_\tau} = \overline{F}\overline{F_\tau} + \overline{F_\tau} = \underbrace{\overline{F}(\overline{F}}+\mathcal{I}d) + \overline{F_\tau}.
$$
In the case of the functor $\overline{\Sigma_\tau}:\mathcal{K}l(\mathcal{P})\to \mathcal{K}l(\mathcal{P})$ from Subsection \ref{subsection:LTS_monad} this part is given by $\overline{\Sigma}\overline{\Sigma}$, i.e. by pairs of visible labels.
\end{rem}

\subsection{Monadic structure on \texorpdfstring{$TF^{*}$}{TF*}}
In the previous subsection we introduced a monadic structure on the functor $TF_\tau $ in a natural way
so that the monadic multiplication deals with silent and visible part of the functor $TF_\tau $ by killing the visible-visible transitions. This was possible thanks to the assumption about existence of zero morphisms in $\mathcal{K}l(T)$. What if, in general, the category $\mathcal{K}l(T)$ does not have this property? The solution we propose here is to consider a free monad $F^{*}$ over $F$. In this subsection instead of imposing a monadic structure on the functor $TF_\tau$, we will embed it into the monad $TF^{*}$.

Since we will only use a direct construction of  a free monad, a curious reader is referred to \cite{Barr} for a detailed definition of this notion. In the rest of the subsection we assume the following:
\begin{itemize}
\item the functor $F$ admits a free $F$-algebra $i_X$ in $\mathsf{C}^F$ (=initial $F(-) +  X$-algebra in $\mathsf{C}^{F(-)+X}$)  for any object $X$.
\end{itemize} 

\begin{lem}\cite{Barr}\label{lemma:free_monad_algebra}
For an object $X$ and a morphism $f:X\to Y$ in $\mathsf{C}$ let $F^{*}X$ denote the carrier of $i_X$ and  let $F^{*}f:F^{*}X\to F^{*}Y$ denote the unique morphism for which the following diagram commutes:
\begin{align}
\xymatrix@-0.7pc{
FF^{*}X+X\ar[rr]^{i_X}\ar@{-->}[d]_{F(F^{*}f)+id} & & F^{*}X\ar@{-->}[d]^{F^{*}f} \\
FF^{*}Y+X\ar[r]_{id+f}& 
FF^{*}Y+Y\ar[r]_-{i_Y} & F^{*}Y
}
\label{diagram:1_lemma_barr}
\end{align}  
The assignment $F^{*}$ is functorial. Define a transformation $m:F^{*}F^{*}\implies F^{*}$, whose $X$-component $m_X:F^{*}F^{*}X\to F^{*}X$ is the unique morphism making the following diagram commute:
\begin{align}
\xymatrix@-0.7pc{
FF^{*}F^{*}X+F^{*}X\ar[rr]^{i_{F^{*}X}}\ar@{-->}[d]_{F(m_X)+id} & &  F^{*}F^{*}X\ar@{-->}[d]^{m_X} \\
FF^{*}X+F^{*}X\ar[rr]_{[i_X\circ \mathsf{inl},id]} & & F^{*}X
}
\label{diagram:2_lemma_barr}
\end{align}
Then the triple $(F^{*},m,e)$, where $e_X:X\to F^{*}X; e_X = i_X \circ \mathsf{inr}$, is a free monad over $F$. \qed
\end{lem}

The remaining part of this subsection is devoted to proving that the functor $F^{*}$ lifts to a free monad over $\overline{F}$ in $\mathcal{K}l(T)$.

\begin{lem}\cite{JacSilSok}\label{lemma:adjuntion_lifting}
The adjunction $\mathsf{C}\rightleftarrows \mathcal{K}l(T)$ lifts to an adjunction $\mathsf{C}^{F}\rightleftarrows \mathcal{K}l(T)^{\overline{F}}$ between categories of algebras. To be more precise, we have the following diagram in which the vertical arrows are the forgetful functors:
$$
\xymatrix{
\mathsf{C}^{F}\ar[dd] \ar@<1.5ex>[r]^-{\hat{^{\sharp}}}\ar@{}[r]|-\perp  & \mathcal{K}l(T)^{\overline{F}}\ar[dd]\ar@<1.5ex>[l]^{\hat{U_T}}\\\\
\mathsf{C}\ar@<1.5ex>[r]^-{^{\sharp}}\ar@{}[r]|-\perp & \mathcal{K}l(T) \ar@<1.5ex>[l]^-{U_T}
}
$$
Here, $\hat{^\sharp}:\mathsf{C}^F\to \mathcal{K}l(T)^{\overline{F}}$ is defined for any object $a:FA\to A$ and a homomorphism $f:A\to B$ between $a:FA\to A$ and $b:FB\to B$ by
$a^{\hat{\sharp}} := \eta_A\circ a$  and  $f^{\hat{\sharp}} := \eta_B\circ f$. \qed
\end{lem}

\begin{lem}
For any object $X\in \mathcal{K}l(T)$  the morphism 
$$
i_X^\sharp:F(F^{*}X)+X\stackrel{  i_X }{\to} F^{*}X\stackrel{\eta_{F^{*}X}}{\to} TF^{*} X
$$
is a free $\overline{F}$-algebra over $X$ in $\mathcal{K}l(T)^{\overline{F}}$ (=initial $\overline{F}(-)+X$-algebra in $\mathcal{K}l(T)^{\overline{F}(-)+X}$).
\end{lem}
\proof
Since $F$ lifts to $\overline{F}$, the functor $F(-)+X$ lifts to $\overline{F}(-)+X:\mathcal{K}l(T)\to\mathcal{K}l(T)$ with the corresponding lifting $\lambda'$ given on its $Y$-component by:
$$
\lambda'_Y:FTY+X \stackrel{\lambda_{Y}+\eta_X}{\to}  TFY+TX\stackrel{[T\mathsf{inl},T\mathsf{inr}]}{\to} T(FY+X).
$$
Here, the reader is once again referred to \cite{HasJacSok} for a discussion on liftings of coproducts of functors. By Lemma \ref{lemma:adjuntion_lifting} and the fact that the initial object is a colimit over the empty diagram together with the fact that any left adjoint preserves colimits we infer that $i^\sharp_X$ is initial in $\mathcal{K}l(T)^{\overline{F}(-)+X}$. This proves the assertion.
\qed
Let $\overline{F}^{*}:\mathcal{K}l(T)\to\mathcal{K}l(T)$ denote the functor obtained by following the guidelines of the construction from Lemma \ref{lemma:free_monad_algebra} using the family $\{i_X^\sharp\}_{X\in \mathcal{K}l(T)}$ of all free $\overline{F}$-algebras. 
\begin{thm}\label{theorem:monad_free}
We have:
\begin{itemize}
\item $F^{*}$ lifts to $\overline{F}^{*}$,
\item $(\overline{F}^{*},m^\sharp,e^\sharp)$ is a free monad over $\overline{F}$. 
\end{itemize}
\end{thm}
\proof
By the definition of $\overline{F}^{*}$ it follows that for any object $X\in \mathsf{C}$ (i.e. $X\in \mathcal{K}l(T)$) we have $\overline{F}^{*}X = F^{*}X$.   Therefore, in order to prove that $F^{*}$ lifts to $\overline{F}^{*}$ it is enough to show that for any $f:X\to Y$ we have $\overline{F}^\ast f^\sharp = (F^{*}f)^\sharp$. We will do this by showing that the morphism $(F^{*}f)^\sharp:F^{*}X\to TF^{*}Y$ makes the diagram (\ref{diagram:1_lemma_barr}) commute in the category $\mathcal{K}l(T)$. We have 
$$
\overline{F}((F^{*}f)^\sharp)+f^\sharp = (FF^{*}f)^\sharp+f^\sharp = [FF^{*}f+f]^\sharp,
$$
where, all coproducts in the above expression except for the last one live in $\mathcal{K}l(T)$. The last coproduct is taken in $\mathsf{C}$. This identity follows by the fact that any functor which is a left adjoint commutes with colimits. 
Hence, the following diagram commutes in $\mathcal{K}l(T)$:
$$
\xymatrix{
FF^{*}X+X\ar[rr]|{\circ}^{i_X^\sharp }\ar[d]|{\circ}_{\overline{F}((F^{*}f)^\sharp)+1_X} \ar[dr]|{(FF^{\ast}f+f)^\sharp } & & F^{*}X\ar[d]|{\circ}^{(F^{*}f)^\sharp} \\
FF^{*}Y+X\ar[r]|{\circ}_{1+f^\sharp}& 
FF^{*}Y+Y\ar[r]|{\circ}_{i_Y^\sharp} & F^{*}Y
}
$$  
Since $\overline{F}^{*}(f^\sharp)$ is the unique morphism for which this diagram commutes, we get that $\overline{F}^\ast f^\sharp = (F^{*}f)^\sharp$. This proves the first statement. We will now prove the second statement. By the definition of $\overline{F}^{*}$ and Lemma \ref{lemma:free_monad_algebra} the functor $\overline{F}^{*}$ is a free monad over $\overline{F}$. In order to complete the proof  we have to show that the unit and the multiplication of the monad $\overline{F}^{*}$ obtained by following the guidelines of Lemma~\ref{lemma:free_monad_algebra} coincide with $e^\sharp$ and $m^\sharp$ respectively. We proceed as in the first part of the proof. We take the morphism $m_X^\sharp$ and show it makes the diagram (\ref{diagram:2_lemma_barr})
commute for $\overline{F}^{*}$ in $\mathcal{K}l(T)$. By uniqueness, we get that $m^\sharp$ is the multiplication of the free monad $\overline{F}^\ast$ and $e^\sharp$ is its unit. This completes the proof.
\qed
Note that in the above proof we did not refer to the distributive law associated with the lifting $\overline{F}^{*}$ of $F^{*}$. We will now give its explicit description. By Subsection~\ref{subsection:monads} this distributive law $\lambda^{*}:F^{*}T\implies TF^{*}$  is given by $\lambda^{*}_X := \overline{F}^{*}(id_{TX})$. It follows directly by the definition of $\overline{F}^{*}$ that $\lambda_{X}^{*}$ is the unique morphism making the following diagram commute in $\mathsf{C}$ (existence and uniqueness of $\lambda^{*}_X$ is guaranteed by the fact that $i_{TX}$ is an initial $F(-)+TX$-algebra):
%$$
%\xymatrix{
%\overline{F}\overline{F}^{*}TX+TX\ar@{-o}[rr]^{i^\sharp_{TX}}\ar@{--o}[d]_{\overline{F}(\lambda^{*})+1} & &  \overline{F}^{*}TX\ar@{--o}[d]^{\lambda^{*} = \overline{F}^{*}(id_{TX})} \\
%\overline{F}\overline{F}^{*}X+TX\ar@{-o}[r]_{1+id_{TX}}&  
%\overline{FF}^{*}X+X\ar@{-o}[r]_-{i^\sharp _{F^{*}X}} & \overline{F}^{*}X
%}
%$$ 
%The diagram above rewritten in terms of the composition in $\mathsf{C}$ is given by:
$$
\xymatrix{
FF^{*}TX+TX\ar[rrr]^{i_{TX}}\ar@{-->}[d]_{F(\lambda^{*}_{TX})+id} & & &  F^{*}TX\ar@{-->}[d]^{\lambda^{*}_{TX}} \\
FTF^{*}X+TX\ar[r]_{\lambda+id}&TFF^{*}X+TX \ar[r]_{[T\mathsf{inl},T\mathsf{inr}]}& 
T(FF^{*}X+X)\ar[r]_-{Ti_{F^{*}X}} & TF^{*}X
}
$$ 

The adjunctions $\mathsf{C}\leftrightarrows \mathcal{K}l(T) \leftrightarrows \mathcal{K}l(\overline{F}^{*})$ yield a monadic structure on $TF^{*}$.
%$$
%\xymatrix{
%\mathsf{C}\ar@/^1.5pc/[r]^{^{\sharp}}\ar@{}[r]|\perp & \mathcal{K}l(T)\ar@/^1.5pc/[r]^{^{\sharp}}\ar@/^1.5pc/[l]^{U_\mathcal{T}}\ar@{}[r]|\perp & \mathcal{K}l(\overline{F}^{\ast})\ar@/^1.5pc/[l]^{U_{\overline{F}^{\ast}}}
%}
%$$
By Subsection \ref{subsection:monads_on_kleisli} the composition $\cdot$ in $\mathcal{K}l(TF^{*})$ is given for $f:X\to TF^{*}Y$ and $g:Y\to TF^{*}Z$ by:
$$
\xymatrix{
X\ar@{-->}[d]_{g\cdot f} \ar[r]^f & TF^*Y \ar[r]^{TF^{*}g} &
TF^{*}TF^{*}Z \ar[r]^{T\lambda^*_{F^{*}Z}} & T^2F^{{*}}F^{{*}}Z \ar[d]^{T^2(m_Z^\sharp)} \ar[dl]|{T^2(m_Z)} \\
 TF^{*}Z & &T^2F^{*}Z\ar[ll]^{\mu_{F^{*}Z}} & T^3F^{*}Z\ar[l]^{T\mu_{F^{*}Z}} 
}
$$
We will now elaborate more on the example that was briefly discussed in Subsection~\ref{subsection:LTS_extended_monad}.\enlargethispage{\baselineskip}
\begin{exa}
Take $T=\mathcal{P}$ and $F = \Sigma\times \mathcal{I}d$. It is easy to see that the functor $F$ satisfies the assumptions from the beginning of this subsection and that Lemma~\ref{lemma:free_monad_algebra} yields $F^\ast = \Sigma^{*}\times \mathcal{I}d$ with $m$ and $e$ defined on their $X$-components as follows:
\begin{align*}
& m_X:\Sigma^{*}\times \Sigma^{*} \times X\to \Sigma^{*}\times X; (s_1,s_2,x) \mapsto (s_1s_2,x),\\
& e_X:X\to \Sigma^{*}\times X; x\mapsto (\varepsilon,x).
\end{align*}
The lifting $\overline{F}^{*}:\mathcal{K}l(\mathcal{P})\to\mathcal{K}l(\mathcal{P})$ is given by:
\begin{align*}
&\overline{F}^{*} X = \Sigma^{*}\times X \text{ for }X\in \mathcal{K}l(\mathcal{P}),\\
&\overline{F}^{*}f:\Sigma^{*}\times X\to \mathcal{P}(\Sigma^{*}\times Y); (s,x) \mapsto\{(s,y)\mid y\in f(x) \} \text{ for }f:X\to \mathcal{P}Y. 
\end{align*}
By Theorem \ref{theorem:monad_free} the triple $(\overline{F}^{*},m^\sharp,e^\sharp)$ is a monad on $\mathcal{K}l(\mathcal{P})$ which, by Subsection~\ref{subsection:monads_on_kleisli}, yields a monadic structure on $\mathcal{P}F^{*} = \mathcal{P}(\Sigma^{*}\times \mathcal{I}d)$. For $f:X\to \mathcal{P}(\Sigma^{*}\times Y)$ and $g:Y\to \mathcal{P}(\Sigma^{*}\times Z)$ we have $g\cdot f:X\to \mathcal{P}(\Sigma^{*}\times Z)$:
$$
g\cdot f (x) = \{(s_1s_2,z)\mid x\stackrel{s_1}{\to}_f y \stackrel{s_2}{\to}_g z \text{ for some }y\in Y\}.
$$
\end{exa}

We end this subsection with the following lemma.

\begin{lem}
There is a natural transformation $\nu$ from $F_\tau = F+\mathcal{I}d$ to $F^{*}$ given for any object $X\in \mathsf{C}$ by 
$$
\nu_X:= FX+X \stackrel{Fe_X +id_X}{\to} F(F^{*}X)+X\stackrel{i_X}{\to} F^{*}X.
$$\qed
\end{lem}
The above result implies that any $TF_\tau$-coalgebra can be naturally translated into a $TF^{*}$-coalgebra. To be more precise for $\alpha:X\to TF_\tau X$ we put 
$$\underline{\alpha} := T\nu_X\circ \alpha:X\to TF^{*}X.$$
Although the functor $TF_\tau$ is not necessarily a monad, any $TF_\tau$-coalgebra  can be turned into a $TF^{*}$-coalgebra. The functor $TF^{*}$ carries a monadic structure that handles internal transitions. Therefore, from now on coalgebras with internal moves will be considered as coalgebras whose type is a monad without referring to their visible and invisible part of the transition explicitly. 

\section{Ordered saturation monads}
\label{section:saturation}
 Recall that for labelled transition systems the saturated map $\alpha^{*}$  can in fact be thought of as the reflexive and transitive closure of $\alpha:X\to \mathcal{P}(\Sigma_\tau \times X)$ w.r.t. the Kleisli composition~$\cdot$ and the partial order $\leqslant$ given in Section \ref{section:LTS_coalgebraically}. It is easy to see that it satisfies the following conditions: $1_X\leqslant\alpha$, $\alpha\leqslant\alpha^{*}$ and $\alpha^{*}\cdot \alpha^{*}\leqslant \alpha^{*}$ and $\alpha^{*}$ is the least morphism with these properties. The aim of this section is to present a definition of an ordered saturation monad which generalizes this idea.

A monad $(T,\mu,\eta)$ on $\mathsf{C}$ whose Kleisli category is order enriched is called \emph{ ordered saturation monad } provided that in $\mathcal{K}l(T)$ for any morphism $\alpha: X \rightdcirc X$ there is a morphism $\alpha^{*}:X\rightdcirc X$ satisfying the following conditions:
\begin{enumerate}
\item \label{axiom:1} $1\leqslant\alpha^{*}$,
\item \label{axiom:2} $\alpha\leqslant\alpha^{*}$,
\item \label{axioms:3} $\alpha^{*}\cdot \alpha^{*}\leqslant \alpha^{*}$,
\item \label{axiom:3.5} if $\beta:X\rightdcirc X$ satisfies $1\leqslant \beta$, $\alpha\leqslant \beta$ and $\beta\cdot \beta\leqslant \beta$ then $\alpha^{*}\leqslant \beta$,
\item \label{axioms:4} for any $f:X\to Y$ in $\mathsf{C}$ and any $\beta:Y\rightdcirc Y$ in $\mathcal{K}l(T)$ we have:
$$
f^{\sharp}\cdot \alpha \Box \beta \cdot f^{\sharp} \implies f^{\sharp}\cdot \alpha^{*} \Box \beta^{*} \cdot f^{\sharp} \text{ for } \Box \in \{\leqslant,\geqslant\}.
$$
\end{enumerate}

From now on, unless stated otherwise, in this section we assume that $(T,\mu,\eta)$ is an ordered saturation monad.

\begin{lem}
\label{lemma:saturation:properties}
For any $\alpha:X\rightdcirc X$ in $\mathcal{K}l(T)$ we have:
\begin{itemize}
\item $\alpha^{**}= \alpha^{*}$,
\item $\alpha^{*} = \alpha^{*}\cdot \alpha^{*}$,
\item $1^{*} = 1$.
\end{itemize}
\end{lem}

\proof
To prove the first assertion we use Condition (\ref{axiom:2}) and see $\alpha^{*}\leqslant \alpha^{**}$. Moreover, by Condition (\ref{axiom:3.5}) since $1\leqslant \alpha^{*}$, $\alpha^{*}\leqslant \alpha^{*}$ and $\alpha^{*}\cdot \alpha^{*}\leqslant \alpha^{*}$ it follows that $\alpha^{**}\leqslant \alpha^{*}$.  
The last two assertions follow easily. 
\qed 

We will often refer to $(-)^{*}$ operator as \emph{saturation} operator and call a structure $\alpha$ \emph{saturated} if $\alpha^{*} = \alpha$ (i.e. if $1\leqslant\alpha$ and $\alpha\cdot \alpha \leqslant \alpha$).

We will now discuss Conditions (\ref{axiom:1})-(\ref{axioms:4}). At first let us focus on the last condition. It says that if a morphism $f$ is a (lax or oplax) homomorphism between two structures $\alpha$ and $\beta$ then it is also a (lax or oplax) homomorphism between their saturations $\alpha^{*}$ and $\beta^{*}$. This technical property proves to be useful in Section \ref{section:weak_bisimulation} when defining weak bisimulation and studying its properties. Its stronger version is also used in Theorem \ref{theorem:saturation_fixed_point} in order to pinpoint the relation between $\alpha^{*}$ and the least fixpoint of the assignment $x\mapsto 1\vee x\cdot \alpha$.  Let us now focus on Conditions (\ref{axiom:1})-(\ref{axiom:3.5}). Saturating a structure $\alpha$ can be thought of as an abstract way to consider its reflexive and transitive closure. Indeed, the first four axioms say that given a coalgebra $\alpha$ the structure $\alpha^{*}$ is the least coalgebra closed under the composition such that $1\leqslant\alpha^{*}$ and $\alpha\leqslant\alpha^{*}$.
However, it may not be instantly clear to the reader why we choose Conditions (\ref{axiom:1})-(\ref{axiom:3.5}) to define the saturator and not any other. These conditions bare some resemblance to the axioms of Kleene algebra $(A,+,\cdot,^{*})$ \cite{Kozen} and Kleene monad \cite{Gov}. We are intentionally not using those, as some of our main examples fit the framework presented above but would not fit the Kleene monad framework. To be more precise, one of the requirements for a monad to be a Kleene monad is that the composition in $\mathcal{K}l(T)$ distributes over finite suprema. However, the monad used to model Segala systems described in Section~\ref{section:segala} gives rise to its Kleisli category where the composition does not satisfy this property.  Here, we would also like to give a categorical interpretation of saturated coalgebras and Conditions (\ref{axiom:1})-(\ref{axiom:3.5}). Any order enriched category is a special type of a $2$-category with the $2$-cell structure imposed by the partial order on its hom-sets. In an arbitrary 2-categorical setting one can introduce the notion of a monad \emph{in} a $2$-category as a $1$-cell $t:X\to X$ with the same domain and codomain whose unit and multiplication are given by $2$-cells $\eta:id_X\implies t$ and $\mu:t\circ t\implies t$ satisfying the usual monad laws  (see e.g. \cite{Lack} for basic definitions and properties from $2$-category theory). In this case any ordinary monad on a category is a monad in $\mathsf{Cat}$, the 2-category of all categories as objects, functors as 1-cells and natural transformations as 2-cells.  Now, if we view the order enriched category $\mathcal{K}l(T)$ as a 2-category then the saturated coalgebras are exactly monads \emph{in} $\mathcal{K}l(T)$ with the unit and the multiplication given by the $2$-cells $1 \leqslant \alpha$ and $\alpha\cdot \alpha \leqslant \alpha$ respectively (the monad laws hold vacuously in this setting). A~careful study of the definition of a free monad over a functor \cite{Barr} leads to a conclusion that by Conditions (\ref{axiom:2}) and (\ref{axiom:3.5}) the saturated coalgebra $\alpha^{*}:X\rightdcirc X$ can be in fact thought of as a free monad over $\alpha$.

The following theorem shows the connection between saturation and the least fixed point of the assignment $x\mapsto 1\vee x\cdot \alpha$. 

\begin{thm}\label{theorem:saturation_fixed_point}
 Assume that hom-sets in $\mathcal{K}l(T)$ admit finite joins and that
 for any $f:X\rightdcirc Y$ and $\beta:Y\rightdcirc Y$ in $\mathcal{K}l(T)$ we have:
\begin{align}
f\cdot \alpha \Box \beta \cdot f \implies f \cdot \alpha^{*} \Box \beta^{*} \cdot f \text{ for } \Box \in \{\leqslant,\geqslant\}. \label{assumption:additional}
\end{align}
Then $\alpha^{*} = \mu x. (1\vee x\cdot \alpha)$.
\end{thm}
\proof
In the first part of the proof we show that for any $\alpha:X\rightdcirc X$ if $\gamma=1\vee \gamma \cdot \alpha$ for a structure $\gamma:X\rightdcirc X$ then $\alpha^{*} \leqslant \gamma$. Assume $\gamma=1\vee \gamma \cdot \alpha$. Then $\gamma \cdot \alpha \leqslant \gamma=1\cdot \gamma$. By implication (\ref{assumption:additional}) we get $\gamma \cdot \alpha^{*}\leqslant 1^{*}\cdot \gamma = 1\cdot \gamma = \gamma$. Since $1\leqslant \gamma$ we can infer that
$$\alpha^{*} = 1\cdot \alpha^{*} \leqslant \gamma \cdot \alpha^{*} \leqslant \gamma.$$
In order to complete the proof  we need to show that  $\alpha^{*} = 1\vee \alpha^{*}\cdot \alpha$.  The proof below is a simplification of a proof generated by Prover9 \cite{McCune}. The following properties hold.
\begin{enumerate}[label=\({\alph*}]
\item $(\alpha^{*} \cdot \alpha)^{*} = \alpha^{*}$. By Condition (\ref{axioms:4}) and Lemma~\ref{lemma:saturation:properties} since $$\alpha\leqslant \alpha^{*}\cdot \alpha \leqslant \alpha^{*}\cdot \alpha^{*} = \alpha^{*}.$$ \label{property:a}
\item $\alpha \cdot \alpha^{\ast} = \alpha^{*}\cdot \alpha$. By the equality $\alpha\cdot \alpha = \alpha\cdot \alpha$ and the implication (\ref{assumption:additional}). \label{property:b}
\item $\alpha^{*}\cdot (1\vee \alpha) = \alpha^{*}$. By the following inequalities:
$$\alpha^{*} = \alpha^{*}\cdot 1 \leqslant  \alpha^{*}\cdot (1\vee \alpha) \leqslant \alpha^{*} \cdot \alpha^{*} = \alpha^{*}.$$  
\item $\alpha\cdot \alpha^{*}\cdot (1\vee \alpha) = \alpha\cdot \alpha^{*}$. By multiplying the previous equality by $\alpha$. \label{property:c}
\item $\alpha^{*}\cdot \alpha \cdot (1\vee \alpha^{*}\cdot \alpha) = \alpha^{*}\cdot \alpha$. Substitute $\alpha$ with $\alpha^{*}\cdot \alpha$ in (\ref{property:c}). We get:
$$
\alpha^{*}\cdot \alpha \cdot (\alpha^{*} \cdot \alpha)^{*}\cdot (1\vee \alpha^{*}\cdot \alpha ) = \alpha^{*}\cdot \alpha \cdot (\alpha^{*} \cdot \alpha)^{*}.
$$ 
We use (\ref{property:a}) and obtain:
$$
\alpha^{*}\cdot \alpha \cdot \alpha^{*} \cdot (1\vee \alpha^{*}\cdot \alpha ) = \alpha^{*}\cdot \alpha \cdot \alpha^{*}.
$$ 
 By applying (\ref{property:b}) and $\alpha^{\ast}\cdot \alpha^{*} = \alpha^{*}$ we get the desired conclusion. \label{property:e}
\item $\gamma \cdot \beta \leqslant \beta \implies  \gamma^{*}\cdot \beta = \beta$. If $\gamma \cdot \beta  \leqslant \beta =  \beta\cdot 1$ then $ \gamma ^{*}\cdot \beta \leqslant \beta$.
Moreover, since $1\leqslant\gamma^{*}$ we get $\beta \leqslant  \gamma^{*}\cdot \beta$ which proves the assertion.  \label{property:f}
\item $1\vee \alpha^{*}\cdot \alpha = \alpha^{*}\cdot (1\vee \alpha^{*}\cdot \alpha)$. By (\ref{property:e}) we get
$$
\alpha^{*}\cdot \alpha\cdot (1\vee \alpha^{*}\cdot \alpha) \leqslant 1\vee \alpha^{*}\cdot \alpha.
$$
By applying (\ref{property:f}) with $\beta = 1\vee \alpha^{*}\cdot \alpha$ and $\gamma = \alpha^{*}\cdot \alpha $ and by using (\ref{property:a}) we get:
$$
\alpha^{*} \cdot (1\vee \alpha^{*}\cdot \alpha) = (\alpha^{*}\cdot \alpha)^{*}\cdot (1\vee \alpha^{*}\cdot \alpha) = 1\vee \alpha^{*}\cdot \alpha.
$$\label{property:g}
\item $\alpha^{*} = 1\vee \alpha^{*}\cdot \alpha$. By (\ref{property:g}) and the fact that $\alpha^{*}\cdot (1\vee \alpha^{*}\cdot \alpha) = \alpha^{*}$. The last equality follows by
$$
\alpha^{*} \leqslant \alpha^{*}\cdot (1\vee \alpha^{*}\cdot \alpha) \leqslant \alpha^{*} \cdot \alpha^{*} =\alpha^{*}.
$$
\end{enumerate}

\qed
\begin{rem}
It is important to see that a monad $T$ is an ordered saturation monad satisfying implication (\ref{assumption:additional}) if and only if the identity monad $\mathcal{I}d$ on the Kleisli category $\mathcal{K}l(T)$ is an ordered saturation monad. 
We do not know if the above theorem remains true if we remove implication (\ref{assumption:additional}) from the assumptions. We leave it as an open problem.
\end{rem}

\subsection{Ordered saturation monads: sufficient conditions}
The purpose of this subsection is to list different sufficient conditions that guarantee a given monad is an ordered saturation monad. The subsection is divided into three independent paragraphs. The results presented here are used in Subsection \ref{subsection:saturation_for_LTS} and Section~\ref{section:segala} in order to prove that the monads associated with labelled transition systems and Segala systems are ordered saturation monads.

\subsubsection{Kleene monads are ordered saturation monads} 
\label{subsection:saturation_Kleene} The purpose of this paragraph is to show that any Kleene monad \cite{Gov} is an ordered saturation monad. A monad $T$ is called \emph{additive} if the category $\mathcal{K}l(T)$ is enriched over join-semilattices with a least element. To be more precise, for any two objects $X,Y\in \mathcal{K}l(T)$ the set $Hom_{\mathcal{K}l(T)}(X,Y)$ is a join-semilattice with the join operation given by $\vee$ and the smallest element $\perp$ satisfying:
\begin{itemize}
\item $(f\vee g)\cdot h = f\cdot h\vee g\cdot h$ for any $f,g:X\rightdcirc Y$ and $h:Z\rightdcirc X$,
\item $h\cdot (f\vee g) = h\cdot f\vee h\cdot g$ for any $f,g:X\rightdcirc Y$ and $h:X\rightdcirc Z$,
\item $f\cdot \perp =\perp \cdot f=  \perp $ for any $f:X\rightdcirc Y$.
\end{itemize} 
An additive monad $(T,\mu,\eta)$ is called \emph{Kleene monad} \cite{Gov} if for any composable morphisms $p,r$ in $\mathcal{K}l(T)$ the assignments:
$$
f\mapsto p\vee f\cdot r \text{ and } f\mapsto p\vee r\cdot f
$$
both have least fixpoints w.r.t. the order on $Hom_{\mathcal{K}l(T)}(X,Y)$ and for any morphism $q$ with a suitable domain and codomain:
\begin{align}
&\mu x .(p\cdot q \vee x\cdot r ) =p \cdot \mu x.(q\vee x\cdot r),\label{identity_kleene:1}\\
&\mu x .(p\cdot q \vee r\cdot x ) = \mu x.(p\vee r\cdot x)\cdot q.\label{identity_kleene:2}
\end{align}
These two conditions can be equivalently restated in terms of a single identity given by \cite{Gov}:
$$
\mu x .(p \vee x\cdot r )\cdot q =p \cdot \mu x.(q\vee r\cdot x).
$$

\begin{thm}\label{theorem:kleene_monad_ordered_saturation}
Any Kleene monad is an ordered saturation monad which additionally satisfies:
\begin{align*}
f\cdot \alpha \Box \beta \cdot f \implies f \cdot \alpha^{*} \Box \beta^{*} \cdot f \text{ for } \Box \in \{\leqslant,\geqslant\} 
\end{align*}
 for any $\alpha:X\rightdcirc X$, $f:X\rightdcirc Y$ and $\beta:Y\rightdcirc Y$.
\end{thm} 
\proof
For any  $\alpha:X\rightdcirc X$ we define $\alpha^{*}:X\rightdcirc X$ to be
$$
\alpha^{*} := \mu x.(1\vee x\cdot \alpha) = \mu x.(1\vee \alpha\cdot x).
$$
Conditions (\ref{axiom:1}) and (\ref{axiom:2}) are obviously true. 
To see (\ref{axioms:3}) holds set $p=\alpha^{*}$, $q=1$ and $r = \alpha$ in identity (\ref{identity_kleene:1}). Then $p\cdot \alpha = \alpha^{*}\cdot \alpha\leqslant \alpha^{*}$ and   
$$
p \cdot \alpha^{*} = p \cdot \mu x.(1\vee x\cdot \alpha) = \mu x .(p \vee x\cdot \alpha ) = p.
$$
Condition (\ref{axiom:3.5}) follows directly from (\ref{axiom:1})-(\ref{axioms:3}) as if $\beta$ satisfies $1\leqslant \beta$, $\alpha\leqslant \beta$ and $\beta \cdot \beta\leqslant \beta$ it follows that $1\vee \beta \cdot (\alpha\vee 1) = \beta$. Since $\alpha^{*} = \mu x. (1\vee x\cdot \alpha)$ is also the least fixpoint of the assignment $x\mapsto 1\vee x\cdot (\alpha\vee 1)$ we get $\alpha^{*}\leqslant \beta$. To prove the last  assertion assume $f\cdot \alpha \leqslant  \beta \cdot f$. We have $f\cdot\alpha\leqslant \beta^{*}\cdot f$. Hence $\beta^{*}\cdot f\cdot \alpha\leqslant \beta^{*}\cdot \beta^{*}\cdot f \leqslant \beta^{*}\cdot f$. Let $p=\beta^{*}\cdot f$. Then $p\cdot \alpha \leqslant p$. By the properties of a Kleene monad we get
$p\cdot \alpha^{*} = p\cdot \mu x.(1\vee x\cdot \alpha) =\mu x. (p\vee x\cdot \alpha) \leqslant p$. Hence, $\beta^{*}\cdot f\cdot \alpha^{*}\leqslant \beta^{*}\cdot f$. Since $1\leqslant \beta^{*}$ we get $f\cdot \alpha^{*}\leqslant \beta^{*}\cdot f$. Analogous reasoning proves the implication for the inverse inequality.
\qed

\begin{exa} \label{example:powerset_kleene_monad}
It is an easy exercise to prove that the powerset monad $\mathcal{P}$ is a Kleene monad (see \cite{Gov} for details). For any $\mathcal{P}$-coalgebra $\alpha:X\to \mathcal{P}X$ the structure $\alpha^{*}~:~X\to \mathcal{P}X$ is given by $\alpha^{*}(x) = \bigcup_{n\in \mathbb{N}\cup \{0\}} \alpha^n (x)$,
where $\alpha^{0} = 1_X$ is the $X$-component of the unit of $\mathcal{P}$.  As we have seen in Example~ \ref{example:kleisli_powerset} any structure $\alpha:X\to \mathcal{P}X$ may be viewed as a relation $R_\alpha\subseteq X\times X$.
The relation $R_{\alpha^{*}}$ associated with $\alpha^{*}:X\to\mathcal{P}X$ is exactly the reflexive and transitive closure of $R_{\alpha}$.
\end{exa}

As a direct consequence of Theorem \ref{theorem:kleene_monad_ordered_saturation} we get the following.
\begin{cor}\label{coro:kleene_identity}
If $T$ is a Kleene monad then the identity monad $\mathcal{I}d$ on $\mathcal{K}l(T)$ is an ordered saturation monad. \qed
\end{cor}

\subsubsection{$\omega$-cpo enriched categories and ordered saturation monads} The aim of this paragraph is to show that if a monad $T$ satisfies the properties listed below it is an ordered saturation monad. 
This fact will be used in Section \ref{section:segala} to prove that the convex distribution monad $\mathcal{CM}$ is an ordered saturation monad.

 We say that a category is \emph{enriched over $\omega$-complete partial orders} or simply \emph{$\omega$-cpo enriched} if each hom-set is a poset in which a supremum of every countable directed family of morphisms $f_1\leqslant f_2\leqslant\ldots$ exists and that the composition preserves these suprema. 

In this paragraph we assume the following:
\begin{itemize}
\item $\mathcal{K}l(T)$ is  $\omega$-cpo enriched,
\item hom-sets in $\mathcal{K}l(T)$ admit finite joins,
\item for any $\alpha:X\rightdcirc X$, any $f:X\to Y$ in $\mathsf{C}$ and $\beta:Y\rightdcirc Y$ the following implications hold:
$$
f^{\sharp}\cdot \alpha \Box \beta \cdot f^{\sharp} \implies f^{\sharp}\cdot (\alpha\vee 1) \Box (\beta \vee 1) \cdot f^{\sharp} \text{ for } \Box \in \{\leqslant,\geqslant\}.
$$
\end{itemize}
Note that although we assume that hom-sets in $\mathcal{K}l(T)$ admit finite joins, the composition does not necessarily distribute over them in general. 
\begin{thm}
\label{theorem:dcpo_enriched_monads_star_monads}
If $T$ is a monad satisfying the above properties then it is an ordered saturation monad, where for any morphism $\alpha:X\rightdcirc X$ the morphism $\alpha^{*}$ is given by $$\alpha^{*} = \bigvee_{n\in \mathbb{N}} (1\vee\alpha)^n.$$
\end{thm}
\proof
The first two axioms of ordered saturation monad follow directly by the definition of $\alpha^{*}$. To see Condition (\ref{axioms:3}) holds note that 
$$\alpha^{*}\cdot \alpha^{*} = \left(\bigvee_n (1\vee\alpha)^n\right)\cdot \left(\bigvee_m (1\vee\alpha)^m\right) \stackrel{\Diamond}{=} \bigvee_{m,n} (1\vee \alpha)^n\cdot (1\vee \alpha)^m = \alpha^{*},$$
where the equality marked with $\Diamond$ follows by the fact that $\mathcal{K}l(T)$ is $\omega$-cpo enriched. Condition~ (\ref{axiom:3.5}) follows directly from the definition of $\alpha^{*}$. Indeed, if $\beta$ satisfies the assumptions listed in Condition~(\ref{axiom:3.5}) then $(1\vee \alpha)\leqslant \beta$. Now by induction hypothesis assume that $(1\vee\alpha)^n\leqslant\beta$. Then $(1\vee\alpha)^{n+1} = (1\vee \alpha)^n\cdot (1\vee\alpha) \leqslant \beta\cdot \beta\leqslant\beta$. This proves that for any $n$ we have $(1\vee\alpha)^n\leqslant\beta$. Hence, $$\alpha^{*}=\bigvee_n (1\vee \alpha)^n \leqslant \beta.$$ Finally, to prove Condition (\ref{axioms:4}) assume $f^\sharp \cdot \alpha \leqslant \beta \cdot f^\sharp$. By our assumptions we infer that $f^\sharp\cdot (\alpha\vee 1) \leqslant (\beta\vee 1)\cdot f^{\sharp}$. By induction hypothesis assume that $f^\sharp\cdot (\alpha\vee 1)^n \leqslant (\beta\vee 1)^n\cdot f^{\sharp}$ and consider 
$$
f^\sharp \cdot (\alpha\vee 1)^{n+1} = f^\sharp \cdot (\alpha \vee 1)^{n}\cdot (\alpha \vee 1) \leqslant  (\beta \vee 1)^n\cdot f^{\sharp}\cdot (\alpha\vee 1)\leqslant (\beta\vee 1)^{n+1}\cdot f^\sharp.
$$
This and the fact that the monad is $\omega$-cpo enriched proves that $f^\sharp\cdot \alpha^{*} \leqslant \beta^{*}\cdot f^\sharp$.
Using a similar argument we prove the assertion for the inverse inequality.
\qed

\subsubsection{Ordered saturation monad $TS$}
In this paragraph we assume the following:
\begin{itemize}
\item $T$ is an ordered saturation monad with saturation denoted by $(-)^{*}$,
\item a functor $S:\mathsf{C}\to\mathsf{C}$ lifts to $\overline{S}:\mathcal{K}l(T)\to \mathcal{K}l(T)$,
\item $(\overline{S},m,e)$ is a monad on the category $\mathcal{K}l(T)$.
\end{itemize}
Note that $S$ does not have to be a monad and we only assume that its lifting $\overline{S}$ is.  By composing two adjunctions $\mathsf{C}\rightleftarrows \mathcal{K}l(T)\rightleftarrows \mathcal{K}l(\overline{S})$ we introduce a monadic structure on $TS:\mathsf{C}\to\mathsf{C}$. The formula for the composition $\bullet$ in $\mathcal{K}l(TS)$ can be found in Subsection \ref{subsection:monads_on_kleisli}. It can be also expressed in terms of the composition $\cdot$ in $\mathcal{K}l(T)$ as follows. For $f:X\to TSY$ and $g:Y\to TSZ$ we have
$$
g\bullet f = m \cdot \overline{S}g \cdot f.
$$ 
The aim of this paragraph is to prove that given some mild assumptions on $\overline{S}$, the monad $TS$ is an ordered saturation monad. We will see examples of application of the theorems below in Subsection \ref{subsection:saturation_for_LTS} and in Section \ref{section:segala}.

\begin{thm}\label{theorem:ordered_saturation_monad_TS}
Assume $\overline{S}$ is locally monotonic and 
\begin{align} 
m_X\cdot \overline{S}[(m_X\cdot \overline{S}\alpha)^{*}\cdot  e_X] = (m_X\cdot \overline{S}\alpha)^{*} \label{identity:distributivity_S_over_star}
\end{align} for any $\alpha:X\rightdcirc \overline{S}X$ in $\mathcal{K}l(T)$.
Then $TS$ is an ordered saturation monad.
\end{thm}
\proof
In the first part of the proof we show that the Kleisli category for the monad $TS$ is order enriched.
For any two objects $X,Y$ we have $Hom_{\mathcal{K}l(TS)}(X,Y) = Hom_{\mathcal{K}l(T)}(X,SY)$. We impose an order on each hom-set $Hom_{\mathcal{K}l(TS)}(X,Y)$ by considering the partial order from $Hom_{\mathcal{K}l(T)}(X,SY)$. Let $f_1,f_2:X\to TSY$ and $g_1,g_2:Y\to TSZ$ be morphisms in $\mathcal{K}l(TS)$ and let $f_1\leqslant f_2$ and $g_1\leqslant g_2$. We have
\begin{align*}
&g_1\bullet f_1  = m\cdot \overline{S}g_1 \cdot f_1 \leqslant m\cdot \overline{S}g_1 \cdot f_2 = g_1\bullet f_2,\\
& g_1\bullet f_1 = m\cdot \overline{S}g_1 \cdot f_1 \leqslant m\cdot \overline{S}g_2 \cdot f_1 = g_2\bullet f_1.
\end{align*}

This proves that $\mathcal{K}l(TS)$ is an order enriched category. We will now show that the monad $TS$ is an ordered saturation monad. Let $\alpha:X\rightdcirc \overline{S}X$. We define $\alpha^{\bigstar}:X\rightdcirc \overline{S}X$ by:
\begin{align}
\alpha^{\bigstar} := (m_X \cdot \overline{S}\alpha)^{*}\cdot e_X . \label{eq:star_equation}
\end{align}
 We will prove that $(-)^{\bigstar}$ satisfies the desired properties from the definition of an ordered saturation monad. Indeed, to see that Condition (\ref{axiom:1}) holds, it is enough to see that $1_{\overline{S}X}\leqslant( m_X\cdot \overline{S}\alpha)^{*}$. Hence, 
$e_X= 1_{\overline{S}X}\cdot e_X \leqslant (m_X \cdot \overline{S}\alpha)^{*}\cdot e_X=\alpha^{\bigstar}$.
To prove that Condition (\ref{axiom:2}) is satisfied, note that $m_X\cdot \overline{S}\alpha\leqslant (m_X\cdot \overline{S}\alpha)^{*}$. Therefore, $\alpha = m_X\cdot \overline{S}\alpha \cdot e_X \leqslant (m_X\cdot \overline{S}\alpha)^{*}\cdot e_X$. 
To see Condition~(\ref{axioms:3}) holds consider
\begin{align*}
&\alpha^\bigstar \bullet \alpha^\bigstar = m_X\cdot \overline{S}((m_X \cdot \overline{S}\alpha)^{*}\cdot e_X)\cdot (m_X \cdot \overline{S}\alpha)^{*}\cdot e_X = \\
& (m_X \cdot \overline{S}\alpha)^{*}\cdot (m_X \cdot \overline{S}\alpha)^{*}\cdot e_X \leqslant (m_X \cdot \overline{S}\alpha)^{*}\cdot e_X = \alpha^\bigstar.
\end{align*}

To see Condition (\ref{axiom:3.5}) is true, take any $\beta:X\rightdcirc \overline{S}X$ and assume $e_X\leqslant \beta$, $\alpha\leqslant \beta$ and $\beta\bullet \beta \leqslant \beta$.   The first two inequalities imply that $m_X\cdot \overline{S}e_X\leqslant m_X\cdot \overline{S}\beta$ (i.e. $1_{\overline{S}X}\leqslant m_X\cdot \overline{S}\beta$)  and $m_X\cdot \overline{S}\alpha\leqslant m_X\cdot \overline{S}\beta$. The third implies that $\overline{S}(\beta\bullet \beta) \leqslant \overline{S}(\beta)$. This precisely means that  $\overline{S}( m_X\cdot \overline{S}(\beta)\cdot \beta) \leqslant \overline{S}\beta$. Hence,
\begin{align*}
&\overline{S}( m_X)\cdot \overline{S}^2(\beta)\cdot \overline{S}(\beta) \leqslant \overline{S}\beta,\\
&m_X\cdot \overline{S}( m_X)\cdot \overline{S}^2(\beta)\cdot \overline{S}(\beta) \leqslant m_X\cdot \overline{S}\beta,\\
& m_X\cdot m_{\overline{S}X} \cdot \overline{S}^2(\beta)\cdot \overline{S}(\beta) \leqslant m_X\cdot \overline{S}\beta,\\
& m_X\cdot \overline{S}\beta \cdot m_X \cdot \overline{S}\beta \leqslant m_X\cdot \overline{S}\beta.
\end{align*}
By the fact that $T$ is an ordered saturation monad we infer $(m_X\cdot \overline{S}\alpha)^{*}\leqslant m_X\cdot \overline{S}\beta$. This means that
$(m_X\cdot \overline{S}\alpha)^{*}\cdot e_X\leqslant m_X\cdot \overline{S}\beta\cdot e_X = \beta$ which proves the assertion.
We will now prove Condition (\ref{axioms:4}) holds. Consider a morphism $f:X\to Y$ in $\mathsf{C}$ and $\beta:Y\rightdcirc \overline{S}Y$. We will show the following implication:
$$
f^{\sharp\sharp}\bullet \alpha \leqslant \beta\bullet f^{\sharp\sharp} \implies f^{\sharp\sharp}\bullet \alpha^\bigstar \leqslant \beta^\bigstar \bullet f^{\sharp\sharp}.
$$
Note that $f^{\sharp\sharp} = e_Y\cdot f^{\sharp}$. We have:
\begin{align*}
f^{\sharp\sharp}\bullet \alpha &\leqslant \beta\bullet f^{\sharp\sharp},\\
m_Y\cdot \overline{S} f^{\sharp\sharp}\cdot \alpha &\leqslant m_Y\cdot \overline{S}\beta\cdot f^{\sharp\sharp},\\
m_Y\cdot \overline{S} (e_Y \cdot f^{\sharp})\cdot \alpha &\leqslant m_Y\cdot \overline{S}\beta\cdot e_Y \cdot f^{\sharp},\\
m_Y\cdot \overline{S} (e_Y) \cdot \overline{S} f^{\sharp}\cdot \alpha &\leqslant m_Y\cdot e_{\overline{S}Y}\cdot \beta \cdot f^{\sharp},\\
\overline{S} f^{\sharp}\cdot \alpha &\leqslant \beta \cdot f^{\sharp}.
\end{align*}
This implies that
\begin{align*}
\overline{S}^2 f^{\sharp}\cdot \overline{S}\alpha &\leqslant  \overline{S}\beta \cdot  \overline{S}f^{\sharp},\\
m_Y\cdot \overline{S}^2 f^{\sharp}\cdot \overline{S}\alpha &\leqslant  m_Y\cdot \overline{S}\beta \cdot  \overline{S}f^{\sharp},\\
\overline{S}f^{\sharp}\cdot m_X \cdot \overline{S}\alpha &\leqslant  m_Y\cdot \overline{S}\beta \cdot  \overline{S}f^{\sharp},\\
(Sf)^{\sharp}\cdot m_X \cdot \overline{S}\alpha &\leqslant  m_Y\cdot \overline{S}\beta \cdot  ({S}f)^{\sharp}.
\end{align*}
Since $T$ is ordered saturation monad we get:
\begin{align*}
\overline{S}f^{\sharp}\cdot (m_X \cdot \overline{S}\alpha)^{*} &\leqslant ( m_Y\cdot \overline{S}\beta)^{*} \cdot  \overline{S}f^{\sharp},\\
m_Y\cdot \overline{S}e_Y\cdot\overline{S}f^{\sharp}\cdot  (m_X \cdot \overline{S}\alpha)^{*} &\leqslant  m_Y\cdot \overline{S}[(m_Y\cdot \overline{S}\beta)^{*}\cdot e_Y] \cdot  \overline{S}f^{\sharp},\\
m_Y\cdot \overline{S}e_Y\cdot\overline{S}f^{\sharp}\cdot (m_X \cdot \overline{S}\alpha)^{*}\cdot e_X &\leqslant  m_Y\cdot \overline{S}(\beta^\bigstar)\cdot \overline{S} f^{\sharp}\cdot e_X,\\
m_Y\cdot \overline{S}e_Y\cdot\overline{S}f^{\sharp}\cdot \alpha^{\bigstar} &\leqslant m_Y\cdot \overline{S}\beta^\bigstar\cdot  e_Y\cdot f^{\sharp},\\
f^{\sharp\sharp}\bullet \alpha^\bigstar &\leqslant \beta^\bigstar \bullet f^{\sharp\sharp}.
\end{align*}
The implication with the inverse inequality is proved analogously.
\qed 

\begin{rem} Identity (\ref{identity:distributivity_S_over_star}) can be intuitively understood as a form of distributivity of $\overline{S}$ and its monadic multiplication over $(-)^{*}$. We will show this law on a more concrete example. The left hand side of this identity spelled out in the case of $T$ being a Kleene monad looks as follows:
$$
m_X\cdot \overline{S}\left (\mu x. ( 1\vee (m_X\cdot \overline{S}\alpha)\cdot x) \cdot e_X\right ) = m_X\cdot \overline{S}\left( \mu x. (e_X \vee m_X\cdot \overline{S}\alpha\cdot x)\right) .
$$  
The right hands side of this identity is
$\mu x. (1\vee m_X\cdot \overline{S}\alpha\cdot x)$.
The next theorem gives sufficient conditions for (\ref{identity:distributivity_S_over_star}) to hold in the setting of $\omega$-cpo enriched category $\mathcal{K}l(T)$ satisfying additional properties. To see the intuition behind identity (\ref{eq:star_equation}) the reader is referred to Example \ref{example_lts_saturation}.
\end{rem}

\begin{thm}
\label{theorem_star_monad_S}
Assume that:
\begin{itemize}
\item the Keisli category $\mathcal{K}l(T)$ is $\omega$-cpo enriched,
\item hom-sets in $\mathcal{K}l(T)$ admit finite joins,
\item $\overline{S}$ is locally continuous, i.e. $\overline{S}\bigvee_i f_i = \bigvee_i \overline{S}f_i$ for any countable directed family of morphisms $f_1\leqslant f_2\leqslant \ldots$,
\item for any $\alpha:X\rightdcirc \overline{S} X$ we have
$1 \vee m_X \cdot \overline{S} \alpha = m_X\cdot \overline{S}(e_X \vee \alpha)$.
\end{itemize}
Then the monad $TS$ is an ordered saturation monad.
\end{thm}
\proof
By our assumptions and Theorem \ref{theorem:dcpo_enriched_monads_star_monads} we have  $\alpha^{*} = \bigvee_{n\in \mathbb{N}} (\alpha\vee 1)^n$ for any $\alpha$.
Since any locally continuous functor is locally monotonic, by Theorem~\ref{theorem:ordered_saturation_monad_TS} we only need to show that
 $
m_X\cdot \overline{S}[(m_X\cdot \overline{S}\alpha)^{*}\cdot e_X] = (m_X\cdot \overline{S}\alpha)^{*}.$
We have $(m_X\cdot \overline{S}\alpha)^{*} =\bigvee_n (m_X\cdot \overline{S}\alpha\vee 1)^n$. Therefore,
\begin{align*}
& m_X\cdot \overline{S}\left [\left( \bigvee_n (m_X\cdot \overline{S}\alpha\vee 1)^n\right) \cdot e_X\right ] = \bigvee_n m_X\cdot \overline{S}\left [ \left ( m_X\cdot \overline{S}(\alpha\vee e_X)\right)^n \cdot e_X\right ] = \\
&\bigvee_n m_X\cdot \overline{S}\left [ \left ( m_X\cdot \overline{S}(\alpha\vee e_X)\right)^n \right ]\cdot \overline{S} e_X =\\
&\bigvee_n m_X\cdot\underbrace{ \left(\overline{S} m_X\cdot \overline{S}^2(\alpha\vee e_X)\right)\cdot  \ldots \cdot \left(\overline{S} m_X\cdot \overline{S}^2(\alpha\vee e_X)\right)}_\text{n}  \cdot \overline{S} e_X
\end{align*}
We have:
\begin{align*}
&\bigvee_n m_X\cdot \overline{S} m_X\cdot \overline{S}^2(\alpha\vee e_X)\cdot  \ldots \cdot\overline{S} m_X\cdot \overline{S}^2(\alpha\vee e_X)  \cdot \overline{S} e_X =\\
&\bigvee_n m_X\cdot m_{\overline{S}X}\cdot \overline{S}^2(\alpha\vee e_X)\cdot  \ldots \cdot\overline{S} m_X\cdot \overline{S}^2(\alpha\vee e_X)  \cdot \overline{S} e_X.
\end{align*}
This implies that:
\begin{align*}
&\bigvee_n m_X\cdot  \overline{S}(\alpha\vee e_X)\cdot m_X\cdot  \ldots \cdot\overline{S} m_X\cdot \overline{S}^2(\alpha\vee e_X)  \cdot \overline{S} e_X = \ldots =\\
&\bigvee_n m_X\cdot  \overline{S}(\alpha\vee e_X)\cdot  \ldots \cdot m_X\cdot \overline{S}(\alpha\vee e_X)  \cdot m_X\cdot \overline{S} e_X = \\
& \bigvee_n (m_X\cdot \overline{S}\alpha\vee 1)^n = (m_X \cdot
  \overline{S}\alpha)^{*}.\rlap{\hbox to 208.6 pt{\hfill\qEd}}
\end{align*}

\subsection{Saturation for LTS and non-deterministic automata}\label{subsection:saturation_for_LTS}
Here, we deal with coalgebras whose type functors are of the form $\mathcal{P}F_\tau$ for $F_\tau = \Sigma_\tau \times \mathcal{I}d$ and $F_\tau = \Sigma_\tau \times \mathcal{I}d+1$. The subsection is divided into two paragraphs. In the first paragraph we revisit the saturation procedure for LTS stated in Subsection~\ref{subsection:LTS_monad} and \ref{subsection:saturation_LTS}. The second paragraph is devoted to non-deterministic automata saturation.

\subsubsection{Labelled transition systems revisited} As noted in Example \ref{example:lifting_of_powerset} the functor $$\Sigma_\tau \times \mathcal{I}d:\mathsf{Set}\to \mathsf{Set}$$ lifts to $\overline{\Sigma_\tau}:\mathcal{K}l(\mathcal{P})\to \mathcal{K}l(\mathcal{P})$ by the corresponding distributive law 
given by the strength $t$ of $\mathcal{P}$.  Moreover, the strength map $t_{X,Y}$ satisfies:
$$
t_{X,Y}(x,\varnothing) = \varnothing \text{ and }t_{X,Y}(x,\bigcup_{i\in I} Y_i) = \bigcup_{i\in I}t_{X,Y}(x,Y_i).
$$
Hence,
for any family $\{f_i:X\to \mathcal{P}Y\}_{i\in I}$ of morphisms we have:
$$\overline{\Sigma_\tau}\bigvee_i f_i = \bigvee_i \overline{\Sigma_\tau} f_i.$$ 
%\begin{thm}
%The functor $\overline{\Sigma}$ preserves the order. Moreover, the monadic structure $(\overline{\Sigma},m,e)$ defined in Fact \ref{fact:LTS_label_monad} satisfies the following. For any $\alpha:X\rightdcirc \overline{\Sigma} X$ ($\alpha:X\to \mathcal{P}(\Sigma\times X)$) we have:
%$$
%m_X\cdot \overline{\Sigma}(\alpha \vee e_X) =  m_X\cdot \overline{\Sigma}\alpha \vee 1_X \text{ and } m_X \cdot \overline{\Sigma}[(m_X\cdot \overline{\Sigma} \alpha)^{*}\cdot e_X] = (m_X\cdot \overline{\Sigma}\alpha)^{*}
%$$
%\end{thm}
Consider the monad $(\overline{\Sigma_\tau},m,e)$ from Subsection \ref{subsection:LTS_monad}. We have the following.
\begin{thm}\label{theorem:LTS_monad_ordered_saturation_monad}
The LTS monad $\mathcal{P}(\Sigma_\tau \times \mathcal{I}d):\mathsf{Set}\to\mathsf{Set}$ is an ordered saturation monad with the saturation operator defined for any $\alpha:X\to \mathcal{P}(\Sigma_\tau\times X)$ by:
$$(m_X\cdot \overline{\Sigma_\tau }\alpha)^{*}\cdot e_X,$$
where $(-)^{*}$ denotes the saturation operator for $\mathcal{P}$-coalgebras. 
\end{thm}
\proof
The proof follows directly by the fact that the category $\mathcal{K}l(\mathcal{P})$ is enriched over complete join-semilattices, by the properties above and Theorem \ref{theorem_star_monad_S} (take $T=\mathcal{P}$ and $S=\Sigma_\tau \times \mathcal{I}d$).
\qed
\begin{exa}\label{example_lts_saturation}
Let $X=\{x,y,z\}$, $\Sigma_\tau =\{\sigma,\tau\}$ and consider the LTS coalgebra $\alpha:X\to \mathcal{P}(\Sigma_\tau \times X)$ whose structure is given by:
$$
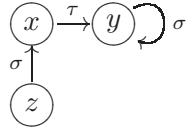
\entrymodifiers={++[o][F-]}
\SelectTips{cm}{}
\xymatrix@-1pc{ 
x\ar[r]^{\tau} & y \ar@(ur,dr)^{\sigma}\\
z\ar[u]^{\sigma}
}
$$
Consider the coalgebras $$\overline{\Sigma_\tau}\alpha:\Sigma_\tau \times X \to \mathcal{P}(\Sigma_\tau\times ( \Sigma_\tau \times X))\text{ and } m_X \cdot \overline{\Sigma_\tau}\alpha:\Sigma_\tau \times X\to \mathcal{P}(\Sigma_\tau \times X).$$
The first structure is an LTS coalgebra with the set of states given by $\Sigma_\tau\times X$. The second can be viewed as a $\mathcal{P}$-coalgebra over the same state space. They are depicted below.
$$
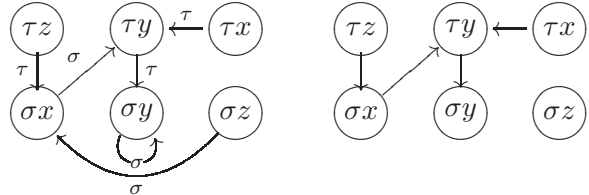
\entrymodifiers={++[o][F-]}
\SelectTips{cm}{} 
\xymatrix@-1pc{
\tau z \ar[d]_{\tau }  & \tau y \ar[d]^{\tau} & \ar[l]_{\tau} \tau x \\
\sigma x \ar[ur]^{\sigma}& \sigma y  \ar@(dl,dr)|{\sigma} & {\sigma z} \ar@/^2pc/[ll]^{\sigma}
}\qquad 
\xymatrix@-1pc{
\tau z \ar[d]  & \tau y \ar[d]  & \ar[l] \tau x \\
\sigma x  \ar[ur]& \sigma y   & {\sigma z}
}
$$
Finally, consider the coalgebras: 
$$(m_X \cdot \overline{\Sigma_\tau}\alpha)^{*}:\Sigma_\tau \times X\to \mathcal{P}(\Sigma_\tau \times X)
\text{ and }(m_X \cdot \overline{\Sigma_\tau}\alpha)^{*}\cdot e_X: X\to \mathcal{P}(\Sigma_\tau \times X).$$
The first structure should be considered a $\mathcal{P}$-coalgebra and it is simply the reflexive and transitive closure of the $\mathcal{P}$-coalgebra $m_X\cdot \overline{\Sigma_\tau} \alpha$. The second is a labelled transition system coalgebra. They are both depicted below.
\\
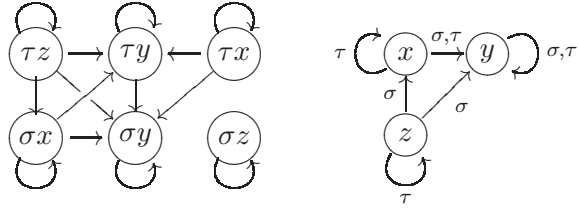
$$
\entrymodifiers={++[o][F-]}
\SelectTips{cm}{} 
\xymatrix@-1pc{
\tau z \ar[r] \ar[dr]|{\phantom{ a }} \ar@(ul,ur) \ar[d]  & \tau y \ar@(ul,ur) \ar[d] & \ar[l] \ar@(ul,ur) \tau x \ar[dl]\\
\sigma x\ar@(dl,dr) \ar[r] \ar[ur]& \sigma y \ar@(dl,dr)  & {\sigma z} \ar@(dl,dr)
}
\qquad 
\xymatrix@-1pc{ 
x\ar[r]^{\sigma,\tau} \ar@(dl,ul)^\tau & y \ar@(ur,dr)^{\sigma,\tau}\\
z\ar[u]^{\sigma} \ar@(dl,dr)_\tau \ar[ur]_\sigma
}
$$
To summarize, since the monadic multiplication of $\overline{\Sigma_\tau}$ commutes with the reflexive and transitive closure $(-)^{*}$ in $\mathcal{K}l(\mathcal{P})$ we can first turn an LTS coalgebra $\alpha$ into an endomorphism $m_X\cdot \overline{\Sigma_\tau}\alpha$ in $\mathcal{K}l(\mathcal{P})$ and saturate it afterwards to obtain the map $(m_X\cdot \overline{\Sigma_\tau}\alpha)^{*}$. Precomposing the saturated map with $e_X$ in $\mathcal{K}l(\mathcal{P})$ yields the desired labelled transition system. 
\end{exa}

\subsubsection{Non-deterministic automata with silent moves} Here, we take 
$$F_\tau = \Sigma_\tau\times \mathcal{I}d + 1.$$  This functor lifts to $\overline{F}_\tau:\mathcal{K}l(\mathcal{P})\to \mathcal{K}l(\mathcal{P})$ with the corresponding distributive law $\lambda:F_\tau \mathcal{P}\implies \mathcal{P}F_\tau$ given in Example \ref{example:NA_monad}. It is easy to see that
for an arbitrary non-empty family $\{f_i:X\to \mathcal{P}Y\}_{i\in I}$ of morphisms we have:
$$
\overline{F}_\tau (\bigvee f_i) = \lambda_Y \circ F_\tau \bigvee_i f_i = \bigvee_i \lambda_Y \circ F_\tau  f_i = \bigvee_i \overline{F}_\tau f_i.
$$
Let $(\overline{F}_\tau,m,e)$ be the monad from Example \ref{example:NA_monad}. By a similar reasoning as we used in the previous paragraph the $\mathsf{Set}$-based monad $\mathcal{P}(\Sigma_\tau \times \mathcal{I}d +1)$ is an ordered saturation monad. For any $\alpha:X\to \mathcal{P}(\Sigma_\tau \times X+1)$ we have:
\begin{align*}
\alpha^{\ast}(x) =  \{(\tau , x)\} \cup \{(\sigma,x')\mid x(\stackrel{\tau }{\to})^{\ast}\circ  \stackrel{\sigma}{\to} \circ (\stackrel{\tau}{\to})^{\ast} x'\} \cup   
  \{\checked \mid x(\stackrel{\tau }{\to})^{\ast} x' \text{ and } \checked \in \alpha(x') \}.  
\end{align*}
It should be noted that, in fact, a stronger condition holds.
\begin{thm}\label{theorem:nondeterministic_automata_kleisli}
The monad $\overline{F}_\tau:\mathcal{K}l(\mathcal{P})\to \mathcal{K}l(\mathcal{P})$ is an ordered saturation monad.
\end{thm}
\proof
Since $\mathcal{P}$ is a Kleene monad,  by Corollary \ref{coro:kleene_identity} the identity monad on $\mathcal{K}l(\mathcal{P})$ is an ordered saturation monad. The remaining part of the proof is similar to  the proof of Theorem \ref{theorem:LTS_monad_ordered_saturation_monad} and hence is omitted.
\qed
\begin{rem}
The small difference between the fact that $\mathcal{P}F_\tau:\mathsf{Set}\to \mathsf{Set}$ is an ordered saturation monad and that $\overline{F_\tau}:\mathcal{K}l(\mathcal{P})\to \mathcal{K}l(\mathcal{P})$ is one lies in Condition~(\ref{axioms:4}) from the definition of ordered saturation monad. By the above theorem this condition now holds for any morphism $f:X\to\mathcal{P}Y$ taken from $\mathcal{K}l(\mathcal{P})$.
\end{rem}

The reader is referred to Section \ref{section:weak_bisimulation_final_semantics} to see an interesting example of an application of the above theorem. We will use it to demonstrate what final weak bisimulation
semantics is for non-deterministic automata considered as $\overline{F_\tau}$-coalgebras in the base category $\mathcal{K}l(\mathcal{P})$. 

%%%%%%%%%%%%%%%%%%%%%%%%%%%%%%%%%%%%%%%%%%%%%%%%%%%%%%%%%%%
%
% Weak bisimulation
%
%%%%%%%%%%%%%%%%%%%%%%%%%%%%%%%%%%%%%%%%%%%%%%%%%%%%%%%%%%%
\section{Weak bisimulation for coalgebras over ordered saturation monads}
\label{section:weak_bisimulation}
The aim of this section is to define weak bisimulation and study its properties in the setting of $\mathsf{Set}$-based coalgebras. Here, we assume that $T$ is an ordered saturation monad on $\mathsf{Set}$. We restrict ourselves to this category intentionally since it is easier for us to formulate the definition of weak bisimulation and study its properties. The next section describes the more general case from the point of view of final weak bisimulation semantics. Note that in this section and in the definition of weak bisimulation we do not consider any visible or silent part of the type functor $T$. We assume this is handled internally by its monadic structure. This section is motivated by our results presented in \cite{Brengos12}. In this case though, a saturator is not an arbitrary closure operator, but it is directly linked to the type $T$ via the theory presented in Section~\ref{section:saturation}.

\begin{defi}
\label{definition:weak_bisimulation}
We say that a symmetric relation $R\subseteq  X\times X$ is a \emph{weak bisimulation} on $\alpha: X\to TX$ if there is a coalgebraic structure $\gamma: R\to TR$ such that in the category $\mathcal{K}l(T)$ we have:
\begin{align*}
&\alpha \cdot \pi_1^{\sharp}  \leqslant \pi_1^{\sharp}\cdot \gamma \text{ and }  
 \pi_2^{\sharp} \cdot \gamma\leqslant \alpha^{*} \cdot \pi_2^{\sharp}.  
\end{align*}
\end{defi}
These two inequalities can be restated in terms of the composition in $\mathsf{Set}$ as follows:
\begin{align*}
&\alpha \circ \pi_1  \leqslant T\pi_1\circ  \gamma \text{ and }  
 T\pi_2\circ \gamma\leqslant \alpha^{*} \circ  \pi_2.  
\end{align*}
They amount to the commutativity of the following (lax) diagram in $\mathsf{Set}$:
$$
	\xymatrix@-0.5pc{ \ar @{} [dr] |{\leqslant }
		X\ar[d]_{\alpha} &  R\ar[d]^{\gamma} \ar[l]_{\pi_1} \ar[r]^{\pi_2} \ar @{} [dr] |{\leqslant} &  X \ar[d]^{\alpha^{*}} & \\
		TX & TR\ar[l]^{T\pi_1} \ar[r]_{T\pi_2} & TX
	}
$$
The assumption about symmetry of $R$ allows us to formulate the definition in a more compact way, with only one diagram, and this is the only reason why we use it. 

\begin{exa}
Let $T=\mathcal{P}(\Sigma_\tau\times \mathcal{I}d)$ be the LTS monad from Subsection \ref{subsection:LTS_monad} and let $\alpha$ be an LTS coalgebra. As we have seen in Subsection \ref{subsection:LTS_monad} and \ref{subsection:saturation_for_LTS} its saturation $\alpha^{*}$ is given by:
$$
\alpha^{*}(x) = \{(\sigma,x')\mid x(\stackrel{\tau}{\to})^{*} \circ \stackrel{\sigma}{\to} \circ (\stackrel{\tau}{\to})^{*} x'\}\cup \{(\tau,x)\}.
$$
It is easy to see that a relation $R$ on $\alpha$ is a weak bisimulation according to the above definition if and only if it is a weak bisimulation in the sense of Definition~\ref{definition:LTS_weak_bisimulation}. To see another example the reader is referred to Section~\ref{section:segala}.
\end{exa}
\begin{lem}
 If a relation $R\subseteq X\times X$ is a strong bisimulation on $\alpha^{*}$ then it is a weak bisimulation on $\alpha$. \qed
\end{lem}
Given a morphism $f:X\to Y$ a \emph{kernel pair} is a pullback of $X\stackrel{f}{\to}Y \stackrel{f}{\leftarrow} X$. 
For more information on kernel pairs and $\mathsf{Set}$-endofunctors weakly preserving kernel pairs the reader is referred to e.g. \cite{GummEl}.
\begin{lem}
\label{lemma:weak_implies_strong}
Let $T$ weakly preserve kernel pairs. If $R\subseteq X\times X$ is an equivalence relation and a weak bisimulation on $\alpha: X\to TX$ then there is a structure $\gamma: R\to TR$ such that the following diagram commutes:
$$
	\xymatrix@-0.7pc{\ar @{} [dr] |{ = }
		X\ar[d]_{\alpha^{*}} & R\ar @{} [dr] |{ = } \ar[d]_\gamma \ar[l]_{\pi_1} \ar[r]^{\pi_2} & X \ar[d]^{\alpha^{*}} \\
		TX & TR\ar[l]^{T\pi_1} \ar[r]_{T\pi_2} & TX
	}
$$
In other words, $R$ is a strong bisimulation on $\alpha^{*}:X\to TX$.
\end{lem}
\proof
Since $R$ is a symmetric relation and a weak bisimulation there is a structure $\gamma:R\to TR$ such that
$\alpha \cdot \pi_1^{\sharp}  \leqslant \pi_1^{\sharp}\cdot \gamma$ and   
$\alpha^{*} \cdot \pi_2^{\sharp}  \geqslant \pi_2^{\sharp} \cdot \gamma$.
By the properties of saturation we can infer that $\alpha^{*} \cdot \pi_1^{\sharp}  \leqslant \pi_1^{\sharp}\cdot \gamma^{*} \text{ and }  
\alpha^{*} \cdot \pi_2^{\sharp}  \geqslant \pi_2^{\sharp} \cdot \gamma^{*}$. Let $p:X\to X/R; x\mapsto [x]_{R}$.  
Since $p\circ \pi_1 = p \circ \pi_2$ we have $p^{\sharp} \cdot \pi_1^{\sharp}\cdot \gamma^{*} = p^{\sharp} \cdot \pi_2^{\sharp}\cdot \gamma^{*}$. Therefore,
$$
p^{\sharp} \cdot \alpha^{*} \cdot \pi_1^{\sharp} \leqslant p^{\sharp} \cdot \pi_1^{\sharp}\cdot \gamma^{*} = p^{\sharp} \cdot \pi_2^{\sharp}\cdot \gamma^{*} \leqslant p^{\sharp} \cdot \alpha^{*} \cdot \pi_2^{\sharp}.
$$ 
Let $k:R\to R;(a,b)\mapsto (b,a)$. Hence,
$$
p^{\sharp} \cdot \alpha^{*} \cdot \pi_1^{\sharp}\cdot k^{\sharp} \leqslant p^{\sharp} \cdot \alpha^{*} \cdot \pi_2^{\sharp}\cdot k^{\sharp}.
$$
Since $\pi_1\circ k = \pi_2$ and $\pi_2\circ k =\pi_1$ we get $p^{\sharp} \cdot \alpha^{*} \cdot \pi_2^{\sharp} \leqslant p^{\sharp} \cdot \alpha^{*} \cdot \pi_1^{\sharp}$. Therefore,  $p^{\sharp} \cdot \alpha^{*} \cdot \pi_2^{\sharp} = p^{\sharp} \cdot \alpha^{*} \cdot \pi_1^{\sharp}$ (i.e. $Tp \circ \alpha^{*} \circ \pi_1 = Tp \circ \alpha^{*} \circ \pi_2$). Since $R$ with $\pi_1$ and $\pi_2$ is a pullback of $X\stackrel{p}{\to} X/R \stackrel{p}{\leftarrow} X$ and since $T$ weakly preserves kernel pairs there is a structure $\gamma':R\to TR$
making the following diagram commute:
$$
\xymatrix@-0.5pc{
& T X \ar[r]^{T p} & T(X/R) \\
& T R \ar[r]^{T \pi_2}\ar[u]_{T \pi_1} & T X \ar[u]_{T p}\\
R\ar@/^/[uur]^{\alpha^{*}\circ \pi_1} \ar@/_/[urr]_{\alpha^{*}\circ \pi_2}\ar@{.>}[ru]|{\gamma'}
}
$$ 
Hence, $\gamma':R\to TR$ is a coalgebra for which $\pi_1,\pi_2:R\to X$ are homomorphisms from $\gamma'$ to $\alpha^{*}$. This proves the assertion. 
\qed
We say that a monad $T$ whose Kleisli category is order enriched \emph{weakly lax preserves pullbacks} provided that for any pullback diagram on the left the diagram on the right is a weak pullback diagram:
$$
\xymatrix{
P\ar[d]_{p_1}\ar[r]^{p_2} & B\ar[d]^g \\
A\ar[r]_f & C
}\qquad 
\xymatrix{
TP\ar[d]_{Tp_1}\ar[r]^{Tp_2} & TB\ar[d]^{Tg} \\
TA\ar[r]_{Tf} & TC
}
$$
such that for any object $X$ with $q_1:X\to TA$, $q_2:X\to TB$ and $Tf\circ q_1 \leqslant Tg\circ q_2$  there is $\xi:X\to TP$ satisfying $q_1\leqslant Tp_1\circ \xi$ and $Tp_2\circ \xi\leqslant q_2 $:
\\
\\
$$
\xymatrix{
X\ar[r]^{q_2} \ar[d]_{q_1}\ar@{}[dr]|\leqslant   & TB\ar[d]^{Tg} \ar@{}[drrr]|\implies & & & X\ar@{}[dr]|\leqslant  \ar[r]|-\xi\ar@(ur,ul)^{q_2} [rr]_\leqslant \ar@(dl,l)_{q_1}[dr]& TP\ar[d]^{Tp_1}\ar[r]_{Tp_2} & TB\ar[d]^{Tg}\\
 TA\ar[r]_{Tf} & TC & & & & TA\ar[r]_{Tf} & TC
}
$$
\begin{exa}
The powerset monad weakly preserves pullbacks and also weakly lax preserves them. To see this consider two maps $A\stackrel{f}{\to}C \stackrel{g}{\leftarrow} B$. Their pullback is given by $P=\{(a,b)\mid f(a)=g(b)\}$, $\pi_1:P\to A$ and $\pi_2:P\to B$. Assume that for an object $X$ and morphisms $q_1:X\to \mathcal{P}A$ and $q_2:X\to \mathcal{P}B$ we have $\mathcal{P}f\circ q_1 \leqslant \mathcal{P}g\circ q_2$. For $x\in X$ let $A_x\subseteq A$ and $B_x\subseteq B$ such that $q_1(x) = A_x$ and $q_2(x)=B_x$. By our assumption we get $f(A_x)\subseteq g(B_x)$. This means that for any $a\in A_x$ there is $b_a\in B_x$ such that $f(a)=g(b_a)$. Define $\xi:X\to \mathcal{P}P$ on $x\in X$ by $\xi(x) = \{(a,b_a) \in P\mid a\in A_x\}$. The map $\xi$ satisfies the desired properties. By a similar argument we can prove that the LTS monads $\mathcal{P}(\Sigma_\tau \times \mathcal{I}d)$ and $\mathcal{P}(\Sigma^{*}\times \mathcal{I}d)$ from Section \ref{section:LTS_coalgebraically} also satisfy this property.
\end{exa}
\begin{lem}
The following assertions are true:
\begin{itemize}
\item Let arbitrary cotupling in $\mathcal{K}l(T)$ be monotonic. If  $\{R_i\}_{i\in I}$ is a family of weak bisimulations on $\alpha:X\to TX$ then $\bigcup_{i\in I} R_i$ is a weak bisimulation on this structure.
\item If $R$ and $S$ are weak bisimulations on $\alpha:X\to TX$ and if $T$ weakly lax preserves pullbacks then $R\circ S$ is a weak bisimulation.
\end{itemize} 
\end{lem}
\proof
We will now prove the first assertion. Let  $\{R_i\}_{i\in I}$ be a family of weak bisimulations on $\alpha:X\to TX$ with suitable structures $\gamma_i:R_i\to TR_i$. Let the coproduct in $\mathsf{Set}_T$ of the family of coalgebras $\{\gamma_i:R_i\to TR_i\}_i$ be denoted by $\gamma:\coprod_i R_i\to T(\coprod_i R_i)$ (such a coproduct always exists for coalgebras over $\mathsf{Set}$ - see e.g. \cite{GummEl,Rutt2000} for details). Since arbitrary cotupling in $\mathcal{K}l(T)$ is monotonic and since the coproducts in $\mathcal{K}l(T)$ come from the base category we have:
$$
\xymatrix@-0.5pc{
& \\
\ar[r]^{\pi_2} R_i\ar@(ur,ul)[rr]^{\mathsf{in}_i} \ar[d]_{\gamma_i} \ar@{}[dr]|\leqslant& X\ar[d]_{\alpha^{*}}\ar@{}[dr]|\geqslant & \coprod_i R_i \ar[l]_{p_2} \ar[d]^\gamma \\
TR_i\ar[r]_{T\pi_2}\ar@(dr,dl)[rr]_{T\mathsf{in}_i}  & TX & T\coprod_i R_i \ar[l]^{Tp_2}
}\qquad 
\xymatrix@-0.5pc{
& \\
\ar[r]^{\pi_1} R_i\ar@(ur,ul)[rr]^{\mathsf{in}_i} \ar[d]_{\gamma_i} \ar@{}[dr]|\geqslant& X\ar[d]_{\alpha}\ar@{}[dr]|\leqslant & \coprod_i R_i \ar[l]_{p_1} \ar[d]^\gamma \\
TR_i\ar[r]_{T\pi_1}\ar@(dr,dl)[rr]_{T\mathsf{in}_i}  & TX & T\coprod_i R_i \ar[l]^{Tp_1}
}
$$
In the above diagrams $\mathsf{in}_i:R_i\to \coprod_i R_i$ denotes the coprojection into the $i$-th component, and $p_1$, $p_2$ the cotuples $[\{\pi_1:R_i\to X\}_{i\in I}]$ and $[\{\pi_2:R_i\to X\}_{i\in I}]$ respectively.  Hence, the following inequalities hold in $\mathcal{K}l(T)$:
\begin{align}
p_1^\sharp \cdot \gamma \geqslant \alpha \cdot p_1^\sharp, \label{inequality:weak_1} \\
p_2^\sharp \cdot \gamma \leqslant \alpha^{*} \cdot p_2^\sharp. \label{inequality:weak_2}
\end{align}
Define 
$p:\coprod_i R_i\to \bigcup_i R_i; (x,y,i)\mapsto (x,y)$ and  consider a right inverse $q$ of this map. Define $\gamma':\bigcup_i R_i\to T(\bigcup_i R_i)$ by:
$$
\gamma' := Tp \circ \gamma \circ q = p^\sharp \cdot \gamma \cdot q^\sharp. 
$$
By the inequality (\ref{inequality:weak_1}) we have $p_1^\sharp \cdot \gamma\cdot q^\sharp \geqslant \alpha \cdot p_1^\sharp\cdot q^\sharp$. Since $p_1 = \pi_1 \circ p$ we have: 
$$\pi_1^\sharp  \cdot \gamma' =  \pi_1^\sharp \cdot p^\sharp \cdot \gamma \cdot q^\sharp = p_1^\sharp \cdot \gamma \cdot q^\sharp \geqslant \alpha \cdot p_1^\sharp\cdot q^\sharp = \alpha \cdot \pi_1^\sharp.$$
By the inequality (\ref{inequality:weak_2}) using a similar argument as above we prove $\pi_2^\sharp \cdot \gamma' \leqslant \alpha^{*}\cdot \pi_2^\sharp$.

To prove the second assertion consider weak bisimulations $R$ and $S$ with structures $\gamma_R:R\to TR$ and $\gamma_S:S\to TS$ respectively. Let $A$ with $p_1:A\to S$ and $p_2:A\to R$ be the pullback of $\pi_2:R\to X$ and $\pi_1:S\to X$. By our assumptions we have

$$
\xymatrix@-0.5pc{
& A\ar@/_0.5pc/[dl]_{p_2} \ar@/^0.5pc/[dr]^{p_1} \\
R\ar@{}[dr]|\leqslant \ar[d]_{\gamma_R^{*}}\ar[r]^{\pi_2} & X\ar@{}[dr]|\leqslant \ar[d]_{\alpha^{*}} & S\ar[d]^{\gamma_S^{*}}\ar[l]_{\pi_1} \\
TR\ar[r]_{T\pi_2} & TX & TS\ar[l]^{T\pi_1}\\
& TA\ar@/^0.5pc/[ul]^{Tp_2} \ar@/_0.5pc/[ur]_{Tp_1}
}\qquad 
\xymatrix@-0.5pc{
& A\ar@{-->}[d]_{\exists \xi} \ar@/^1pc/[ddr]^{\gamma_S^{*}\circ p_1} \ar@/_1pc/[ddl]_{\gamma_R^{*}\circ p_2}  \ar@{}[ddl]|\leqslant\ar@{}[ddr]|\leqslant  \\
& TA\ar[dr]_{Tp_1} \ar[dl]^{Tp_2} \\
TR\ar[dr]_{T\pi_1} & & TS\ar[dl]^{T\pi_1}\\
& TX
}
$$
Note that the set $A$ is given by $A=\{((a,b),(b,c))\mid (a,b)\in R \text{ and } (b,c)\in S\}$. Define the map $p:A\to R\circ S;((a,b),(b,c)) \mapsto (a,c)$. Surjectivity of $p$ allows us to introduce a coalgebraic structure on $R\circ S$ as follows 
$$\gamma:R\circ S\to T(R\circ S); \gamma = T(p)\circ \xi \circ q,$$ 
where $q$ is a right inverse of $p$. Moreover,
we have:
\begin{align*}
&\pi_1^\sharp \cdot \gamma = \pi_1^\sharp \cdot p^\sharp \cdot \xi \cdot q^\sharp =  \pi_1^\sharp \cdot p_1^\sharp \cdot \xi \cdot q^\sharp\geqslant \pi_1^\sharp \cdot \gamma^{*}_R \cdot p_1^\sharp \cdot q^\sharp  \\
&\geqslant \alpha\cdot \pi_1^\sharp \cdot p_1^\sharp \cdot q^\sharp  = \alpha\cdot \pi_1^\sharp \cdot p^\sharp \cdot q^\sharp = \alpha\cdot \pi_1^\sharp.
\end{align*}
Similarily we prove that $\pi_2^\sharp \cdot \gamma \leqslant \alpha^{*}\cdot \pi_2^\sharp$.
\qed
The following lemma and theorem are a direct consequences of the results above.
\begin{lem}
Assume that arbitrary cotupling in $\mathcal{K}l(T)$ is monotonic. Then the greatest weak bisimulation $\approx_{*}$ on a coalgebra $\alpha:X\to TX$ exists. Moreover, if additionally $T$ weakly lax preserves pullbacks then $\approx_{*}$ is an equivalence relation. \qed
\end{lem}

\begin{thm}\label{theorem:weak_bisim_eq_strong_on_saturated}
Assume that arbitrary cotupling in $\mathcal{K}l(T)$ is monotonic and let $T$ weakly lax preserve pullbacks. The greatest weak bisimulation $\approx_{*}$ on a coalgebra $\alpha:X\to TX$ coincides with the greatest strong bisimulation on $\alpha^{*}:X\to TX$. \qed
\end{thm}

 Assume that the category $\mathsf{Set}_T$ admits the terminal object $\zeta:Z\to TZ$ and let $\mathsf{beh}_{\alpha}$  denote the unique homomorphism from a coalgebra $\alpha:X\to TX$ to the final coalgebra $\zeta$.  By Theorem \ref{theorem:weak_bisim_eq_strong_on_saturated} and the usual coinduction principle \cite{Rutt2000} weak coinduction rule can be stated as follows. If the assumptions of Theorem \ref{theorem:weak_bisim_eq_strong_on_saturated} are met then for any $\alpha:X\to TX$ we have:
$$
x\approx_{\ast} y \iff \mathsf{beh}_{\alpha^{*}}(x) = \mathsf{beh}_{\alpha^{*}}(y).
$$ 
However, the lemma below says that in order to calculate $\mathsf{beh}_{\alpha^{*}}$ it is enough to compose $\mathsf{beh}_{\alpha}$  with $\mathsf{beh}_{\zeta^{*}}$. 
\begin{lem} \label{lemma:beh_map} For any $\alpha:X\to TX$ we have:
$\mathsf{beh}_{\alpha^{*}} = \mathsf{beh}_{\zeta^{*} }\circ \mathsf{beh}_\alpha.$
\end{lem}\enlargethispage{\baselineskip}
\proof
By the definition of an ordered saturation monad the map $\mathsf{beh}_\alpha$ is also a homomorphism from $\alpha^{*}$ to $\zeta^{*}$. Therefore, $\mathsf{beh}_{\zeta^{*} }\circ \mathsf{beh}_\alpha$ is a homomorphism from $\alpha^{*}$ to $\zeta$. By uniqueness we can infer the desired equality.
\qed

\begin{thm}
\label{theorem:weak_coinduction}
Assume that arbitrary cotupling in $\mathcal{K}l(T)$ is monotonic and let $T$ weakly lax preserve pullbacks. Then for any $x,y\in X$ we have
$$
x\approx_{*} y \iff \mathsf{beh}_{\zeta^{*} }\circ \mathsf{beh}_\alpha(x) = \mathsf{beh}_{\zeta^{*} }\circ \mathsf{beh}_\alpha(y).
$$ \qed
\end{thm}

\section{Weak bisimulation via final semantics}
\label{section:weak_bisimulation_final_semantics}
This section is devoted to defining weak bisimulation semantics via terminal object in a general setting of coalgebras whose type is a monad $T$ with $\mathcal{K}l(T)$ an order enriched category. We  define the following categories. Let $\mathsf{C}_{T,\leqslant}$ denote the category of all $T$-coalgebras and lax homomorphisms between them. To be more precise if $\alpha:X\to TX$ and $\beta:Y\to TY$ are $T$-coalgebras then a morphism $f:X\to Y$ is a \emph{lax homomorphism} if $T(f)\circ\alpha \leqslant \beta \circ f$. Let $\mathsf{C}^{*}_{T,\leqslant}$  denote a full subcategory of $\mathsf{C}_{T,\leqslant}$ consisting of $T$-coalgebras $\alpha:X\to TX$ satisfying $1\leqslant \alpha$ and $\alpha\cdot \alpha \leqslant \alpha$.
We can restate the above two conditions using the composition in $\mathsf{C}$ and order in $\mathcal{K}l(T)$ as follows:
$$
\xymatrix{
TX & TX\ar[l]_{id}\\
& X\ar[u]_{\alpha}\ar@/^1pc/[lu]^{\eta_X}\ar@{}[lu]|{\phantom{aa}\leqslant}
}\qquad 
\xymatrix{
TX\ar[d]_{T\alpha} & X\ar[l]_{\alpha}\ar[d]^\alpha\ar@{}[ld]|\leqslant \\
TTX \ar[r]_\mu & TX
}
$$
It is worth noting that, although  these two conditions are not the conditions that define Eilenberg-Moore algebras for a monad \cite{MacLane}, they are somewhat partly dual to them. Indeed, only some arrows are reversed and partial order introduced to the diagrams. Such categories have been considered in modern mathematical literature under a name of \emph{$T$-monoids} or \emph{Kleisli $T$-algebras}  (see e.g.  \cite{Gahler,HofMySeal,Seal}). If $T$ is the filter monad $\mathcal{F}$ on $\mathsf{Set}$ then the category $\mathsf{C}_{\mathcal{F},\leqslant}^{*}$ is equivalent to the category $\mathsf{Top}$ of topological spaces and continuous maps \cite{Gahler}.
Let $\mathsf{C}_T^{*}$ denote the category with objects from $\mathsf{C}_{T,\leqslant}^{*}$ and morphisms being standard homomorphisms from $\mathsf{C}_T$. In particular, the category $\mathsf{C}_T^{*}$ is a full subcategory of $\mathsf{C}_T$. 
 Now assume the following:
\begin{enumerate}[label=\({\alph*}]
\item the inclusion functor $\mathsf{C}^{*}_{T,\leqslant}\to \mathsf{C}_{T,\leqslant}$ has a left adjoint denoted by $(-)^{*}$,\label{adjunction_a}
$$
\xymatrix@-0.5pc{
\mathsf{C}_{T,\leqslant}\ar@<1.5ex>[r]^-{(-)^{*}}\ar@{}[r]|-\perp &  \mathsf{C}_{T,\leqslant}^{*}\ar@<1.5ex>[l]
}
$$
\item the functor $(-)^{*}:\mathsf{C}_{T,\leqslant}\to \mathsf{C}^{*}_{T,\leqslant}$ lifts to $(-)^{*}:\mathsf{C}_{T}\to \mathsf{C}^{*}_{T}$, i.e. \label{adjunction_b}
$$
\xymatrix@-0.5pc{
\mathsf{C}_{T}\ar@{-->}[r]^{(-)^{*}}\ar[d] & \mathsf{C}_{T}^{*}\ar[d]\\
\mathsf{C}_{T,\leqslant}\ar[r]_{(-)^{*}} & \mathsf{C}_{T,\leqslant}^{*}
}
$$
\item the functor $(-)^{*}:\mathsf{C}_{T,\leqslant}\to \mathsf{C}^{*}_{T,\leqslant}$ is the identity on morphisms. \label{adjunction_c}
\end{enumerate}

\begin{exa}
If $T$ is an ordered saturation monad then the functor $\mathsf{C}_{T,\leqslant}\to \mathsf{C}^{*}_{T,\leqslant}$ that assigns to any coalgebra $\alpha:X\to TX$ the saturated coalgebra $\alpha^{*}$ and which is the identity on morphisms satisfies (\ref{adjunction_a}), (\ref{adjunction_b}) and (\ref{adjunction_c}). This follows directly by the definition of ordered saturation monads and that $\beta\in \mathsf{C}_{T}^{*}$ iff $\beta^{*} = \beta$. 
\end{exa}

Assume the category $\mathsf{C}_T$ admits a final object $\zeta:Z\to TZ$. Let $\mathsf{beh}_\alpha$ denote the unique homomorphism from $\alpha:X\to TX$ to $\zeta:Z\to TZ$.  We define \emph{weak bisimulation semantics morphism} for any $T$-coalgebra $\alpha:X\to TX$ by
$$
\mathsf{wbeh}_{\alpha} := \mathsf{beh}_{\alpha^{*}}.
$$
Using a similar argument as in the proof of Lemma \ref{lemma:beh_map} we show the following.
\begin{thm}\label{theorem_wbeh}
For any $\alpha:X\to TX$ we have:
$$
\mathsf{wbeh}_{\alpha} = \mathsf{beh}_{\alpha^{*}} = \mathsf{beh}_{\zeta^{*}}\circ \mathsf{beh}_{\alpha} = \mathsf{wbeh}_{\zeta}\circ \mathsf{beh}_{\alpha}.
$$ \qed
\end{thm}
We end this section with an example of weak bisimulation semantics for coalgebras considered in a category different from $\mathsf{Set}$. 
\begin{exa}
Consider the monad $\overline{F_\tau}:\mathcal{K}l(\mathcal{P})\to \mathcal{K}l(\mathcal{P})$ from Example \ref{example:NA_monad}. As stated in Theorem \ref{theorem:nondeterministic_automata_kleisli} it is an ordered saturation monad. Let the $\mathcal{P}F_\tau$-coalgebra $\zeta$ be defined by:
$$\zeta:\Sigma_\tau^{*} \to \mathcal{P}(\Sigma_\tau \times \Sigma_\tau ^\ast+1)=\mathcal{P}F_\tau (\Sigma_\tau ^\ast);\left \{ \begin{array}{ccc}\sigma s &\mapsto & \{(\sigma,s)\}, \\ \varepsilon & \mapsto & \{\checked\}. \end{array} \right. $$
This coalgebra considered in $\mathcal{K}l(\mathcal{P})$ as an $\overline{F_\tau}$-coalgebra $\zeta:\Sigma_\tau^{\ast} \rightdcirc \overline{F_\tau} \Sigma_\tau ^\ast$ is the final object in $\mathcal{K}l(\mathcal{P})_{\overline{F_\tau}}$ \cite{HasJacSok}. For any non-deterministic automaton coalgebra $\alpha:X\to \mathcal{P}(\Sigma_\tau \times X~+~1)$ considered in $\mathcal{K}l(\mathcal{P})$ as a coalgebra $\alpha:X\rightdcirc \overline{F_\tau} X$ the unique homomorphism $\mathsf{beh}_\alpha:X\rightdcirc \Sigma^{*}_\tau$
is given by the map $\mathsf{beh}_\alpha:X\to \mathcal{P}(\Sigma_\tau^{*})$ which assigns to any state $x$ the set of words it accepts \cite{HasJacSok}. For a word $w\in \Sigma_\tau ^\ast$ we have:
$$
w\in \mathsf{beh}_\alpha(x) \iff \left\{ \begin{array}{c} \text{ if } w=\sigma_1 \ldots \sigma_n \text{ s.t. } x\stackrel{\sigma_1}{\to}\circ \ldots  \circ \stackrel{\sigma_n}{\to} x' \text{ and }\checked \in \alpha(x'), \\ 
\text{ if }w= \varepsilon \text{ and }\checked \in \alpha(x). \end{array} \right. 
$$
By Subsection \ref{subsection:saturation_for_LTS} and Theorem \ref{theorem_wbeh} it is easy to see that the weak bisimulation semantics morphism $\mathsf{wbeh}_\alpha = \mathsf{beh}_{\alpha^\ast}$ is in this case given by the following.
For a word $w\in \Sigma_\tau^{*}$ we have $w\in \mathsf{wbeh}_{\alpha} (x)$ provided that
$$
\left \{ \begin{array}{cc} w\in\tau^{*} &\text{ if } x(\stackrel{\tau}{\to})^{*} x' \text{ with } \checked \in \alpha(x'), \\
w\in \tau^{*}\sigma_1 \tau^{*} \ldots \tau^{*} \sigma_n \tau^{*}, \text{ for } \sigma_i \in \Sigma &\text{ if } x(\stackrel{\tau }{\to})^{*}\circ\stackrel{\sigma_1}{\to}\circ  (\stackrel{\tau}{\to})^{*}\ldots \circ\stackrel{\sigma_n}{\to}\circ  (\stackrel{\tau}{\to})^{*}x' \\
 & \text{ with } \checked \in \alpha(x').
\end{array}\right. 
$$
It is easy to see that for $x,x'\in X$ we have $\mathsf{wbeh}_\alpha (x) = \mathsf{wbeh}_\alpha (x')$ if and only if 
\begin{align}
\Sigma^{*}\cap \mathsf{wbeh}_\alpha (x) = \Sigma^{*} \cap \mathsf{wbeh}_\alpha (x'). \label{identity:trace_equivalence}
\end{align}
Any coalgebra $\alpha:X\to\mathcal{P}(\Sigma_\tau \times X+1)$ may be viewed as a non-deterministic automaton with silent transitions \cite{HasJacSok_jap,SilWester}. By identity (\ref{identity:trace_equivalence}) we can infer that $\mathsf{wbeh}_\alpha (x) = \mathsf{wbeh}_\alpha (x')$ if and only if $x$ and $x'$ admit the same weak traces \cite{HopUll}, i.e. if and only if $\text{tr}_\alpha (x) = \text{tr}_\alpha(x')$, where $\text{tr}_\alpha:X\to \mathcal{P}(\Sigma^{*})$ is given as follows. We have $w\in \text{tr}_\alpha(y)$ provided that:
$$
 \left \{ \begin{array}{cc} w=\varepsilon &\text{ if } y(\stackrel{\tau}{\to})^{*} y' \text{ with } \checked \in \alpha(y'), \\
w=\sigma_1  \ldots\sigma_n, \text{ for } \sigma_i \in \Sigma &\text{ if } y(\stackrel{\tau }{\to})^{*}\circ\stackrel{\sigma_1}{\to}\circ \ldots \circ\stackrel{\sigma_n}{\to}\circ  (\stackrel{\tau}{\to})^{*}y'
 \text{ with } \checked \in \alpha(y').
\end{array}\right. 
$$ 
\end{exa}
\section{Saturation and weak bisimulation for Segala systems}
\label{section:segala}
(Simple) probabilistic systems \cite{Segala, SegalaThesis}, known in the coalgebraic literature under the name of (simple) Segala systems, are modelled as coalgebras of the type $\mathcal{P}(\Sigma\times~\mathcal{D})$ and $\mathcal{P}\mathcal{D}(\Sigma\times \mathcal{I}d)$ respectively \cite{Sokolova} and require extra care. Although the powerset functor $\mathcal{P}$ and the distribution functor $\mathcal{D}$ are endowed with natural monadic structures, the combination $\mathcal{P}\mathcal{D}$ lacks one as there is no distributive law $\mathcal{DP}\implies \mathcal{PD}$ between the monads \cite{Varacca}. In order to deal with this obstacle, we adopt a variant of one of the approaches proposed in \cite{Varacca}  that was further generalised in \cite{Jacobs08}. Later in this section we introduce the monad $\mathcal{CM}$ which is inspired by the work of Jacobs \cite{Jacobs08}. As will be demonstrated, $\mathcal{CM}$ is suitable for modelling the combination of possibilistic and probabilistic observations. In particular, we will show that the monad $\mathcal{CM}(\Sigma_\tau \times \mathcal{I}d)$  can be used to model simple Segala systems and their probabilistic weak  bisimulations \cite{Segala,SegalaThesis}. 

Although Lynch and Segala in \cite{Segala} and Segala in \cite{SegalaThesis} present and study probabilistic systems in their full generality,  the notion of a probabilistic weak bisimulation is defined in their work only for simple systems. Therefore, we start the section by recalling the definitions proposed in \cite{Segala,SegalaThesis} concerning simple Segala systems only. 
For a simple Segala system coalgebra $\alpha:X\to \mathcal{P}(\Sigma_\tau \times \mathcal{D}X)$, a state $x\in X$ and $\sigma\in \Sigma_\tau$ we write $x\stackrel{\sigma}{\to}\mu$ if $(\sigma,\mu)\in \alpha(x)$. For a state $x\in X$ and a measure $\nu\in \mathcal{D}(\Sigma_\tau \times X)$ a pair $(x,\nu)$ is called a \emph{step} in $\alpha$  if there is $\sigma \in \Sigma_\tau$ and $\mu\in \mathcal{D}X$ such that $x\stackrel{\sigma}{\to}\mu$ and $\nu(\sigma,x') = \mu(x')$ for any $x'\in X$. A \emph{combined step} in $\alpha$ is a pair $(x,\nu)$, where $x\in X$  and $\nu\in \mathcal{D}(\Sigma_\tau \times X)$ for which there is a countable family of positive numbers $\{p_i\}_{i \in I}$ such that $\sum_{i\in I}p_i = 1$ and a countable family of steps $\{(x,\nu_i)\}_{i\in I}$ in $\alpha$ such that $\nu = \sum_{i\in I}p_i\cdot\nu_i$. Note that for the sake of simplicity and clarity of notation the definition of a combined step is a slight modification of a similar definition presented in \cite{Segala}. To be more precise, Segala also considers a possibility of a deadlock in a combined step. However, for simple Segala systems deadlocks are not taken into account. Hence, the notion of weak arrows $\stackrel{\sigma}{\leadsto}$ presented here remains the same regardless of this small difference between the two definitions and is defined as follows.  For any natural number $n$ we define $\stackrel{\sigma}{\leadsto}_n\subseteq X\times \mathcal{D}X$ inductively by:
\begin{itemize}
\item $\stackrel{\tau}{\leadsto}_0 = \{(x,\delta_x)\mid x\in X\}$ and $\stackrel{\sigma}{\leadsto}_0 = \varnothing$ for $\sigma\neq \tau$,
\item $x\stackrel{\sigma}{\leadsto}_{n+1} \mu$ if there is a combined step $(x,\nu)$ in $\alpha:X\to \mathcal{P}(\Sigma_\tau \times \mathcal{D}X)$ such that if $(\sigma',x')\notin \{\sigma,\tau\}\times X$ then $\nu(\sigma',x')=0$ and $$\mu = \sum_{(\sigma',x')\in \{\sigma,\tau\}\times X} \nu(\sigma',x')\cdot \mu_{(\sigma',x')}$$
and if $\sigma'=\sigma$ then $x'\stackrel{\tau}{\leadsto}_n \mu_{(\sigma',x')}$ otherwise $\sigma'=\tau$ and $x'\stackrel{\sigma}{\leadsto}_n \mu_{(\sigma',x')}$.
\end{itemize}
For any $\sigma \in \Sigma_\tau$ we write $a\stackrel{\sigma}{\leadsto} \mu$ whenever $a\stackrel{\sigma}{\leadsto}_n \mu$ for some $n$.

\begin{defi}\cite{Segala}\label{definition:prob_weak_bisimulation_segala}
We say that an equivalence relation $R$ on $X$ is a \emph{probabilistic weak bisimulation} on $\alpha$ if the following condition is satisfied:
$$
(x,y)\in R\text{ and } x\stackrel{\sigma}{\to} \mu \text{ implies } y\stackrel{\sigma}{\leadsto} \mu' \text{ and } \mu \equiv_{\mathcal{D}R} \mu',
$$
where $\equiv_{\mathcal{D}R} = (\mathcal{D}\pi_1\times\mathcal{D} \pi_2)(\mathcal{D}R)$.
\end{defi}

\begin{rem}
	Segala in \cite{SegalaThesis} also introduces a standard (i.e. not probabilistic) bisimulation. In this definition one does not consider a convex combination of measures. Segala argues that the probabilistic version is more suitable for probabilistic systems (see   \cite[Example~8.3.1]{SegalaThesis} for more details). 
\end{rem}
In this section we only work with finitary simple Segala systems, i.e. simple Segala systems for which the functor $\mathcal{D}$ is replaced with the functor $\mathcal{D}_{fin}$ which assigns to any set $X$ the set $\mathcal{D}_{fin} X$ of measures on $X$ with finite support. In this case we also modify the definition of a combined step where, instead of a countable family of positive numbers, we assume the family $\{p_i\}_{i\in I}$ to be finite.
The main reason for this simplification is that in order to put Segala systems into our framework, we need to provide the type functor with a monadic structure. The suitable monad to do so is the monad $\mathcal{CM}$ which is described below. As we will see, this monad only deals with measures with finite (not countable) support and their finite convex combinations.  We leave developing the theory describing an analogous monad which deals with countable measures and countable convex combinations as an open problem. To summarize, from now on we assume the following:
\begin{itemize}
\item in the definition of a combined step presented above we consider only a finite family of non-negative numbers $p_1,\ldots,p_n$ and a finite family of steps. With this change we also alter the definition of $\stackrel{\sigma}{\leadsto}$. 
\end{itemize} Hence, if we refer to Definition \ref{definition:prob_weak_bisimulation_segala} then we refer to the version with finitary combined steps.

 The rest of this section is devoted to presentation of the monad $\mathcal{CM}$, introducing a monadic structure on $\mathcal{CM}(\Sigma_\tau \times \mathcal{I}d)$ (Subsection~\ref{subsection:monad_CM})  and studying its properties from the point of view of definition of weak bisimulation from Section~\ref{section:weak_bisimulation}. We end this section with Theorem~\ref{theorem:segala_final} which claims that Segala's  definition of weak bisimulation and our approach coincide for simple Segala systems.

\subsection{The monad \texorpdfstring{$\mathcal{CM}$}{CM} and its properties}
\label{subsection:monad_CM}

The aim of this subsection is to present the monad $\mathcal{CM}$ which is highly inspired by the work of Jacobs \cite{Jacobs08}. A part of the results below come from \cite{Jacobs08}. However, a part of the construction of this monad diverges slightly from Jacobs' original construction \cite{Jacobs08}. See Remark \ref{remark:diverge_CM} below for a detailed discussion.

By $[0,\infty)$ we denote the semiring $([0,\infty),+,\cdot)$ of non-negative real numbers with ordinary addition and multiplication. By a $[0,\infty)$-\emph{semimodule} we mean a commutative monoid with actions $[0,\infty)\times (-)\to (-)$ satisfying axioms listed in e.g.  \cite{Golan}. For a set $X$ and a mapping $f:X\to Y$ put
\begin{align*}
& \mathcal{M} X  = \{\phi:X\to [0,\infty) \mid \mathsf{supp}(\phi) \text{ is finite}\},\\
& \mathcal{M} f:\mathcal{M} X \to \mathcal{M} Y; \mathcal{M} f(\phi)(y) = \sum_{x\in f^{-1}(y)}\phi(y).
\end{align*}
We will often denote elements $\phi \in \mathcal{M}X$ using the formal sum notation by $\sum_x \phi(x) \cdot x$ or simply by $\sum_{i=1,\ldots,n} \phi(x_i)\cdot x_i$ if $\mathsf{supp}(\phi) = \{x_1,\ldots,x_n\}$. The set $\mathcal{M} X$ carries a monoid structure via pointwise operation of addition, and $[0,\infty)$-action via
$$(a\cdot \phi) (x) := a\cdot \phi(x),$$
which turn $\mathcal{M} X$ into a free semimodule over $X$ (see e.g. \cite{Golan,Jacobs08} for details). Let the category of all $[0,\infty)$-semimodules and homomorphisms be denoted by $\mathsf{SMod_{[0,\infty)}(Set)}$. 
We have the following adjunction which yields a monadic structure on $\mathcal{M}$ (the left arrow is the forgetful functor):
$$
\xymatrix{
\mathsf{Set}\ar@<1.5ex>[r]^-{\mathcal{M}}\ar@{}[r]|-\perp &  \mathsf{SMod}_{[0,\infty)}(\mathsf{Set}).\ar@<1.5ex>[l]
}
$$
For any $[0,\infty)$-semimodule $M$ and any non-empty subset $U\subseteq M$ define its convex closure by:
$$
\overline{U} := \{ a_1\cdot x_1 +\ldots +a_n\cdot x_n \mid x_i\in U, a_i\in [0,\infty) \text{ s.t. } \sum_{i=1}^na_i  =1\}.
$$
The operator $\overline{(-)}$ is a closure operator \cite{Jacobs08}. We call a subset $U\subseteq M$ \emph{convex} if $\overline{U} = U$. Put
$
\mathcal{C}(M) = \{U\subseteq M\mid U\text{ is non-empty and convex} \}.
$
The set $\mathcal{C}(M)$ ordered by inclusion forms an \emph{affine complete lattice} (i.e. a poset with joins of all non-empty subsets) with joins over non-empty index sets $I$ given by $
\bigvee_i U_i = \overline{\bigcup_i U_i}$.
Following \cite{Jacobs08} we define:
\begin{itemize}
\item $U+V := \{x+y\mid x\in U, y\in V\}$ for $U,V\in \mathcal{C}(M)$,
\item $0:=\{0\}$,
\item for any $a\in [0,\infty)$ and $U\in \mathcal{C}(M)$ put $a\cdot U :=\{a\cdot x\mid x\in U\}$.
\end{itemize}

\begin{rem}\label{remark:diverge_CM}
The remaining part of the construction of the monad $\mathcal{CM}$ diverges slightly from the construction proposed in Jacobs' work \cite{Jacobs08}. To be more precise, in order to present a general definition of the monad $\mathcal{CM}$, Jacobs considers the category $\mathsf{SMod}_{S}(\mathsf{ACL})$ of semimodules which are affine complete lattices over an arbitrary zero sum-free semifield\footnote{A semiring is \emph{zero sum-free} if the condition $a+b=0$ implies $a=b=0$. It is a \emph{semifield} if all non-zero elements have multiplicative inverses.}  $S$ which is itself an affine complete lattice. The monad $\mathcal{CM}$ from \cite{Jacobs08} is then obtained by composing two adjunctions $\mathsf{Set} \rightleftarrows \mathsf{SMod}_{S}(\mathsf{Set})\rightleftarrows \mathsf{SMod}_{S}(\mathsf{ACL})$, with the second adjunction yielding the convex combinations monad $\mathcal{C}$. However, the semiring $[0,\infty)$ we consider here does not satisfy the desired properties. Although it is a zero sum-free semifield, it is not an affine complete lattice.  We can turn the semiring $[0,\infty)$ into an affine complete lattice by extending it with the greatest element $\infty$. However,  the new structure $[0,\infty]$ is not a semifield anymore. Since we are not aware of any construction of a zero sum-free semifield which is an affine complete lattice that contains $[0,\infty)$ as a subalgebra we deviate from \cite{Jacobs08} and build the convex combinations monad $\mathcal{C}$ directly on the category of $[0,\infty)$-semimodules. Note that, strictly speaking, by considering $[0,\infty)$-semimodules we go outside of the scope of Jacobs' setting. Nevertheless, many results stated in \cite{Jacobs08} hold in our setting as the assumption about $S$ being an affine complete lattice is only used in the definition of the category $\mathsf{SMod}_S(\mathsf{ACL})$.
\end{rem}

The proof of the statement below is direct and goes along the lines of the proof of \cite[Lemma 4.2]{Jacobs08} and hence we omit it. 

\begin{thm} We have:
\begin{itemize}
\item $\mathcal{C}(M)$ with the operations above is a $[0,\infty)$-semimodule.
\item The semimodule operations of $\mathcal{C}(M)$ preserve arbitrary non-empty joins.
\item For a semimodule homomorphism $f:M\to N$  put $$\mathcal{C}(f):\mathcal{C}(M) \to \mathcal{C}(N); U\mapsto f(U).$$ The assignment $\mathcal{C}$ is an endofunctor $\mathcal{C}:\mathsf{SMod}_{[0,\infty)}(\mathsf{Set})\to \mathsf{SMod}_{[0,\infty)}(\mathsf{Set})$.\qed
\end{itemize}
\end{thm}

For a $[0,\infty)$-semimodule $M$ consider the following maps 
$$\eta_M: M\to \mathcal{C}(M);x\mapsto \{x\} \text{ and } \mu_M: \mathcal{C}^2(M) \to \mathcal{C}(M);U\mapsto \bigcup U.$$ It is easy to see that these are well defined $[0,\infty)$-semimodule homomorphisms. Moreover, they induce natural transformations $\eta:\mathcal{I}d\implies \mathcal{C}$ and $\mu:\mathcal{C}^2\implies~\mathcal{C}$ between suitable endofunctors on $\mathsf{SMod}_{[0,\infty)}(\mathsf{Set})$.  
The proof of the following theorem is a straightforward verification of monad axioms. 
\begin{thm}
The triple $(\mathcal{C},\mu,\eta)$ is a monad on $\mathsf{SMod}_{[0,\infty)}(\mathsf{Set})$. \qed
\end{thm}
\noindent The composition of the following two adjunctions yields a monad $\mathcal{CM}:\mathsf{Set}\to\mathsf{Set}$:
$$
\xymatrix{
\mathsf{Set}\ar@<1.5ex>[r]^-{\mathcal{M}}\ar@{}[r]|-\perp &  \mathsf{SMod}_{[0,\infty)}(\mathsf{Set})\ar@<1.5ex>[l] \ar@<1.5ex>[r]^-{^\sharp} \ar@{}[r]|-\perp &  \mathcal{K}l(\mathcal{C}) \ar@<1.5ex>[l]^-{U_\mathcal{C}}.		 
}
$$
 For a set $X$ and a map $f:X\to Y$ we have:
\begin{align*}
& \mathcal{CM}X = \{U\subseteq \mathcal{M}X \mid U \text{ is convex and non-empty}\}, \\
& \mathcal{CM}f:\mathcal{CM}X\to \mathcal{CM}Y; U \mapsto {\mathcal{M}f(U)}.
\end{align*}
The unit and the multiplication of $\mathcal{CM}$ are given on their $X$-components by:
\begin{align*}
& X\to \mathcal{CM}X;x\mapsto \{1\cdot x\} \text{ and }\\
&\mathcal{CM}^2X\to \mathcal{CM}X;U\mapsto \bigcup_{\phi \in U} \sum_{V\in \mathcal{CM}X}\{\phi(V)\cdot \psi \mid \psi \in V\}. 
\end{align*}
The formula for the composition in $\mathcal{K}l(\mathcal{CM})$ is the same as the one given in \cite{Jacobs08} for Jacobs' monad. For $f:X\to \mathcal{CM}Y$ and $g:Y\to \mathcal{CM}Z$ we have:
\begin{align*}
& g\cdot f:X\to \mathcal{CM} Z;  x\mapsto \bigcup_{\phi\in f(x)}\sum_{y \in {supp} (\phi)}\{ \phi(y)\cdot \psi \mid \psi \in g(y)\}.
\end{align*}

The proof of the lemma below is a direct translation of  \cite[Section~6]{Jacobs08} and hence we omit it.
\begin{lem}
\label{lemma:basic_properties_of_CM}
We have the following:
\begin{itemize}
\item The Kleisli category $\mathcal{K}l(\mathcal{CM})$ is  enriched over directed complete partial orders with the order on hom-sets given by
$$
f\leqslant g \iff f(x)\subseteq g(x) \text{ for any }x\in X.
$$
\item If $\{f_i:X\rightdcirc Y\}_{i\in I}$ is a non-empty family of morphisms then $\bigvee_i f_i$ exists and is given by $\bigvee_i f_i(x) := \overline{\bigcup_i f_i(x)}$. Moreover, for any $g:Y\rightdcirc Z$ we have $g\cdot \left (\bigvee_i f_i\right) = \bigvee_i g\cdot f_i.$
\item Let $0_{X,Y}:X\to \mathcal{CM}Y;x\mapsto\{0\}$. Then for any $f:X\rightdcirc Y$ in $\mathcal{K}l(\mathcal{CM})$ we have $0\cdot f = f \cdot 0 =0$.\qed
\end{itemize} 
\end{lem}

\begin{thm}
 The monad $\mathcal{CM}$ is an ordered saturation monad. 
\end{thm}
\proof
By the fact that $\mathcal{K}l(\mathcal{CM})$ is enriched over directed complete partial orders and by Theorem \ref{theorem:dcpo_enriched_monads_star_monads} it follows that we only need to show the implication below. For any $\alpha:X\rightdcirc X$ and $\beta:Y\rightdcirc Y$ in $\mathcal{K}l(\mathcal{CM})$ and any $f:X\to Y$ in $\mathsf{Set}$ we have:
$$
f^{\sharp}\cdot \alpha \Box \beta\cdot f^{\sharp}\implies f^{\sharp}\cdot (\alpha\vee 1)\Box (\beta\vee 1)\cdot f^{\sharp} \text{ for }\box \in \{\leqslant,\geqslant\}.
$$ 
Indeed, by Lemma \ref{lemma:basic_properties_of_CM} it follows that $f^{\sharp}\cdot(\alpha\vee 1)=f^{\sharp}\cdot \alpha\vee f^{\sharp}$ for any $f:X\to Y$. We will now prove that $(\beta\vee 1)\cdot f^{\sharp} = \beta\cdot f^{\sharp}\vee f^{\sharp}$. Since we always have $(\beta\vee 1)\cdot f^{\sharp} \geqslant \beta\cdot f^{\sharp}\vee f^{\sharp}$, it is enough to show  that $(\beta\vee 1)\cdot f^{\sharp} \leqslant \beta\cdot f^{\sharp}\vee f^{\sharp}$. For $x\in X$ assume $\phi\in (\beta\vee 1)\cdot f^{\sharp}(x)$. We have $\phi\in (\beta\vee 1)(f(x))$. 
%Therefore, $\phi$ is of the form
%$$
%\phi = a_1\cdot \phi_1+\ldots a_{n-1} \cdot \phi_{n-1}+ (1-a_1-\ldots-a_{n-1})\cdot f(x), 
%$$
%where $\phi_k\in \beta(f(x))$ and $a_1,\ldots,a_{n-1}\in [0,1]$ such that $\sum_k a_k \leqslant 1$. 
This precisely means that $\phi\in \overline{\beta(f(x))\cup\{1\cdot f(x)\}}$. Hence, $\phi\in (\beta\cdot f^{\sharp}\vee f^{\sharp})(x)$ which proves the assertion. Now to prove the required implications assume $f^{\sharp}\cdot \alpha \Box \beta\cdot f^{\sharp}$ for $\Box \in\{\leqslant,\geqslant\}$. We have
$$
f^{\sharp}\cdot(\alpha\vee 1_X)=f^{\sharp}\cdot \alpha\vee f^{\sharp} \Box  \beta\cdot f^{\sharp}\vee f^{\sharp}= (\beta\vee 1_Y)\cdot f^{\sharp}.\eqno{\qEd}
$$

By the above result and Theorem \ref{theorem:dcpo_enriched_monads_star_monads} it follows that for any $\alpha:X\rightdcirc X$ in $\mathcal{K}l(\mathcal{CM})$ we have:
$$
\alpha^{*} = \bigvee_{n\in \mathbb{N}} (\alpha\vee 1)^n.
$$
Let us introduce some notation. For a coalgebra $\alpha:X\to \mathcal{CM} X$ and $x\in X$ we write $x\to_{\alpha} \phi$ if we have $\phi\in \alpha(x)$. Let $\Rightarrow_\alpha^n \subseteq X \times \mathcal{M}X$ denote the relation inductively defined as follows:
\begin{itemize}
\item $\Rightarrow_{\alpha}^0 = \{(x, 1\cdot x)\mid x\in X\}$,
\item if $x\to_\alpha \psi$ where $\psi = r_1\cdot x_1+\ldots +r_n\cdot x_n $ and $x_1\Rightarrow^n_{\alpha} \psi_1,\ldots, x_n\Rightarrow^n_\alpha \psi_n$ and $x\Rightarrow^n_{\alpha} \psi'$ then
$$
x\Rightarrow^{n+1}_{\alpha} p\cdot  (r_1\cdot \psi_1+\ldots +r_n\cdot \psi_n) + (1-p)\cdot \psi' \text{ for any }p\in [0,1].
$$
\end{itemize}
We have $\Rightarrow_{\alpha}^n \subseteq \Rightarrow_{\alpha}^{n+1}$ for any natural number $n$. By induction on $n$ we prove the following lemma.
\begin{lem} We have $x\Rightarrow_{\alpha}^n \psi$ if and only if  $\psi \in (\alpha\vee 1)^n(x)$. \qed
\end{lem}
Define  $\Rightarrow_{\alpha} := \bigcup_{n}\Rightarrow_{\alpha}^n.$
\begin{thm}
We have $x\Rightarrow_{\alpha} \psi$ if and only if $\psi \in \alpha^{*}(x)$.
\end{thm}
\proof
This follows directly by:
$$
\alpha^{*}(x) = \bigvee_n (\alpha\vee 1)^n(x) = \overline{\bigcup_n (\alpha\vee 1)^n(x)} =\bigcup_n (\alpha \vee 1)^n(x). 
$$
The last equality holds since $\{(\alpha\vee 1)^n(x)\}_{n\in \mathbb{N}}$ is an ascending family of convex sets.
\qed

The monad $\mathcal{CM}$ comes with strength $t_{X,Y}:X\times \mathcal{CM}Y\to \mathcal{CM}(X\times Y)$ 
given by
$$
t(\sigma,U) = \{\sum_{x\in supp \phi}\phi(x)\cdot  (\sigma,x)\mid \phi \in U\}.
$$
This yields a lifting $\overline{\Sigma_\tau}:\mathcal{K}l(\mathcal{CM})\to \mathcal{K}l(\mathcal{CM})$ of the functor $\Sigma_\tau \times \mathcal{I}d:\mathsf{Set}\to \mathsf{Set}$. Since  $\mathcal{K}l(\mathcal{CM})$ is a category with zero morphisms we may introduce a monadic structure on $\overline{\Sigma_\tau} \cong \overline{\Sigma}+\mathcal{I}d:\mathcal{K}l(\mathcal{CM}) \to \mathcal{K}l(\mathcal{CM})$
as in Subsection \ref{subsec:monad_on_TF}. The unit $e$ and  the multiplication $m$ of $\overline{\Sigma_\tau}$ are given on their $X$-components by:
\begin{align*}
& e_X:X\to \mathcal{CM}(\Sigma_\tau \times X); x\mapsto \{1\cdot (\tau,x)\},\\
& m_X:\Sigma_\tau \times \Sigma_\tau \times X\to \mathcal{CM}(\Sigma_\tau \times X); (\sigma_1,\sigma_2,x)\mapsto \left\{ \begin{array}{cc} \{1\cdot (\sigma_1,x)\} & \text{ if }\sigma_2=\tau,\\
\{1\cdot (\sigma_2,x)\} & \text{ if }\sigma_1=\tau,\\
\{0\} & \text{ otherwise.} \end{array} \right.
\end{align*}

\begin{thm}
The monad $\mathcal{CM}(\Sigma_\tau \times \mathcal{I}d)$ obtained by composing two adjunctions $$
\xymatrix{
\mathsf{Set}\ar@<1.5ex>[r]\ar@{}[r]|-\perp &  \mathcal{K}l(\mathcal{CM}) \ar@<1.5ex>[l] \ar@<1.5ex>[r] \ar@{}[r]|-\perp &  \mathcal{K}l(\overline{\Sigma}_\tau) \ar@<1.5ex>[l].		 
}
$$is an ordered saturation monad.
\end{thm}
\proof
We will prove that all assumptions of Theorem \ref{theorem_star_monad_S} are satisfied for $T= \mathcal{CM}$ and $\overline{S}=\overline{\Sigma_\tau}$. The first two assumptions hold by Lemma \ref{lemma:basic_properties_of_CM}. We will now show that $\overline{\Sigma_\tau}$ is locally continuous. We see that for any non-empty family $\{f_i:X\rightdcirc Y\}_{i\in I}$ of morphisms in $\mathcal{K}l(\mathcal{CM})$ we have $t_{\Sigma_\tau,Y}\circ \left( id_{\Sigma_\tau}\times \bigvee_i f_i\right) = \bigvee_i t_{\Sigma_\tau ,Y}\circ (id_{\Sigma_\tau}\times f_i)$. This means that $
\overline{\Sigma_\tau}\bigvee_i f_i = \bigvee_i \overline{\Sigma_\tau}f_i$. Finally, we will show that the last assumption of Theorem \ref{theorem_star_monad_S} holds, namely:
$
1 \vee m_X\cdot \overline{\Sigma_\tau } \alpha = m_X\cdot \overline{\Sigma_\tau}(e_X\vee \alpha).
$
Indeed, by Lemma~\ref{lemma:basic_properties_of_CM} we have
$$
1 \vee m_X\cdot \overline{\Sigma_\tau } \alpha =m_X\cdot \overline{\Sigma_\tau } e_X \vee m_X\cdot \overline{\Sigma_\tau } \alpha = m_X\cdot (\overline{\Sigma_\tau }e_X\vee \overline{\Sigma_\tau }\alpha) =  m_X\cdot \overline{\Sigma_\tau}(e_X
\vee \alpha).\eqno{\qEd}
$$

Following the guidelines of the proof of Theorem \ref{theorem:ordered_saturation_monad_TS} the formula for saturation of a coalgebra $\alpha:X\to \mathcal{CM}(\Sigma_\tau \times X)$  is given by 
$\alpha^{\bigstar} = (m_X\cdot \overline{\Sigma_\tau }\alpha)^{*}\cdot e_X$. Our aim now will be to describe the structure $m_X\cdot \overline{\Sigma_\tau} \alpha :\Sigma_\tau \times X\to \mathcal{CM}(\Sigma_\tau \times X)$ and its transitions. The lemmas below will be used in the next subsection. Note that the coalgebra $m_X\cdot \overline{\Sigma_\tau} \alpha$ is in fact a $\mathcal{CM}$-coalgebra with a state-space given by $\Sigma_\tau \times X$. Hence we adopt the notation for $\mathcal{CM}$-coalgebras introduced in this subsection.
From now on, in order to avoid heavy notation, we will  denote $\to_{m_X\cdot \overline{\Sigma_\tau}\alpha}$ and $\Rightarrow_{m_X\cdot \overline{\Sigma_\tau}\alpha}$ by $\to$ and $\Rightarrow$ respectively.  
\begin{lem} \label{lemma:m_sigma_1}
Assume $\sigma\neq \tau$. We have:
\begin{itemize}
\item $(\tau,x) \to \psi$ if and only if $\psi \in \alpha(x)$,
\item $(\sigma,x)\to \psi$ if and only if the following conditions are met:
\begin{enumerate}
\item  $\psi(\sigma',x')=0$ for any $\sigma'\neq \sigma$ and $x'\in X$,
\item $(\tau,x)\to \sum_{x'} \psi(\sigma,x')\cdot (\tau,x') + \phi$ for some $\phi$ such that $\phi(\tau,x')=0$ for any $x'$.
\end{enumerate}
\end{itemize} 

\end{lem}
\proof
It follows directly by the definition of the monad $(\overline{\Sigma_\tau},m,e)$.
\qed

\begin{lem}\label{lemma:tau_sigma_segala}
For any $n$ and $\sigma\in \Sigma_\tau$ if $(\tau,x)\Rightarrow^n \sum_i r_i\cdot~(\tau,x_i)$ then $$(\sigma,x)\Rightarrow^n \sum_i r_i\cdot (\sigma,x_i).$$
\end{lem}
\proof
The assertion follows directly by induction and Lemma \ref{lemma:m_sigma_1}.
\qed

\begin{lem}
\label{lemma:one_letter_saturated}
Assume $\sigma\neq \tau$. If $(\sigma,x)\Rightarrow\psi$ then $\psi(\sigma',x')=0$ for $\sigma'\neq \sigma $ and $x'\in X$.
\end{lem}
\proof
We will prove the assertion for $\Rightarrow^n$ by induction. The assertion is true for $\Rightarrow^0$. Assume it holds for $\Rightarrow^n$ and take $(\sigma,x)\Rightarrow^{n+1} \psi$. This means that there is $(\sigma,x)\to r_1 \cdot  (\sigma_1,x_1)+\ldots +r_m\cdot (\sigma_m,x_m)$ and $$(\sigma_1,x_1)\Rightarrow^n \psi_1,\quad \ldots, \quad  (\sigma_m,x_m)\Rightarrow^n \psi_m, \quad  (\sigma,x)\Rightarrow^n \psi''$$ such that
$$
\psi= p\cdot (r_1\cdot \psi_1+\ldots +r_m\cdot \psi_m) + (1-p)\cdot \psi'' \text{ for some }p\in [0,1].
$$
By Lemma \ref{lemma:m_sigma_1} it follows that $\sigma_1=\ldots=\sigma_m = \sigma$. By induction hypothesis it follows that for any $i=1,\ldots,m$ we have $\psi_i(\sigma',x')=0=\psi''(\sigma',x')$ for $\sigma'\neq \sigma$, $x'\in X$. Hence, $\psi(\sigma',x')=0$ for $\sigma'\neq \sigma$ and $x'\in X$.
\qed

\subsection{Segala and simple Segala systems as
  \texorpdfstring{$\mathcal{CM}(\Sigma_\tau \times
    \mathcal{I}d)$}{CM(Sigma-tau x Id)}-Coalgebras}
Consider a Segala system $\alpha:X\to \mathcal{P}\mathcal{D}_{fin}(\Sigma_\tau \times X)$ and define a $\mathcal{CM}(\Sigma_\tau \times \mathcal{I}d)$-coalgebra $\underline{\alpha}:~X\to \mathcal{CM}(\Sigma_\tau \times X)$ as follows:
$$
\underline{\alpha}(x) =\overline{\{0\}\cup  \alpha(x)}=\{p_1\cdot \mu_1 +\ldots p_n\cdot \mu_n \mid \sum_{i} p_i \leqslant 1 \text{ and }\mu_i\in \alpha(x)\}.
$$
Since any simple Segala system can also be considered a Segala system the above construction is applicable to simple Segala systems. From now on a $\mathcal{CM}(\Sigma_\tau \times \mathcal{I}d)$-coalgebra $\alpha:X\to \mathcal{CM}(\Sigma_\tau \times X)$ is called \emph{(simple) Segala system} whenever it is obtained via the above construction from a (simple) Segala system.
\begin{exa}
Consider a simple Segala system for $\Sigma = \{a,b\}$ whose state space is $X=\{x_1,x_2,x_3\}$ and whose structure is given by:
$$
\begin{array}{ccc}
x_1 &\mapsto &\{\frac{1}{3}\cdot (a,x_2) + \frac{2}{3} \cdot (a,x_3), 1\cdot (b,x_3)\},\\
x_2 &\mapsto &\{ 1\cdot  (a,x_1) \},  \\
 x_3 &\mapsto &\varnothing.
\end{array}
\qquad  \vcenter{\vbox{
\entrymodifiers={++[o][F-]}
\SelectTips{cm}{} 
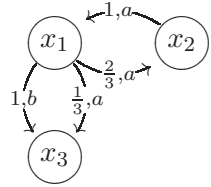
\xymatrix{
x_1\ar@/_1pc/[r]|{\frac{2}{3},a} \ar@/_1pc/[d]|{1,b} \ar@/^1pc/[d]|{\frac{1}{3},a} & x_2 \ar@/_1pc/[l]|{1,a}\\
x_3
}
}}
$$
Then the $\mathcal{CM}(\Sigma_\tau \times \mathcal{I}d)$-coalgebra associated with it has the structure given by:
\begin{align*}
\begin{array}{ccc}
x_1 & \mapsto & \{p_1 \cdot \frac{1}{3}\cdot (a,x_2) + p_1\cdot  \frac{2}{3}\cdot  (a,x_3) +p_2 \cdot (b,x_3)\mid p_1+p_2\in [0,1]\},\\
x_2 & \mapsto & \{ p \cdot (a,x_1) \mid p\in [0,1] \},  \\
x_3 & \mapsto & \{0\},
\end{array}
\end{align*}
which is depicted in the following diagram:
$$
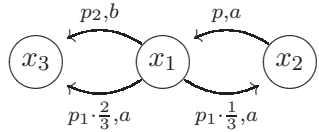
\entrymodifiers={++[o][F-]}
\SelectTips{cm}{} 
\xymatrix{
x_3 & x_1\ar@/_1pc/[r]_{p_1\cdot \frac{1}{3},a} \ar@/_1pc/[l]_{p_2,b} \ar@/^1pc/[l]^{p_1\cdot \frac{2}{3},a} & x_2 \ar@/_1pc/[l]_{p,a}
}
$$
\end{exa}
 The aim of this subsection is to prove that the probabilistic weak bisimulation for simple Segala systems from Definition \ref{definition:prob_weak_bisimulation_segala} coincides with the coalgebraic weak bisimulation from Section \ref{section:weak_bisimulation} for these systems considered as $\mathcal{CM}(\Sigma_\tau \times \mathcal{I}d)$-coalgebras.
 
 Assume $\alpha:X\to \mathcal{CM}(\Sigma_\tau \times X)$ is a simple Segala system obtained from a $\mathcal{P}(\Sigma_\tau \times \mathcal{D}_{fin})$-coalgebra whose strong and weak arrows, defined at the beginning of this section, are denoted by $\to$ and $\leadsto$ respectively. Note that we also use the symbol $\to$ to denote the relation $\to_{m_X\cdot \overline{\Sigma_\tau} \alpha}$. However, since the state space of the $\mathcal{CM}$-coalgebra $m_X\cdot \overline{\Sigma_\tau}\alpha:\Sigma_\tau \times X\ \to \mathcal{CM}(\Sigma_\tau \times X)$ is different from the state space of the simple Segala system taken into consideration this symbol overloading should not lead to any ambiguity.

\begin{lem}
\label{lemma:prob_weak_implies_weak}
 For any $x\in X$ if $x\stackrel{\sigma}{\leadsto} \mu$ then
 $(\tau,x) \Rightarrow \sum_{x}\mu(x)~\cdot~(\sigma,x)$. Moreover, if $x\stackrel{\tau}{\leadsto} \mu$ then $(\sigma,x)\Rightarrow\sum_{x}\mu(x)\cdot (\sigma,x)$ for any $\sigma\in \Sigma_\tau$.
\end{lem}
\proof

We will show the above statement holds for $\leadsto_n$ and $\Rightarrow^n$ for any $n$. Indeed, the two conditions hold for $\leadsto_0$ and $\Rightarrow^0$.
Now assume that both assertions are true for $n$. Let $x\stackrel{\sigma}{\leadsto}_{n+1} \mu$ for $\sigma\in \Sigma_\tau$. This means that there is a combined step $(x,\nu)$ such that for $(\sigma',x')\notin \{\sigma,\tau\}\times X$ we have $\nu(\sigma',x')=0$ and $$\mu = \sum_{(\sigma',x')\in \{\sigma,\tau\}\times X} \nu(\sigma',x')\cdot \mu_{(\sigma',x')}.$$
In the above, if $\sigma'=\sigma$ then $x'\stackrel{\tau}{\leadsto}_n \mu_{(\sigma',x')}$ otherwise $\sigma'=\tau$ and $x'\stackrel{\sigma}{\leadsto}_n \mu_{(\sigma',x')}$. The fact that $\nu$ is a combined step with the above properties implies that
$$
\nu = \sum_{i=1,\ldots,n} p_i \cdot \sum_{x'} \nu_i(x')\cdot (\sigma,x') + \sum_{j=1,\ldots,m} q_j \cdot\sum_{x'} \nu'_j(x')\cdot (\tau,x'),
$$
where $x\stackrel{\sigma}{\to} {\nu}_i$ and $x\stackrel{\tau}{\to} {\nu'}_j$ and $\sum_i p_i + \sum_j q_j = 1$. Hence,  by Lemma \ref{lemma:m_sigma_1} it follows that $(\tau,x)\to \nu$. By induction hypothesis we have 
\begin{itemize}
\item $(\sigma,x')\Rightarrow^n \sum_{x\in X}\mu_{(\sigma,x')}(x)\cdot (\sigma,x)$,
 \item  $(\tau,x')\Rightarrow^n \sum_{x\in X}\mu_{(\tau,x')}(x)\cdot (\sigma,x)$. 
\end{itemize}  Moreover,
\begin{align*}
&\mu = \sum_{(\sigma',x')\in \{\sigma,\tau\}\times X} \nu(\sigma',x')\cdot \mu_{(\sigma',x')} = \\
&\sum_{i=1,\ldots,n} p_i\cdot \sum_{x'\in X}  {\nu_i}(x') \cdot  \mu_{(\sigma,x')}+ \sum_{j=1,\ldots,m}q_i\cdot  \sum_{x'\in X}  {\nu_j'}(x')\cdot \mu_{(\tau,x')}.
%\\
%&\sum_{i=1,\ldots,n} \sum_{x'\in X} p_i\bullet \overline{\nu_i}(x') \bullet \mu_{(\sigma,x')}+\sum_{j=1,\ldots,m} \sum_{x'\in X} q_i\bullet 	\overline{\nu_j'}(x')\bullet \mu_{(\tau,x')}= \\
%&\sum_{i=1,\ldots,n}  p_i\bullet \sum_{x'\in X} \overline{\nu_i}(x') \bullet \mu_{(\sigma,x')}+\sum_{j=1,\ldots,m} q_i\bullet \sum_{x'\in X}  	\overline{\nu_j'}(x')\bullet \mu_{(\tau,x')}=\\
%&\sum_{i=1,\ldots,n}  p_i\bullet \sum_{x'\in X} \nu_i(\sigma,x') \bullet \mu_{(\sigma,x')}+\sum_{j=1,\ldots,m} q_i\bullet \sum_{x'\in X}  	\nu_j'(\tau,x')\bullet \mu_{(\tau,x')}.
\end{align*}
Hence, by the definition of $\Rightarrow^{n+1}$ and the fact that $(\tau,x)\to \nu$  we can infer that 
 $$(\tau,x)\Rightarrow^{n+1} \sum_{x}\mu(x)\cdot (\sigma,x).$$
Now, to prove that the condition $x\stackrel{\tau}{\leadsto}_{n+1} \mu$ implies $(\sigma,x)\Rightarrow^{n+1}\sum_{x}\mu(x)\cdot (\sigma,x)$
it is enough to see that by the previous condition we have $(\tau,x)\Rightarrow^{n+1}\sum_{x}\mu(x)\cdot (\tau,x)$. Then by Lemma~\ref{lemma:tau_sigma_segala} we get
$$(\sigma,x)\Rightarrow^{n+1}\sum_{x}\mu(x)\cdot (\sigma,x).\eqno{\qEd}$$

An expression $\psi\in \mathcal{M}(\Sigma_\tau\times X)$ for which there is a letter $\sigma\in \Sigma_\tau$ such that $\sum_{x}\psi(\sigma,x) = 1$ and $\psi(\sigma',x')=0$ for any $\sigma'\neq \sigma$ and $x'\in X$ is called \emph{simple probabilistic expression over} $\sigma$. 

\begin{lem}
\label{lemma:simple_prob_exp_arrow}
 If $(\tau,x)\Rightarrow  \psi$ and $\psi = \sum_i p_i\cdot (\sigma,x_i)$ is a simple probabilistic expression over $\sigma$ then $x\stackrel{\sigma}{\leadsto} \sum_{i} p_i \cdot x_i$. 
\end{lem}

\proof
We will again prove the assertion for relations $\Rightarrow^n$ and $\leadsto_n$ by induction. It is true for $\Rightarrow^0$ and $\leadsto_0$. Now assume it holds for  $\Rightarrow^n$ and  $\leadsto_n$ and take $(\tau,x) \Rightarrow^{n+1} \psi$,  where $\psi$ is a simple probabilistic expression over $\sigma$. By the definition of $\Rightarrow^{n+1}$ this means that 
\begin{itemize}
\item $(\tau,x)\to  r_1\cdot (\sigma_1,x_1)+\ldots +r_n\cdot(\sigma_n,x_n)$ and 
\item $(\sigma_1,x_1)\Rightarrow^n \psi_1,\ldots, (\sigma_n,x_n)\Rightarrow^n\psi_n$ and
\item $(\tau,x)\Rightarrow^n \psi'$ such that
\end{itemize}  
$$
\psi =  p\cdot (r_1\cdot \psi_1+\ldots +r_n\cdot \psi_n) + (1-p)\cdot \psi' \text{ for some }p\in [0,1].
$$
Since $\psi$ is a simple probabilistic expression over $\sigma$ it follows that  $\psi_i$ for any $i=1,\ldots,n$ and $\psi'$ are simple probabilistic expressions over $\sigma$ and $r_1+\ldots +r_n = 1$. This, together with Lemma \ref{lemma:one_letter_saturated}, means that the letters $\sigma_i$ satisfy $\sigma_i \in \{\sigma,\tau\}$. Therefore,  by induction hypothesis and the definition of $\leadsto_{n+1}$ it follows that $x\stackrel{\sigma}{\leadsto}_{n+1} \sum_{i} p_i~\cdot~x_i$.
\qed

\begin{thm}\label{theorem:segala_final}
Assume $R\subseteq X\times X$ is an equivalence relation. The following conditions are equivalent:
\begin{itemize}
\item $R$ is a weak bisimulation in the sense of Definition \ref{definition:prob_weak_bisimulation_segala},
\item $R$ is a weak bisimulation on $\alpha$ in the sense of Definition \ref{definition:weak_bisimulation}.
\end{itemize}
\end{thm}

\proof
Assume $R$ is a weak bisimulation the sense of Definition \ref{definition:prob_weak_bisimulation_segala}. This means that for any pair $(x,y)\in R$ the following holds. If for a letter $\sigma\in \Sigma_\tau$ we have $x\stackrel{\sigma}{\to}_{} \mu$ then $y\stackrel{\sigma}{\leadsto} \mu'$ and 
$\mu \equiv_{\mathcal{D}_{fin} R} \mu'$. Since $\mathcal{D}_{fin}$ is a subfunctor of $\mathcal{M}$ we have $$\sum_{x'} \mu(x')\cdot  (\sigma,x')\equiv_{\mathcal{M}(\Sigma_\tau\times R)} \sum_{x'}\mu'(x')\cdot (\sigma,x').$$ 
This precisely means that there is an element $\vec{r}_{(x,y)}^{\sigma,\mu}\in \mathcal{M}(\Sigma_\tau \times R)$ such that 
\begin{itemize}
\item $\mathcal{M}(\Sigma_\tau \times \pi_1)(\vec{r}_{(x,y)}^{\sigma,\mu}) = \sum_{x'} \mu(x')\cdot (\sigma,x')$ and 
\item  $\mathcal{M}(\Sigma_\tau \times \pi_2)(\vec{r}_{(x,y)}^{\sigma,\mu}) = \sum_{x'} \mu'(x')\cdot (\sigma,x')$.
\end{itemize}
 Define $\gamma:R\to \mathcal{CM}(\Sigma_\tau \times R)$ as follows:
 $$
 \gamma(x,y) = \overline{\{\vec{r}_{(x,y)}^{\sigma,\mu}\mid x\stackrel{\sigma}{\to} \mu \text{ for }\sigma\in \Sigma_\tau \}\cup\{0\}}.
 $$
 It is easy to show that by Lemma \ref{lemma:prob_weak_implies_weak} the structure $\gamma$ satisfies the desired properties of Definition \ref{definition:weak_bisimulation}. Now conversely, let $\gamma:R\to \mathcal{CM}(\Sigma_\tau \times R)$ be a structure satisfying the conditions in Definition \ref{definition:weak_bisimulation} for $\alpha$. Consider $(x,y)\in R$ and let $x\stackrel{\sigma}{\to} \mu$. Then $\sum_{x'}\mu(x')\cdot (\sigma,x')\in \alpha(x)$. Let $\vec{r}_{(x,y)}\in \gamma(x,y)$ denote the element such that $$\mathcal{M}(\Sigma_\tau \times \pi_1)(\vec{r}_{(x,y)}) = \sum_{x'} \mu(x')\cdot (\sigma,x')$$ 
 and put $\psi := \mathcal{M}(\Sigma_\tau \times \pi_2)(\vec{r}_{(x,y)})$. Since $\mathcal{M}(\Sigma_\tau \times \pi_1)(\vec{r}_{(x,y)})$ is a simple probabilistic expression over $\sigma$ then so is $\psi$. Since  $(\tau,y) \Rightarrow \psi$ by Lemma \ref{lemma:simple_prob_exp_arrow} it follows that $y\stackrel{\sigma}{\leadsto} \psi$. Hence, $R$ is a weak bisimulation in the sense of Definition~\ref{definition:prob_weak_bisimulation_segala}.
\qed

\section{Summary and future work}
\label{section:summary}
This paper presents a general setting in which it is possible to define and study properties of weak bisimulation for coalgebras. In this setting we require from the type of coalgebras we consider to be a monad whose Kleisli category is order enriched. However, not all monads satisfying this condition fit into our framework as not always the adjunction $\mathsf{C}_{T,\leqslant}\rightleftarrows \mathsf{C}_{T,\leqslant}^{*}$ described in Section \ref{section:weak_bisimulation_final_semantics} exists. For instance, consider the subdistribution monad $\mathcal{D}_{\leqslant 1}$. Since the class of objects of $\mathsf{C}_{\mathcal{D}_{\leqslant 1},\leqslant}^{*}$ consists only of coalgebras $1_X:X\to \mathcal{D}_{\leqslant 1} X;x\mapsto \delta_x$ for any set $X$, there fails to be a left adjoint to the inclusion functor $\mathsf{C}_{\mathcal{D}_{\leqslant 1},\leqslant}^{*}\to \mathsf{C}_{\mathcal{D}_{\leqslant 1},\leqslant}$.  It is worth noting that in \cite{SokViWor} a coalgebraic weak bisimulation for $\mathcal{D}_{\leqslant 1}(\Sigma_\tau \times \mathcal{I}d)$-coalgebras has been successfully defined and studied. However, the authors extend the type they consider and work with coalgebras whose type is the following functor:
\begin{align*}
&\mathcal{G}_{\Sigma_\tau} (X) = (\mathcal{P}\Sigma_\tau \times \mathcal{P}X \to [0,1]) \text{ for any object } X, \\
&\mathcal{G}_{\Sigma_\tau} f(\nu) = \nu \circ (id_{\mathcal{P}\Sigma_\tau }\times f^{-1}) \text{ for } f:X\to Y \text{ and }\nu \in \mathcal{G}_{\Sigma_\tau} X.
\end{align*}
 We believe that a similar approach can be adopted here to fit these coalgebras into our setting. Nevertheless, we leave it as an open problem.

Recall that in Section \ref{section:hiding_invisible_transitions} given a functor $TF_\tau $ we proposed two ways to handle the invisible transition by a monadic structure. Although these methods lead to two different notions of saturation, at least for LTS weak bisimulation coincides for both of them. It would be very interesting to see how does the notion of weak bisimulation behave for these two approaches in general.

We also plan to investigate the properties of the category $\mathsf{C}_{T,\leqslant}^{*}$ of saturated $T$-coalgebras that is used in Section~\ref{section:weak_bisimulation_final_semantics} to express saturation via existence of a certain adjunction. 

\subsubsection*{Acknowledgements.}
I would like to thank Alexandra Silva for valuable remarks on an early version of this paper. I am very grateful to the anonymous referees for various comments that have hopefully made this paper more interesting and more pleasant to read. Finally, I would like to thank Tony Barrett and Bartek Jab\l o\'nski for their linguistic and technical support.

\vspace{-40 pt}
\end{document}